\documentclass[10pt, pre, amsmath, onecolumn, showpacs, superscriptaddress]{revtex4-1}
\usepackage{graphicx,color}
 \graphicspath{{./}{./figures/}}
\usepackage{ams math}
\usepackage{enumerate}
\usepackage{mathbbol}
\usepackage{amsfonts}
\usepackage{natbib}

\usepackage{bm}
\usepackage{hyperref}
\hypersetup{
    colorlinks=true,       
    linkcolor=red,          
    citecolor=blue,        
    filecolor=magenta,      
    urlcolor=cyan           
}

\usepackage{color}
\usepackage[dvipsnames]{xcolor}

\usepackage{cancel}

\usepackage[caption=false]{subfig}
\usepackage{bbm}

\newcommand{\sgn}{{\mathrm{sgn}}}

\newcommand{\be}{\begin{equation}}
\newcommand{\ee}{\end{equation}}
\newcommand{\bea}{\begin{eqnarray}}
\newcommand{\eea}{\end{eqnarray}}
\newcommand{\beq}{\begin{eqnarray}}
\newcommand{\eeq}{\end{eqnarray}}

\newcommand{\Ai}{{\rm Ai}}

\newlength{\bilderlength}

\usepackage{subfig}

\begin{document}

\title{Fluctuations in the active Dyson Brownian motion and the overdamped Calogero-Moser model} 

\author{L\'eo \surname{Touzo}}
\affiliation{Laboratoire de Physique de l'Ecole Normale Sup\'erieure, CNRS, ENS and PSL Universit\'e, Sorbonne Universit\'e, Universit\'e Paris Cit\'e,
24 rue Lhomond, 75005 Paris, France}
\author{Pierre Le Doussal}
\affiliation{Laboratoire de Physique de l'Ecole Normale Sup\'erieure, CNRS, ENS and PSL Universit\'e, Sorbonne Universit\'e, Universit\'e Paris Cit\'e,
24 rue Lhomond, 75005 Paris, France}
\author{Gr\'egory \surname{Schehr}}
\affiliation{Sorbonne Universit\'e, Laboratoire de Physique Th\'eorique et Hautes Energies, CNRS UMR 7589, 4 Place Jussieu, 75252 Paris Cedex 05, France}

\date{\today}

\begin{abstract} 
Recently, we introduced the active Dyson Brownian motion model (DBM), in which $N$ run-and-tumble particles interact via a logarithmic repulsive potential in the presence of a harmonic well. We found that in a broad range of parameters the density of particles converges at large $N$ to the Wigner semi-circle law, as in the passive case. In this paper, we provide an analytical support for this numerical observation, by studying the fluctuations of the positions of the particles in the nonequilibrium stationary state of the active DBM, in the regime of weak noise and large persistence time. In this limit, we obtain an analytical expression for the covariance between the particle positions for any $N$
from the exact inversion of the Hessian matrix of the system. 
We show that, when the number of particles is large $N \gg 1$, the covariance matrix takes scaling forms that we compute explicitly both
in the bulk and at the edge of the support of the semi-circle. In the bulk, the covariance scales as $N^{-1}$, while at the edge, it scales as $N^{-2/3}$.  
Remarkably, we find that these results can be transposed directly to an equilibrium model, the overdamped Calogero-Moser model 
in the low temperature limit, providing an analytical confirmation of the numerical results { by Agarwal, Kulkarni and Dhar} \cite{Agarwal2019}. For this model, our method also allows us to obtain the equilibrium two-time correlations and their dynamical scaling forms both in the bulk and at the edge. In the bulk the dynamics exhibits an anomalous diffusion regime $\sim t^{1/4}$. 
Our predictions at the edge are reminiscent of a recent result in the mathematics literature {by Gorin and Kleptsyn}
\cite{GorinInfiniteBeta} on the (passive) DBM. {That result can be recovered by the present methods, and also,
as we show, using the stochastic Airy operator.} 
Finally, our analytical predictions are confirmed by precise numerical simulations, in a wide range of parameters.  
\end{abstract}

\maketitle

\tableofcontents

\section{Introduction}

An important open problem in active matter is the characterisation of the collective behavior of many interacting active
particles. A paradigmatic model of active systems is the so-called run-and-tumble particle (RTP), a motion exhibited by E. Coli bacteria \cite{Berg2004,TailleurCates},
driven by telegraphic noise \cite{HJ95,W02,ML17}. 
Even in one dimension, and for such a stylized model, there are only a few cases where exact results can be obtained for a large number of RTP's \cite{nonexistence,KunduGap2020,SinghChain2020,PutBerxVanderzande2019,Metson2022,MetsonLong,Dandekar2020,Thom2011}. 
Recently, we introduced an active version of the Dyson Brownian motion where RTP's interact via a repulsive
logarithmic potential in the presence of a quadratic external potential. We 
showed that this model is amenable to analytical treatment, at least in some regimes \cite{ADBM1}. 
In 
the regime of weak noise the RTP's form a well ordered state. In fact 
in the limit of zero (active and passive) noise, the equilibrium positions of the particles are those
of the ground state of the log gas. These coincide with the zeros of the Hermite polynomials, as
is well known in random matrix theory~\cite{Forrester_book,Mehta_book}. As a consequence, for a large number of particles $N \gg 1$, 
the equilibrium density of the positions converges to the Wigner semi-circle. An outstanding
question is to describe the fluctuations around this equilibrium density profile. 

For interacting particle systems submitted to passive noise, the equilibrium measure is a Gibbs-Boltzmann weight.
It is then possible to compute the weak noise (i.e., low temperature)
fluctuations using an expansion of the energy functional in small displacements, 
such as phonons, or spin waves~\cite{Ashcroft}. In the case of active systems the stationary measure is usually 
non-trivial and not of Gibbs-Boltzmann type, so there is no energy functional and the corresponding calculation is much more challenging. 

The aim of this paper is to provide an explicit calculation of the weak noise fluctuations in the
case of the active DBM. Remarkably, we show that the method developed here also allows to treat 
an a priori unrelated system, namely the overdamped Langevin dynamics of the Calogero-Moser (CM) system in one dimension~\cite{Calogero75,Moser76}. 
This CM system has been much studied in the case of Hamiltonian dynamics due to its integrability properties~\cite{Calogero75,Moser76,BGS09,KP17,Poly06,OP81}.
Much less is known in the case of the overdamped Langevin dynamics. Recently, the latter was studied numerically, with some
emphasis on the regime of low temperature \cite{Agarwal2019}. Here we obtain analytical results in that
regime. The reason why the small fluctuations in the two systems are similar is that (i) the equilibrium positions
are the same and (ii) their Hessian 
matrices are related, as we will explain below. For both systems, active DBM and passive CM, we obtain the correlations
both in the bulk and at the edge of the Wigner semi-circle, where they take markedly different scaling forms at large $N$. 
Remarkably, we find that the scaling function which describes the edge behavior in our models
share some similarities with the one for the passive DBM which
was obtained recently in the math literature~\cite{GorinInfiniteBeta}. 

Let us now describe the two models that we will study in this paper. The first model is the so-called {\it active DBM} \cite{ADBM1}. It describes the dynamics of $N$ run and tumble particles (RTPs) in one dimension,
described by their positions $x_i(t)$. Each particle can be in two internal states $\sigma_i(t)=\pm 1$ of velocities respectively $\pm v_0$, and flips its sign
with a constant rate $\gamma$. In addition each particle is submitted to a confining potential $V(x)=\frac{\lambda}{2} x^2$ which ensures
that the system reaches a stationary state at large time (which is non Gibbsian and non trivial). The evolution equations read~\cite{ADBM1}
\bea \label{model0} 
 \dot x_i(t) &=& - \lambda x_i(t) +  \frac{2\,g }{N} \sum_{j \neq i} 
\frac{1}{x_i(t)-x_j(t)} +  v_0 \sigma_i(t)  + \sqrt{\frac{2 T}{N}} \xi_i(t) \quad {\rm for} \ i=1,2, \cdots, N.
\eea
The last term represents a thermal noise at temperature $T/N$, where the $\xi_i(t)$'s are independent standard white noises, of zero mean and delta correlations $\langle \xi_i(t) \xi_j(t') \rangle = \delta_{ij} \delta(t-t')$. 
The particles interact via a repulsive pairwise logarithmic potential (i.e. a $1/x$ force) of strength
$g/N$. In the case $v_0=0$ this model is the celebrated Dyson Brownian motion, whose stationary measure 
for any $N$ describes the statistics of the eigenvalues of the Gaussian beta-ensemble with Dyson index $\beta_{\rm DBM}=2 g/T$ \cite{Forrester_book,bouchaud_book}. In that case
the average density at large $N$ converges to the Wigner semi-circle density
with edges at $\pm 2 \sqrt{g/\lambda}$ \cite{Forrester_book,Mehta_book}. 
Numerous results exist in the math literature on random matrix theory concerning
the universality of the behavior, e.g., for the level spacing distribution, both in the bulk and at the edge~\cite{Erdos,ErdosYau,Pastur,Soshnikov,Sosh99,Tao_book}.
Here we will focus
on the purely active case $v_0>0$ and $T=0$. Note that, since the interactions diverge at contact, the particles in that case can never cross, hence their
ordering is preserved under the dynamics. Here we will choose $x_1(t)>x_2(t)>\cdots>x_N(t)$. 

The second model is the overdamped dynamics of the Calogero-Moser (CM) model for $N$ particles at positions $X_i(t)$ in one dimension \cite{Calogero75,Moser76}, which belongs to the more general family of Riesz gases \cite{Riesz,Lewin,riesz3}. The particles interact via a $1/X^2$ potential of strength $\tilde g^2/N^2$ and are subjected to a quadratic confining potential
which ensures that the system reaches Gibbs equilibrium. It is described by the equations of motion
\begin{equation}
 \dot X_i(t) = - \lambda X_i(t) +  \frac{8\, \tilde g^2}{N^2} \sum_{j \neq i} 
\frac{1}{(X_i(t)-X_j(t))^3} + \sqrt{\frac{2 T}{N}} \xi_i(t) \quad {\rm for} \ i=1,2, \cdots, N.
\label{Calogero}
\end{equation}
Here again the $\xi_i(t)$'s are independent standard white noises, and one can show that the particles
cannot cross (see Appendix \ref{NoCrossingCalogero}). Here also, we choose $X_1(t)>X_2(t)>\cdots > X_N(t)$. In both models, the scaling with $N$ of the different terms has been chosen such that the support of the densities is finite, i.e., independent of $N$, for large $N$. 

In this paper we will study the fluctuations of the positions of the particles in the steady state of each model.
We will obtain analytical results and also perform numerical simulations of Eqs. \eqref{model0} (at $T=0$) and \eqref{Calogero} (at finite $T$), the steady state averages being obtained by time averaging of the observables. It turns out that the weak noise limit of both models are very similar and can be studied in the same framework.
The first model, the active DBM, has been studied in \cite{ADBM1}. As mentionned in that work, there are two dimensionless parameters 
\be 
\frac{v_0^2}{g \lambda} \quad {\rm and} \quad \frac{\gamma}{\lambda} \;.
\ee 
The weak noise limit corresponds to a small value of the parameter $\frac{v_0^2}{g \lambda}$.
In that limit, the particles remain close to their equilibrium positions which we denote $x_{\rm eq,i}$,
and one can thus write
\be \label{eq_adbm}
x_i = x_{\rm eq,i} + \delta x_i   \quad , \quad x_{\rm eq,i} = \sqrt{\frac{2g}{\lambda \, N}}\, y_i \quad , \quad H_N(y_i) = 0\;,
\ee 
where the $\delta x_i$'s are the small deviations from equilibrium which vanish as $v_0 \to 0$. A remarkable property 
is that the scaled equilibrium positions $y_1 > y_2 > \cdots > y_N$ are the zeros of the Hermite polynomial of degree $N$, i.e., the roots of $H_N(y_i) = 0$.
At large $N$ the particles thus form a "crystal", and from the properties of the Hermite polynomials one can show that
the mean density $\rho_N^{\rm eq}(x)$ of particles at equilibrium (i.e. for $v_0=0$) in the quadratic well is a Wigner semi-circle, as is also the case for the DBM. Indeed, one has  \cite{Mehta_book}
\bea \label{Hermite_Wigner}
\rho_N^{\rm eq}(x) = \frac{1}{N} \sum_{i=1}^N \delta(x-x_{\rm eq,i}) \underset{N \to \infty}{\longrightarrow} \sqrt{\frac{\lambda}{2g}} \, \rho_{\rm sc}\left( \sqrt{\frac{\lambda}{2g}}\, x\right) \quad, \quad \rho_{\rm sc}(z) = \frac{1}{\pi} \sqrt{2-z^2} \;.
\eea
Here we will study the fluctuations of the $\delta x_i$ around this equilibrium state in the weak noise limit.
In addition for technical reason the analytical calculations will be performed in the persistent limit of
small $\gamma$. This allows to show that the semi-circle mean density
persists in a broad range of parameters for this model, as observed numerically in \cite{ADBM1}. In addition
we will quantify the long-range order of the crystal. 

In the CM model, it turns out, quite remarkably, that the equilibrium positions (i.e., for $T=0$) are also given in terms of the zeros of the Hermite polynomial $H_N(y)$, namely one has
\be \label{eq_CM}
X_{\rm eq,i}  = \frac{1}{\lambda^{1/4}}\sqrt{\frac{2 \tilde g}{ \, N}}\, y_i \;,
\ee 
which leads to the $T=0$ equilibrium density
\bea \label{CM_Wigner}
\rho_N^{\rm eq}(X) = \frac{1}{N} \sum_{i=1}^N \delta(X-X_{\rm eq,i}) \underset{N \to \infty}{\longrightarrow} {\frac{\lambda^{1/4}}{\sqrt{2\tilde g}}} \, \rho_{\rm sc}\left( {\frac{\lambda^{1/4}}{\sqrt{2\tilde g}}}\, x\right) \quad, \quad \rho_{\rm sc}(z) = \frac{1}{\pi} \sqrt{2-z^2} \;.
\eea
Here we will study the fluctuations of the particle positions around this $T=0$ equilibrium. 
In this model there is only a single dimensionless parameter
\begin{equation}
    \frac{T}{\tilde{g}\sqrt{\lambda}}
\end{equation}
and we will study the regime where this parameter is small. These fluctuations
were studied numerically in \cite{Agarwal2019}. Here we will obtain analytical results for any $N$
and we will compare with the numerical results. 


\section{Main results} \label{sec:results}

In this section we summarize the main results of this paper. The detailed derivations will be presented in
Sections \ref{sec:finiteN}-\ref{sec:dyn}. Section \ref{sec:dbm} discusses the application of our method to the passive DBM.

\subsection{Active Dyson Brownian Motion} \label{ADBM_main}

Let us start with the active DBM. The main idea to compute the statistics of the deviations $\delta x_i=x_i-x_{\rm eq,i}$ is to
consider the small $\gamma$ limit. In that limit the system has enough time to relax to a stable fixed point corresponding 
to a given $\vec{\sigma}(t) \approx \vec \sigma$ in \eqref{model0}, before its state changes again. Hence
one can write to lowest order in $v_0$ (and for $\gamma = 0^+$) and $1 \leq i \leq N$
\begin{eqnarray}
\delta {x}_i  = \frac{v_0}{\lambda} \sum_{j=1}^N(\mathcal{H}^{-1})_{ij} {\sigma}_j + O(v_0^2) \;,
\label{small_dx_fp_ADBM}
\end{eqnarray}
where $\lambda \mathcal{H}$ is the Hessian matrix given below [see Eq. (\ref{eqHessian})]. It is important to note that the matrix $\mathcal{H}$ is independent of the model parameters. In the stationary state the system explores all the possible $\vec{\sigma}$ (i.e. all the possible fixed points) and a meaninful description of the system is thus obtained by averaging over all these possible fixed points, with equal weight. One then obtains the
moments 
\begin{eqnarray}
&&\langle \delta x_{i} \rangle = O(v_0^2)\; , \nonumber \\
&&\langle \delta x_{i} \delta x_{j} \rangle = \frac{v_0^2}{\lambda^2} (\mathcal{H}^{-2})_{ij} + O(v_0^3)\; . \label{covHessian_ADBM}
\end{eqnarray}
It turns out that the Hessian can be diagonalized exactly for any $N$ in terms of Hermite polynomials \cite{eigenvectors, Agarwal2019}. This allows to
evaluate its inverse and obtain the more explicit formula
\begin{equation}
\langle \delta x_i \delta x_j \rangle = 
\frac{v_0^2}{\lambda^2} \sum_{k=1}^N \frac{1}{k^2} \frac{u_k(y_i)u_k(y_j)}{\sum_{l=1}^N u_k(y_l)^2} + O(v_0^3)  \quad {\rm with} \quad u_k(y)=\frac{H_N^{(k)}(y)}{H_N'(y)}  \;,
\label{cov_dx_ADBM}
\end{equation}
where $H_N^{(k)}(y)$ denotes the $k$-th derivative of $H_N(y)$ and we recall that $y_i$ is the $i$-th largest zero of $H_N(y)$. 

The exact result \eqref{cov_dx_ADBM} can be analyzed in the large $N$ limit. Let us recall that, in this limit, the mean equilibrium density has 
a finite support $[-x_e,x_e]$ with two edges at $x=\pm x_e$ with $x_e= 2 \sqrt{g/\lambda}$. One must thus distinguish
between the bulk region { , i.e. far from the boundary of the support $\pm x_e$,} and the edge region{ , i.e. close to $\pm x_e$}. The width of the edge region is found to be $x_{\rm eq,i}-x_e = O(N^{-2/3})$.
\\

{\bf Bulk}.
In the bulk region one finds
\begin{equation}
\langle \delta x_i \delta x_j \rangle \simeq \frac{v_0^2}{\lambda^2 N} \mathcal{C}_b\left( \frac{x_{{\rm eq},i}}{2\sqrt{g/\lambda}}, \frac{x_{{\rm eq},j}}{2\sqrt{g/\lambda}} \right) \quad {\rm with} \quad \mathcal{C}_b(x,y) 
= \sum_{k=1}^\infty \frac{1}{k^2} \frac{\sin(k \arccos x)}{\sqrt{1-x^2}} \frac{\sin(k \arccos y)}{\sqrt{1-y^2}} \;, 
\label{cov_largeN_ADBM}
\end{equation}
where the index '$b$' refers to 'bulk'. The sum over $k$ can be performed explicitly, see Appendix~\ref{app:alternative},
which leads to the more explicit expression of the scaling function $\mathcal{C}_b(x,y)$ 
\begin{equation}
\mathcal{C}_b(x,y) = \frac{\pi \arccos(\max(x,y)) - \arccos (x) \arccos (y)}{2\sqrt{1-x^2} \sqrt{1-y^2}} \;.
\label{newCbIntro}
\end{equation}
Note that this result, as well as all the other bulk results presented below, can be rewritten using that at large $N$ $\frac{x_{{\rm eq},i}}{2\sqrt{g/\lambda}} = \frac{y_{{\rm eq},i}}{\sqrt{2N}} \simeq G^{-1}(i/N)$ where $G(x)=\int_{-1}^x du \frac{2\sqrt{1-u^2}}{\pi}$ is the cumulative of the semi-circle density (see e.g. \cite{ORourke2010}).




For the variance of the displacement of a single particle this becomes
\begin{equation}
\langle \delta x_i^2 \rangle \simeq \frac{v_0^2}{\lambda^2 N} \mathcal{V}_b\left(\frac{x_{{\rm eq},i}}{2\sqrt{g/\lambda}}\right) \quad {\rm with} \quad \mathcal{V}_b(x) = \mathcal{C}_b(x,x)=\sum_{k=1}^\infty \frac{1}{k^2} \frac{\sin^2(k \arccos x)}{1-x^2}  = \frac{\arccos (x) (\pi-\arccos (x))}{2(1-x^2)} \; .
\label{var_largeN_ADBM}
\end{equation}
Inside the bulk, $\mathcal{V}_b\left(\frac{x_{{\rm eq},i}}{2\sqrt{g/\lambda}}\right)$ is of order 1, and thus the variance of particle displacements scales as $1/N$.


The result for the covariance allows us to compute the variance of the gap between particles $i$ and $i+n$ in the bulk, for $n \gg 1$. In the intermediate regime $1\ll n \ll N$ we obtain, e.g. for $i=N/2$,
\begin{equation}
    \langle (\delta x_i - \delta x_{i+n})^2 \rangle \simeq \frac{\pi^2}{4} \frac{v_0^2}{\lambda^2 N^2} n\; .
    \label{gapvariance_largeN_ADBM}
\end{equation}
The linear behavior in $n$ of the variance can be understood qualitatively as arising from the $1/k^2$ factor in the sum over eigenmodes in Eq. (\ref{cov_largeN_ADBM}) and can be obtained by an approximate calculation which neglects the space dependence of the mean density, see Appendix \ref{Bouchaud_approach}.
\\

{\bf Edge}. The equation \eqref{cov_dx_ADBM} can also be used to obtain the covariance of the particle displacements in the edge region, again in the large $N$ limit. 
In that region the equilibrium positions take the large $N$ scaling form near the right edge $x_e=2 \sqrt{g/\lambda}$ for $i \geq 1$ and $i=O(1)$
\begin{eqnarray}
x_{{\rm eq},i} = 2\sqrt{\frac{g}{\lambda}} \left( 1  + \frac{a_i}{2} N^{-2/3} + O(N^{-1}) \right) \;,
\label{hermite_roots_edge_app}
\end{eqnarray}
where $a_i$ is the $i^{th}$ zero of the Airy function, with, e.g., $a_1=-2.3811$ and for large $i$, $a_i = - (\frac{3 \pi}{8} (4i-1))^{2/3} + O(i^{-4/3})$.
The $\sim i^{2/3}$ power law behavior is consistent with the fact that the equilibrium density vanishes as a square root at the edge. 
We obtain the following result for the position fluctuations at the edge
\begin{equation}
\langle \delta x_i \delta x_j \rangle \simeq \frac{v_0^2}{\lambda^2 N^{2/3}} {\cal C}_e(a_i,a_j) \;\;, \;\; {\cal C}_e(a_i,a_j) =\frac{1}{\Ai'(a_i) \Ai'(a_j)} \int_0^{+\infty} dx \ \frac{\Ai(a_i + x)\Ai(a_j + x)}{x^2} \;,
\label{cov_edge_integral_ADBM}
\end{equation}
where $\Ai(x)$ is the Airy function and $a_i$ is its $i^{th}$ zero. This shows that the variance of the particle positions scales as $N^{-2/3}$ at the edge. Interestingly, this expression \eqref{cov_edge_integral_ADBM} is reminiscent of a similar result previously obtained for the (passive) Dyson Brownian motion \cite{GorinInfiniteBeta}. 
Note that one can check that our formula at the edge matches correctly the formula in the bulk \eqref{cov_largeN_ADBM}, as we show below in Section \ref{sec:edge}. 

\begin{figure}
    \centering
    \includegraphics[width=0.8\linewidth, trim={0 2.7cm 0 8.1cm}, clip]{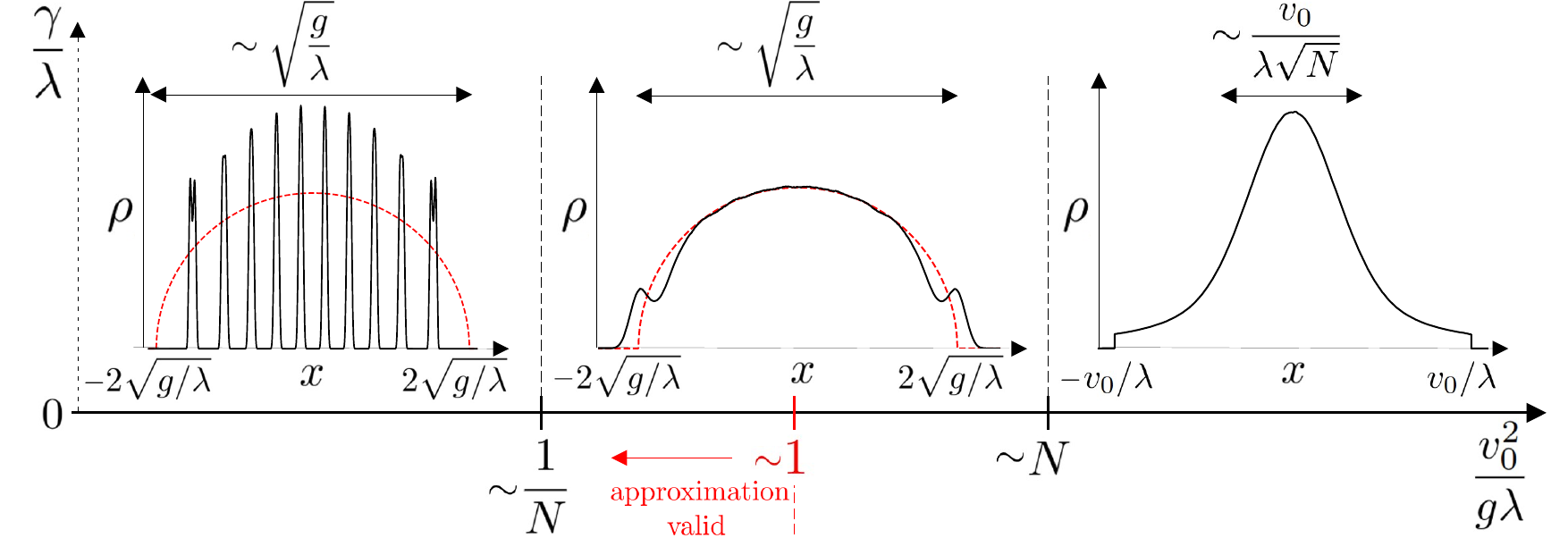}
    \caption{Shape of the total density $\rho$ in the active DBM (model II in Ref. \cite{ADBM1}) as a function of the parameter $v_0^2/g\lambda$ showing the different regimes at large $N$. The dashed red line shows the semi-circle. In the middle regime the wings which are visible near the edges disappear
    in the limit $N \to +\infty$, and can be related to edge effects. The spatial extension of the density as a function of the parameters is also shown in the different regimes. Its behavior for $v_0^2/g\lambda \gg 1$ { (right panel)}, was obtained from numerical simulations of an effective model, see
    Ref \cite{ADBM1}. 
    The results were derived for $\gamma \to 0$ but simulations suggest that they are valid beyond this limit. }
    \label{phase_diagram}
\end{figure}

Let us now turn to the implications of these results on the particle density in the bulk of the active DBM. 
In Fig.~\ref{phase_diagram} we are showing the various regimes as the dimensionless parameter $v_0^2/(g \lambda)$ is varied,
which were discussed in \cite{ADBM1}. The results obtained here allow to describe the left part of the figure (i.e. to the left of the red line in Fig.~\ref{phase_diagram}). The important consequence of \eqref{var_largeN_ADBM} is that the variance of the position fluctuations of any particle in the bulk scales as $v_0^2/(\lambda^2 N)$ at large $N$. More precisely one can compare the root mean squared displacement to the typical separation between particles $\sqrt{(g/\lambda)}/N$
by considering the dimensionless ratio
\begin{equation} \label{ratio1} 
    \frac{\sqrt{\langle \delta x_i^2 \rangle}}{\langle x_i - x_{i+1} \rangle} \sim \frac{v_0}{\sqrt{g\lambda}} \sqrt{N} \;,
\end{equation}
which is thus small compared to unity when $v_0^2/(g \lambda) \ll 1/N$. This defines the regime represented
on the left in Fig.~\ref{phase_diagram} where the crystal is very well ordered and the density exhibits peaks around the equilibrium positions $x_{{\rm eq},i}$. The second regime in Fig.~\ref{phase_diagram} can be identified by looking at the dimensionless ratio 
\begin{equation} \label{ratio2} 
    \frac{\sqrt{\langle \delta x_i^2 \rangle}}{x_e} \sim \frac{v_0}{\sqrt{g \lambda N}} \;.
\end{equation}
Clearly the semi-circle density can hold only when this ratio is small compared to unity, which means $v_0^2/(g \lambda) \ll N$.
In this second regime $\frac{1}{N} \ll v_0^2/(g \lambda) \ll N$, the fluctuations are larger but the semi-circle
density still holds. Note that although the results described above were obtained strictly in the limit $\gamma \to 0$, 
numerical simulations strongly suggest that the scalings obtained above hold for any value of $\gamma$. 

It is important to note that the approximation \eqref{small_dx_fp_ADBM}
and the above results are, strictly speaking, only valid
when the typical variations of the distance between successive particles $\sqrt{\langle (\delta x_i - \delta x_{i+1})^2 \rangle}$ is much smaller than the average distance $\langle \delta x_i - \delta x_{i+1} \rangle$. Although \eqref{gapvariance_largeN_ADBM} is valid only for $n \gg 1$, we still expect it to give the correct order of magnitude for $n=1$. Thus we get that our results should be valid when the 1D Lindemann-like ratio is small, i.e.
\begin{equation}  \label{ratio3} 
  c_L =   \frac{\sqrt{\langle (\delta x_i - \delta x_{i+1})^2 \rangle}}{\langle x_i - x_{i+1} \rangle} \sim \frac{v_0/(\lambda N)}{\sqrt{(g/\lambda)}/N} = \frac{v_0}{\sqrt{g\lambda}} \ll 1 \; .
\end{equation}
This defines an additional line in the Fig. \ref{phase_diagram} which falls in the middle of the intermediate regime, i.e,
our detailed predictions (\ref{covHessian_ADBM})-(\ref{cov_edge_integral_ADBM}) are valid for $v_0^2/(g \lambda) \lesssim 1$. 
However, we expect our theory to describe the system beyond this line, at least qualitatively as order of magnitude estimates. We have indeed checked numerically that this is the case in Ref. \cite{ADBM1}. 

Let us now briefly describe the implications of our result in the edge region, i.e. 
Eq. \eqref{cov_edge_integral_ADBM}. The fluctuations are larger in that
region and the scaling with $N$ is different. Indeed, the dimensionless ratio defined in \eqref{ratio2} 
is $\sim N^{-1/3}$ at the edge, instead of $\sim N^{-1/2}$ in the bulk.
However one can again consider the {\it relative fluctuations} by comparing with the distance
between particles which is also larger, i.e. consider the ratios defined in 
Eqs. \eqref{ratio1}, \eqref{ratio3}. Estimating any function of the $a_i$ to be of order unity,
one finds that these relative fluctuations are small when $v_0^2/(g \lambda) \ll 1/N^{2/3}$,
which also gives the condition for our detailed predictions \eqref{cov_edge_integral_ADBM} to be valid. 

{Finally, we have tested some of the above predictions for the active DBM numerically and have observed a very good agreement. 
This is discussed later in the paper, see sections \ref{sec:bulk} and \ref{sec:edge}
and Figs. \ref{gammaEffect}, \ref{variancefig}, \ref{variancefig2}, \ref{FigcovarianceADBM}, \ref{variancefigedge} (see also
some discussion at the end of the next section in the context of the CM model).
}



\subsection{Overdamped Calogero-Moser model}

The dynamics of the CM model defined in \eqref{Calogero} converges at large time towards a Gibbs-like equilibrium state which however 
retains the same ordering as the particles in the initial state, see Appendix \ref{NoCrossingCalogero}. Choosing
$X_1>\dots>X_N$ in the initial state, the joint PDF of the
positions of the particles in this equilibrium state can be written as
\be 
{\cal P}[X] \sim e^{- \frac{N}{T} ( \frac{\lambda}{2} \sum_i X_i^2 + \frac{4 \tilde g^2}{N^2} \sum_{i<j} \frac{1}{(X_i-X_j)^2} ) } 
\theta(X_1>X_2>\dots>X_N) \;. \label{CM_Gibbs}
\ee 
We are interested here in the correlation functions with respect to this joint PDF. Note that for the observables
which are symmetric in the labels of the particles, the ordering is immaterial. This is however not the
case for correlations with specified particle labels. Let us summarize our main results, both in the 
low temperature regime $T/(\tilde g \sqrt{\lambda}) \ll N$ [where at large $N$ the support of the density remains finite and takes the
Wigner semi-circular form (\ref{CM_Wigner})] 
and the high temperature regime $T/(\tilde g \sqrt{\lambda}) \gg N$ where the support is unbounded.
\\ 

\noindent{\bf Low temperature regime}. Let us define $\delta X_i = X_i - X_{\rm eq,i}$ where $X_{\rm eq,i}$ are the equilibrium positions given in 
\eqref{eq_CM}. For the CM model, it was shown in \cite{Agarwal2019} that the two-point correlation function takes the form at low temperature
\begin{equation}
\langle \delta X_i \delta X_j \rangle = \frac{T}{\lambda N} (\mathcal{H}^{-2})_{ij} \;,
\label{CalogeroCov_intro}
\end{equation}
with the same matrix $\mathcal{H}$ as in Eq.~(\ref{small_dx_fp_ADBM}) above. Quite remarkably, the relation \eqref{CalogeroCov_intro} has exactly the same form as \eqref{covHessian_ADBM}, and therefore the results above can be transposed directly to the CM model by simply changing the prefactor. 

In the large $N$ limit, we recall that the mean equilibrium density of the CM model is given by the Wigner semi-circle~(\ref{CM_Wigner}) which has a finite support $[-X_e,X_e]$ with two edges at $X=\pm X_e$ with $X_e= 2 \sqrt{\tilde g}/\lambda^{1/4}$.
Furthermore, from the above results, we obtain that in the bulk
\begin{equation}
\langle \delta X_i \delta X_j \rangle \simeq\frac{T}{\lambda N^2} \mathcal{C}_b\left( \frac{ \lambda^{1/4} X_{{\rm eq},i}}{2\sqrt{\tilde g}}, \frac{\lambda^{1/4} X_{{\rm eq},j}}{2\sqrt{\tilde g}} \right) \;,
\label{cov_largeN_CM}
\end{equation}
where the function ${\cal C}_b(x,y)$ is given in Eq. (\ref{cov_largeN_ADBM}). At the right edge 
the equilibrium positions take the form 
\begin{eqnarray}
X_{{\rm eq},i} = 2 \frac{\sqrt{\tilde g}}{\lambda^{1/4}} \left( 1  + \frac{a_i}{2} N^{-2/3} + O(N^{-1}) \right) \;,
\label{hermite_roots_edge_app2}
\end{eqnarray}
and we obtain that their fluctuations obey
\begin{equation}
\langle \delta X_i \delta X_j \rangle \simeq \frac{T}{\lambda N^{5/3}} \mathcal{C}_e(a_i,a_j) \;,
\label{cov_edge_integral_CM}
\end{equation}
in the edge region, where $\mathcal{C}_e(a_i,a_j)$ is defined in \eqref{cov_edge_integral_ADBM}. 
\\

Finally, we have computed by the same method the temporal correlations
$\langle \delta X_i(t) \delta X_j(t') \rangle$ in the stationary state and at low temperature for the CM model.
The results are given in Section \ref{sec:dyn} in terms of dynamical scaling functions which 
generalize the equal time ones given above. In particular we find that
the time dependent displacement of a particle in the bulk exhibits anomalous diffusion,
i.e. $\langle (\delta X_i(t)-\delta X_i(0))^2 \rangle \sim  t^{1/2} $
in an intermediate time window.
\\

\noindent{\bf High temperature regime}. At high temperature the interaction becomes formally irrelevant 
when $T/(\tilde g \sqrt{\lambda}) \gg N$, although the order of the particle is still retained (see Appendix \ref{app_CM_highT}). Hence the
Gibbs measure \eqref{CM_Gibbs}, expressed in terms of scaled variables, converges to
\be
P(X_1, ..., X_N) = \frac{1}{Z_N} e^{-\frac{1}{2}\sum_i \tilde X_i^2} \theta(\tilde X_1>\tilde X_2>...>\tilde X_N) \quad , \quad X_i = \sqrt{ \frac{T}{N \lambda}}  \tilde X_i \;,
\ee
where $Z_N= 1/(N! (2 \pi)^{N/2})$. In other words, at a given time the particle positions $X_1,\cdots,X_N$ are i.i.d centered Gaussian variables with variance $\sqrt{T/(\lambda N)}$, ordered from the largest to the smallest. When computing the particle density the ordering is irrelevant, so that the density in this regime is a Gaussian with variance $\sqrt{T/(\lambda N)}$. In addition, we can use existing results on the order statistics of i.i.d. Gaussian variables \cite{galambos,nagaraja,sm14,TheseBertrand}, see 
Appendix \ref{app_CM_highT} where some additional results are derived {(and some numerical tests are presented)}, 
to obtain the mean and two-point covariance of the positions of
the particles. In the large $N$ limit they take the form, in the bulk, i.e. $i,j = O(N)$
\bea \label{CM_highT}
&&\langle X_i \rangle \simeq \sqrt{\frac{T}{\lambda N}} \, Q^{-1}\left(\frac{i}{N}\right) \quad , \quad 
\langle X_i X_j \rangle_c \simeq 2 \pi \frac{T}{\lambda N^2} \frac{i}{N}\left(1-\frac{j}{N}\right) \, e^{\frac{1}{2} [Q^{-1}(\frac{i}{N})]^2 + \frac{1}{2} [Q^{-1}(\frac{j}{N})]^2}  \\
{\rm where} \quad &&Q(x)= \int_x^{+\infty} dy \frac{e^{-y^2/2} }{\sqrt{2 \pi}} = \frac{1}{2} {\rm erfc}\left(\frac{x}{\sqrt{2}}\right) \;.
\eea
Note that the one-point variance $\langle X_i^2 \rangle_c$ is given by the same formula setting $i=j$. 
In particular, one finds that the higher order cumulants are subdominant, so that the distribution of the
rescaled positions $N( X_i - \langle X_i \rangle)$ are Gaussian in the limit of large $N$. 
On the other hand, at the (right) edge of the gas, i.e. $i,j=O(1)$ one obtains instead, for $j \geq i $
\bea 
&& X_j = \sqrt{\frac{T}{\lambda N}} \sqrt{2 \log N} \left(1 + \frac{\zeta_j +c_N}{2 \log N} + \dots \right) \\
&& 
\langle \zeta_i \rangle = - \psi_0(i) \quad , \quad \langle \zeta_i \zeta_j  \rangle_c = \psi_1(j) = \langle \zeta_j^2 \rangle_c \;,
\eea  
where $\psi_0(x) = \Gamma'(x)/\Gamma(x)$ and $\psi_1(x) = \psi_0'(x)$ are the digamma and trigamma functions respectively
and $c_N = - \log( \sqrt{4 \pi \log N} )$. As shown in Appendix \ref{app_CM_highT} the two forms (edge and bulk) match correctly. 
As detailed in the Appendix \ref{app_CM_highT}, the position of the edge particle $X_1$ follows a Gumbel distribution,
and furthermore the distributions of the gaps $X_i-X_{i+1}$ are exponential both in the bulk and near the edge.

Note that we expect this high temperature regime to be more general than the current setting. Indeed, the only information that remains about the interaction being the ordering constraint, this should remain valid for any interaction potential of the form $1/|x_i-x_j|^{\alpha}$ with $\alpha>0$ (see Appendix \ref{NoCrossingCalogero}). Moreover, if the external potential $V(x)$ is not quadratic but is instead any type of confining potential, so that an equilibrium measure exists, then the particle positions will be the ordered set of $N$ i.i.d. random variables drawn from the single-particle Gibbs measure $e^{-N V(x)/T}$. 
\\

\begin{figure}
    \centering
    \includegraphics[width=0.8\linewidth, trim={0 2.7cm 0 7.8cm}, clip]{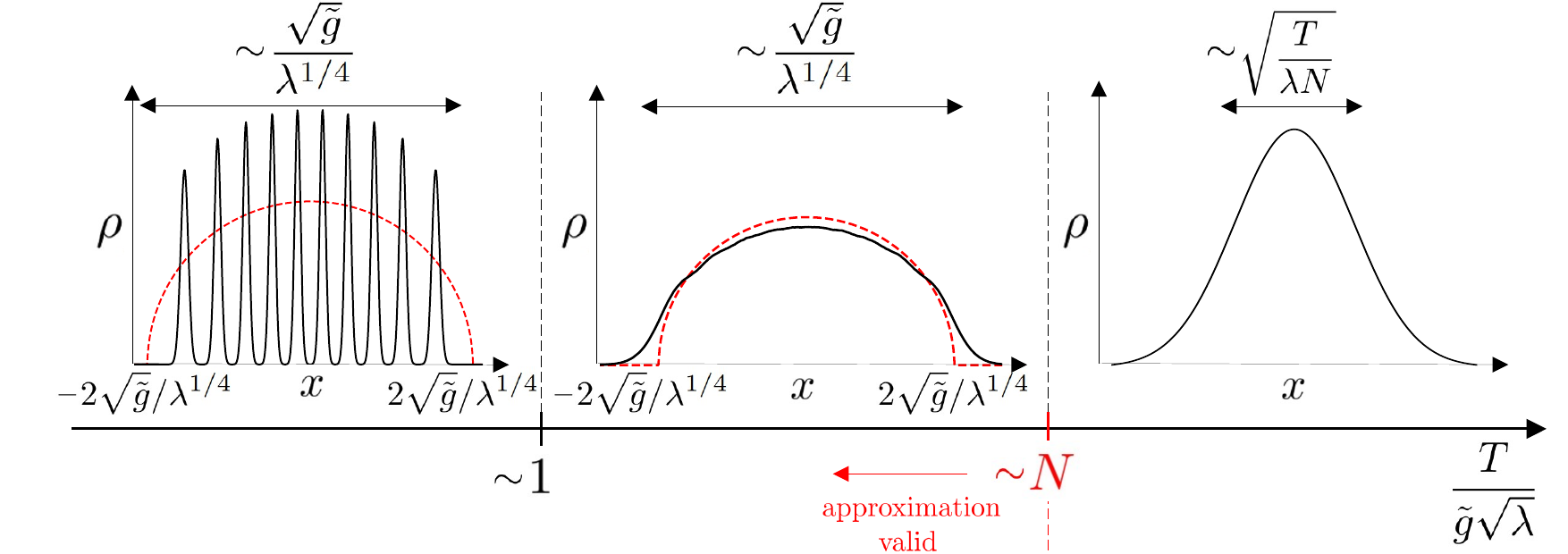}
    \caption{Shape of the total density $\rho$ in the Calogero-Moser model as a function of the parameter $T/\tilde{g}\lambda^{1/4}$ showing the different regimes at large $N$. The dashed red line shows the semi-circle.}
    \label{phase_diagram_CM}
\end{figure}

\noindent{\bf Discussion of the crossovers}. These results allow to discuss the different regimes for the CM model as the dimensionless parameter $T/(\tilde g \sqrt{\lambda})$
is varied, as represented in Fig \ref{phase_diagram_CM}. At variance with the active DBM, the stationary state is a Gibbs equilibrium, which makes the 
discussion somewhat easier.
There are thus again three regimes. The very low temperature well ordered crystal, with peaks in the density (left panel in Fig \ref{phase_diagram_CM}), is obtained when the dimensionless ratio 
\begin{equation} \label{dimless_CG1}
    \frac{\sqrt{\langle \delta X_i^2 \rangle}}{\langle X_i - X_{i+1} \rangle} \sim \sqrt{\frac{T}{\tilde g}} \frac{1}{\lambda^{1/4}}  
\end{equation}
is much smaller than unity, i.e. when $T/(\tilde g \sqrt{\lambda}) \ll 1$. 
For higher temperature $1 \ll T/(\tilde g \sqrt{\lambda}) \ll N$ the 
fluctuations are larger but the semi-circle density still holds. This is the intermediate temperature regime
shown in the middle panel in Fig. \ref{phase_diagram_CM}. Finally there is the 
high temperature regime 
shown in the right panel in Fig. \ref{phase_diagram_CM}.
In that regime the width of the gas behaves as $\sim \sqrt{T/(\lambda N)}$. 
It matches the width of the semi-circle $\sim \sqrt{\tilde{g}}/\lambda^{1/4}$
when $T/(\tilde g \sqrt{\lambda}) \sim N$. Hence we find that the 
boundary between the semi-circle and the high temperature regime 
occurs for $T/(\tilde g \sqrt{\lambda}) \sim N$ and 
coincides with the domain of validity of the 
above analytical results, i.e., when 
\begin{equation}
        \frac{\sqrt{\langle (\delta X_i - \delta X_{i+1})^2 \rangle}}{\langle X_i - X_{i+1} \rangle} \sim \frac{\sqrt{(T/\lambda)} \ N^{-3/2}}{\sqrt{\tilde g} \ \lambda^{-1/4}N^{-1}} = \frac{1}{\lambda^{1/4}} \sqrt{\frac{T}{\tilde g}} \frac{1}{\sqrt{N}} \ll 1 \; .
\end{equation}


\medskip

\begin{figure}[t]
    \centering
    \includegraphics[width=0.325\linewidth,trim={0cm 0 1cm 0.5cm},clip]{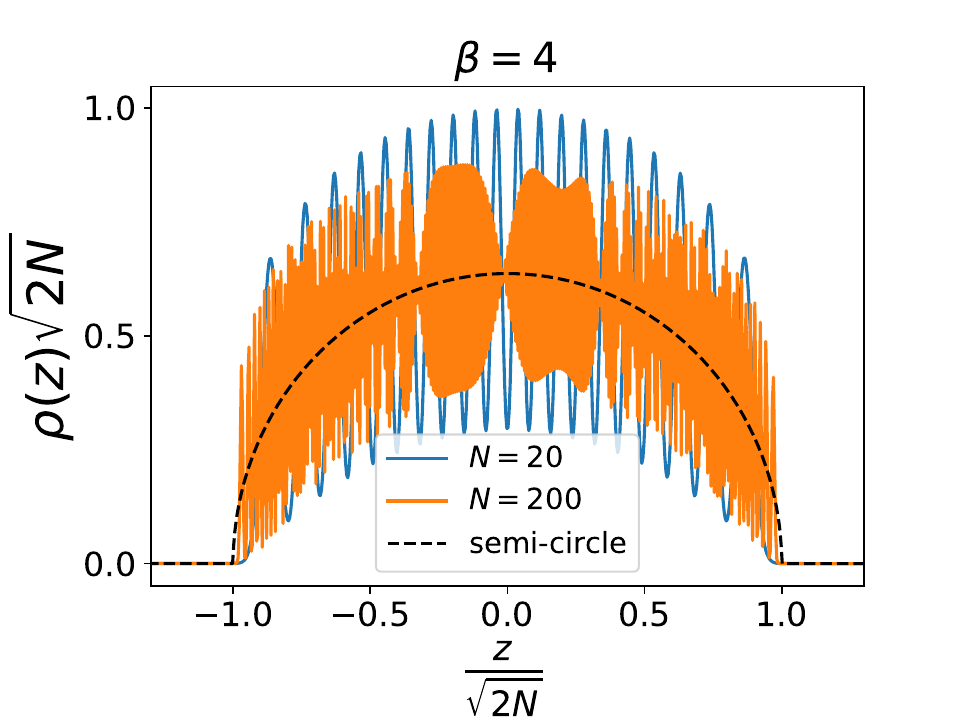}
    \includegraphics[width=0.325\linewidth,trim={0cm 0 1cm 0.5cm},clip]{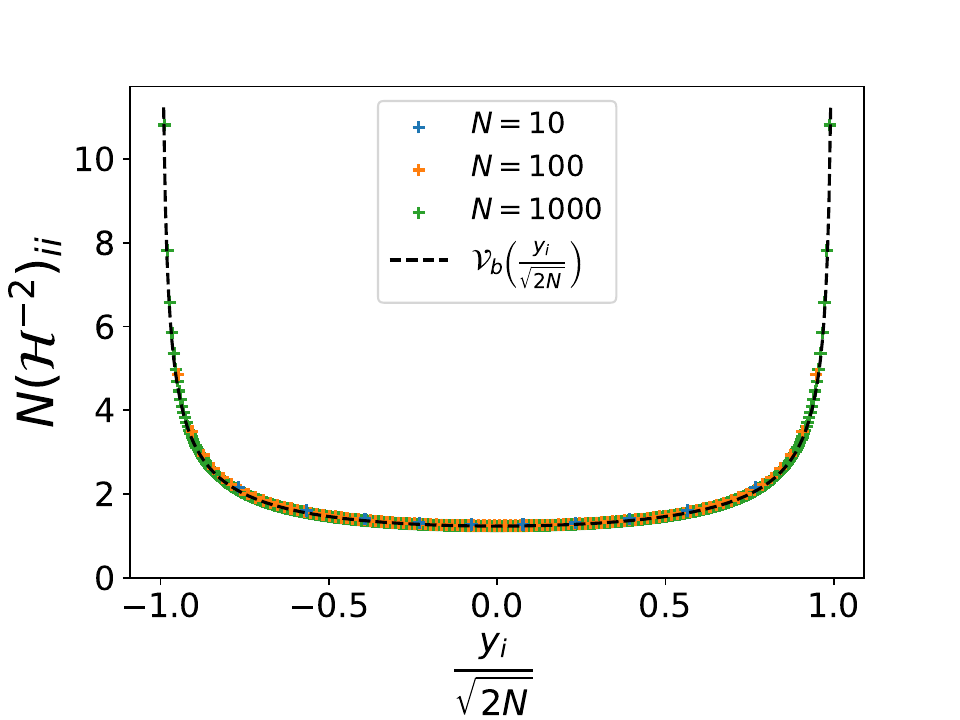}
    \caption{
    {\bf Left:} The left figure shows the density for the CM model for $\beta=4$, obtained by direct simulation of the Langevin dynamics (\ref{Calogero}). 
    We use the notations summarized in the text around \eqref{def_zi}. The peaks are clearly visible
    as predicted for $\beta > 1$, see left panel of Fig. \ref{phase_diagram_CM}. It is analogous to Fig.~1a in \cite{Agarwal2019}, where however the peaks cannot
    be seen for the same values of $N$. {\bf Right:} The right figure shows a comparison between (i) 
the scaled variance in the bulk \eqref{CalogeroCov_intro} 
as a function of the scaled equilibrium position $y_i/\sqrt{2 N}$ computed by inverting the Hessian for different values of $N$
and (ii) the scaling form predicted at in large $N$ in \eqref{cov_largeN_CM}
(dashed black line), recalling that ${\cal V}_b(x)={\cal C}_b(x,x)$. This figure should be compared with 
the Fig.~8a in \cite{Agarwal2019}. 
    }
    \label{AKDcomparison}
\end{figure}

\noindent{\bf Comparison with Ref. \cite{Agarwal2019}}. We can compare our results, {in particular Eqs. \eqref{cov_largeN_CM} and \eqref{cov_edge_integral_CM},}
with the ones obtained numerically on the CM model in~\cite{Agarwal2019}.
In that work they study the same displacement correlations either by (i) numerical evaluation of the Hessian
or (ii) by direct Monte-Carlo simulations, and they compare both methods, with a good agreement for lower temperature 
Their conventions are different from ours, so we start by giving
the corresponding dictionary. Denoting for convenience $z_i$ the positions of the particles
denoted $x_i$ in \cite{Agarwal2019} (they set to unity their parameter $g$) we can identify
\be  \label{def_zi}
\frac{z_i}{\sqrt{2 N}} = \frac{\lambda^{1/4}}{2 \sqrt{\tilde g}} X_i \;,
\ee 
both sides being dimensionless and of order unity in the bulk at equilibrium. In fact
one has $z_{\rm eq,i}=y_i$, where $y_i$ is the $i$-th root of the Hermite polynomial $H_N(y)$. This then leads to the following identification for
the parameter $\beta$ (analogous to the Dyson index for the DBM) 
\be 
\beta = \frac{2 \tilde g \sqrt{\lambda}}{T} \;.
\ee 
In the bulk, our result then leads to, in their notations 
\be 
\langle \delta z_i \delta z_j \rangle = \frac{1}{N \beta} C_b \left(\frac{z_{{\rm eq},i}}{\sqrt{2 N}} , \frac{z_{{\rm eq},j}}{\sqrt{2 N}} \right) \;,
\ee 
where we recall that the function ${\cal C}_b(x,y)$ is given in Eq. (\ref{cov_largeN_ADBM}). Forming the dimensionless ratios we see that 
our prediction is that in their conventions the three regimes in the bulk in Fig. \ref{phase_diagram_CM} are 
(i) the regime with peaks for $\beta \gtrsim 1$, (ii) the semi-circle regime for 
$1/N \lesssim \beta \lesssim 1 $, and the high temperature regime for $\beta \lesssim \frac{1}{N}$. 
Similarly at the edge we predict
\begin{equation}
\langle \delta z_i \delta z_j \rangle \simeq \frac{1}{\beta N^{2/3}} \mathcal{C}_e(a_i,a_j) \;,
\label{cov_edge_integral_CM_Kulkarni}
\end{equation}
where $\mathcal{C}_e$ is given in \eqref{cov_edge_integral_ADBM}.


In the left panel of Fig. \ref{AKDcomparison}, we show the mean particle density for the CM model for $\beta=4$
obtained by direct simulation of the Langevin dynamics (\ref{Calogero}). The density exhibits clearly visible peaks, 
in agreement with our theoretical prediction for $\beta =4$ (see the left panel of Fig. \ref{phase_diagram_CM}). We note that this is qualitatively different from the numerical results shown in Ref. \cite{Agarwal2019} (see their Fig. 1a), which shows instead a rather smooth density profile. 

\begin{figure}[t]
    \centering
    \includegraphics[width=0.325\linewidth,trim={0cm 0 1cm 1cm},clip]{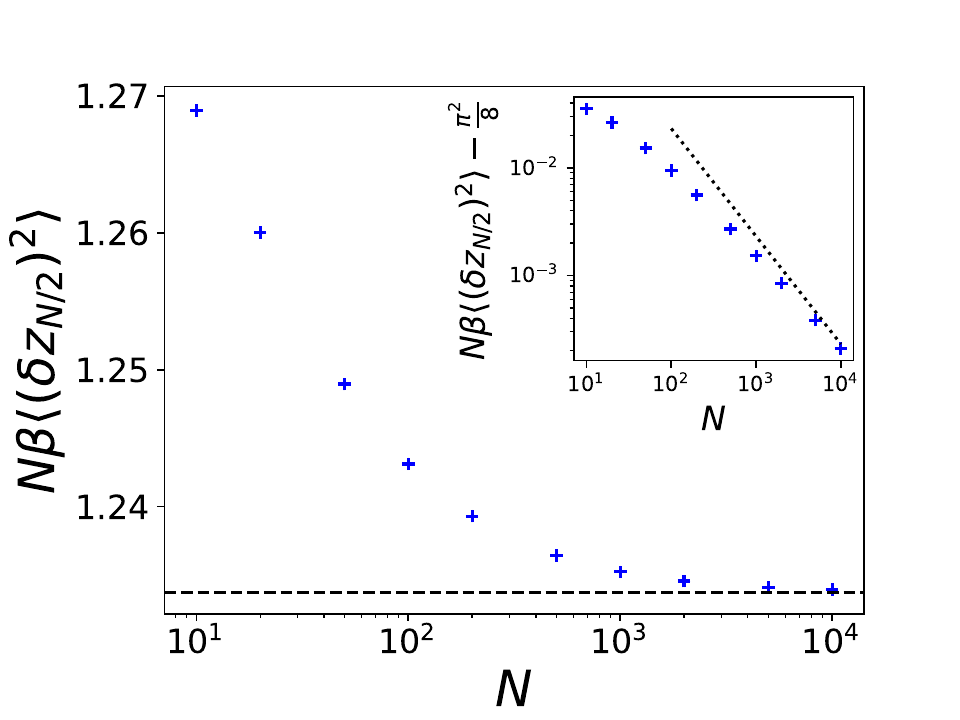}
    \includegraphics[width=0.325\linewidth,trim={0cm 0 1cm 1cm},clip]{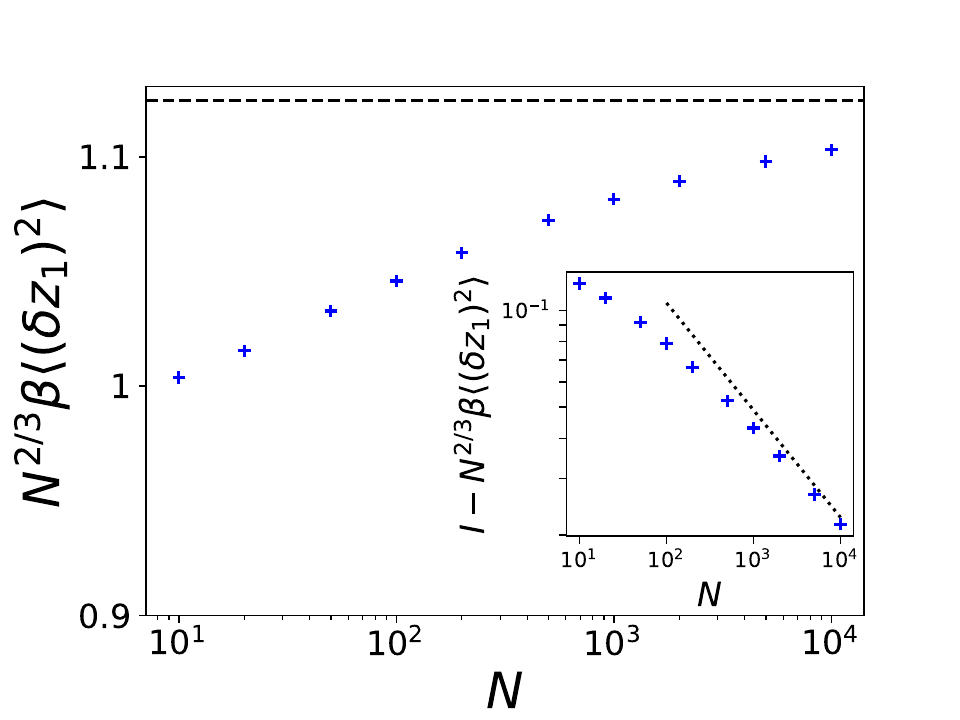}
    \caption{{\bf Left:} The left figure shows the scaled variance of the midpoint position $z_{i=N/2}$ for the CM model for various
    values of $N$. 
    We use the notations summarized in the text around \eqref{def_zi}. 
    The blue crosses correspond to the numerical inversion of the Hessian. The dashed black line is our analytic prediction
    of the value ${\cal V}(0)=\pi^2/8$ in the large $N$ limit, see Eq. \eqref{v_smallx}. 
    The inset shows the difference, with the dotted black line showing a $1/N$ decay.
This figure corresponds to the Fig. 6a in \cite{Agarwal2019}. {\bf Right:} The right figure shows a similar plot for $z_{i=1}$ 
and represents the scaled variance of the edge particle. 
    The dashed black line correspond to our prediction $N^{2/3} \beta \langle \delta z_1^2 \rangle = I =\frac{1}{\Ai'(a_1)^2} \int_0^{+\infty} dx \ \frac{\Ai(a_1 + x)^2}{x^2} \simeq 1.12481...$ (see Eq. \eqref{cov_edge_integral_ADBM}). The inset shows the difference, with the dotted black line showing a $1/N^{1/3}$ decay.
    This figure corresponds to their Fig. 3a. }
    \label{AKDcomparison2}
\end{figure}

\begin{figure}[t]
    \centering
    \includegraphics[width=0.325\linewidth,trim={0cm 0 1cm 1cm},clip]{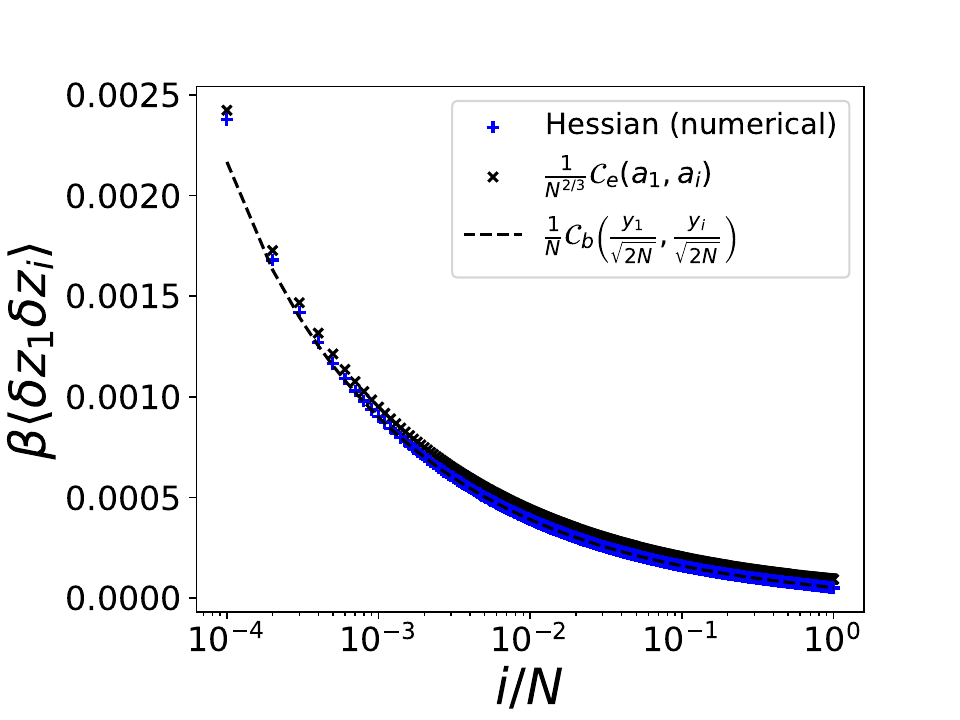}
    \includegraphics[width=0.325\linewidth,trim={0cm 0 1cm 1cm},clip]{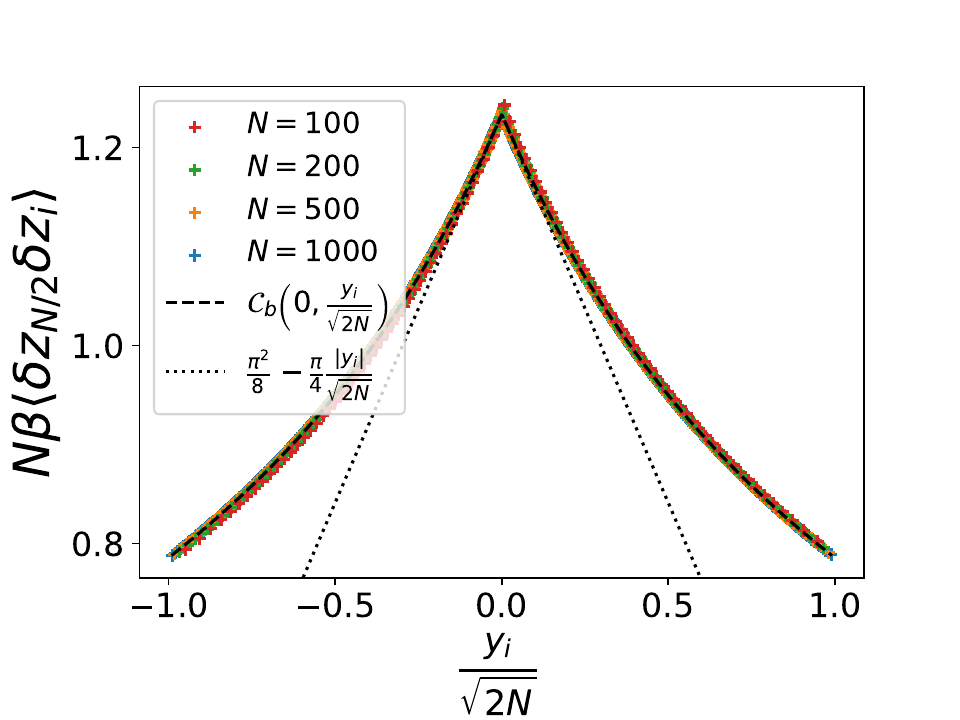}
    \caption{
    {\bf Left:} The left figure shows the covariance between the edge particle 1 and particle $i$ as a function
    of $i/N$ for $N=10^4$. The 
    edge and bulk predictions at large $N$ are also shown.
    For $i=O(1)$ it matches quite well with the edge analytical result, while for $i=O(N)$ it matches well with the bulk result. This figure corresponds to Fig.~10a in  \cite{Agarwal2019}. {\bf Right:} The right figure shows a similar plot versus the scaled equilibrium positions $y_i/\sqrt{2 N}$ of the scaled covariance between the central particle and particle $i$. It corresponds to Fig. 10b in \cite{Agarwal2019}. 
    }
    \label{AKDcomparison3}
\end{figure}

In the right panel of Fig. \ref{AKDcomparison} and in the Figs. \ref{AKDcomparison2} and \ref{AKDcomparison3}
we have compared some of our large $N$ theoretical predictions with the results from an explicit numerical calculation
of the Hessian. We find that the convergence in $N$ is very fast in the bulk ($\propto N^{-1}$) and slower at the edge ($\propto N^{-1/3}$). 
Our large $N$ analytical predictions can also be compared with the results of \cite{Agarwal2019},
more precisely with their Figs. 3a, 6a, 8a, 10a and 10b. For instance we provide the
exact analytical value for the scaled variance of the midpoint in the large $N$ limit,
which is $\pi^2/32 \simeq 0.308425$ for $\beta=4$ which turns out to be amazingly close to
the quoted measured value $0.3084$ (see their Fig. 6a and our Fig. \ref{AKDcomparison2} (left panel)). 

Finally, note that the Figs. \ref{AKDcomparison}, \ref{AKDcomparison2} and \ref{AKDcomparison3}
contain comparison of our large $N$ prediction with the results from an explicit numerical calculation
of the Hessian, hence they also provide a check of our predictions for the active DBM
presented in the previous section.

\section{Derivation of the results for finite $N$} \label{sec:finiteN}

\subsection{Active Dyson Brownian motion}

The goal of this section is to derive equation \eqref{covHessian_ADBM} which relates the covariance of particle displacements in the active DBM with respect to their equilibrium position, $\delta x_i = x_i - x_{{\rm eq},i}$, to the Hessian of the potential. 
We start by recalling that the vector of equilibrium positions $\vec{x}_{{\rm eq}}$ is the solution of the equation
\begin{eqnarray}
\frac{\partial V^{ADBM}}{\partial \vec{x}} (\vec{x}_{\rm eq}) = 0 \quad {\rm with} \quad V^{ADBM}(\vec{x}) = \frac{\lambda}{2} \sum_i x_i^2 -\frac{2g}{N}\sum_{i<j} \log |x_i-x_j|\; .
\label{eqpotential}
\end{eqnarray}
which satisfies $x_{{\rm eq},1}>...>x_{{\rm eq},N}$. It is well known that the solution of this system is given by the roots of the Hermite polynomial $H_N(x)$ as \cite{HermiteZeros,ADBM1}
\begin{equation}
x_{{\rm eq},i} = \sqrt{\frac{2g}{\lambda N}} \ y_i \quad {\rm with} \quad H_N(y_i)=0\; .
\end{equation}
We will denote $\delta \vec{x}$ the vector formed by the displacements $\delta x_i$, and $\vec{\sigma}$ the vector of $\sigma_i$. The first assumption that we make is that 
the typical fluctuations of the gaps $\delta x_i - \delta x_{i+1}$ are small compared to the typical size of the gaps $x_{{\rm eq},i} - x_{{\rm eq},i+1} \sim \sqrt{g/\lambda}/N$. This allows us to linearize the dynamics around $\vec{x}_{\rm eq}$:
\begin{eqnarray}
\frac{\partial \delta \vec{x}}{\partial t} = -\frac{\partial V^{ADBM}}{\partial \vec{x}} (\vec{x}_{{\rm eq}} + \delta \vec{x}) + v_0 \vec{\sigma}(t) \simeq - \lambda \mathcal{H} \delta \vec{x} + v_0 \vec{\sigma}(t) \;,
\label{small_dx_dynamics}
\end{eqnarray}
with the Hessian matrix
\begin{eqnarray}
\lambda \mathcal{H}_{ij} = \frac{\partial^2 V^{ADBM}}{\partial x_i \partial x_j} (\vec{x}_{\rm eq}) &=& \delta_{ij}\left(\lambda+\frac{2g}{N} \sum_{k\neq i} \frac{1}{(x_{{\rm eq},i}-x_{{\rm eq},k})^2}\right) - (1-\delta_{ij}) \frac{2g}{N} \frac{1}{(x_{{\rm eq},i}-x_{{\rm eq},j})^2} \nonumber \\
&=& \lambda\left[\delta_{ij}\left(1+ \sum_{k\neq i} \frac{1}{(y_i-y_k)^2}\right) - (1-\delta_{ij}) \frac{1}{(y_i-y_j)^2}\right]\; .
\label{eqHessian}
\end{eqnarray}
Note that the matrix $\mathcal{H}$ is only a function of the Hermite roots $y_i$, and is thus independent of the model parameters.

We then make a second assumption by considering the limit $\gamma \to 0$. In \cite{ADBM1} we showed that, as long as $\vec{\sigma}$ remains fixed, \eqref{small_dx_dynamics} has a unique fixed point, towards which the system converges (with a relaxation time $\sim 1/\lambda$) until  one of the $\sigma_i$ changes, at which time the system starts to converge to a new fixed point. Thus, if $\gamma$ is small enough (typically $\gamma \ll \lambda/N$), the system will spend most of its time close to a fixed point. Therefore we can write that at any given time
\begin{eqnarray}
\delta \vec{x} \simeq \frac{v_0}{\lambda} \mathcal{H}^{-1} \vec{\sigma} \;,
\label{small_dx_fp}
\end{eqnarray}
where $\vec{\sigma}$ is drawn uniformly among all possible values (since all the $\vec{\sigma}$ have the same probability to be visited over a sufficiently large time window), i.e. the $\sigma_i$'s are independent and take the value $\pm 1$ with equal probability. From \eqref{small_dx_fp} we can compute the mean and covariance of $\delta x_i = x_i - x_{{\rm eq},i}$ (using $\langle \sigma_i \rangle = 0$ and $\langle \sigma_i \sigma_j \rangle = \delta_{ij}$), leading to
\begin{eqnarray}
&&\langle \delta x_{i} \rangle \simeq 0\; , \\
&&\langle \delta x_{i} \delta x_{j} \rangle \simeq \frac{v_0^2}{\lambda^2} \sum_{k,l} (\mathcal{H}^{-1})_{ik} (\mathcal{H}^{-1})_{jl} \langle \sigma_{k} \sigma_{l} \rangle = \frac{v_0^2}{\lambda^2} (\mathcal{H}^{-2})_{ij}\; . \label{covHessian}
\end{eqnarray}
The relation in Eq. \eqref{covHessian} is the first important result of this paper. The rest of the paper will mainly focus on the inversion of the matrix $\mathcal{H}$ to obtain more explicit expressions for the covariance and related quantities. Before that, let us make two important comments on this result.

\begin{figure}
    \centering
    \includegraphics[width=0.325\linewidth,trim={0cm 0 1cm 1cm},clip]{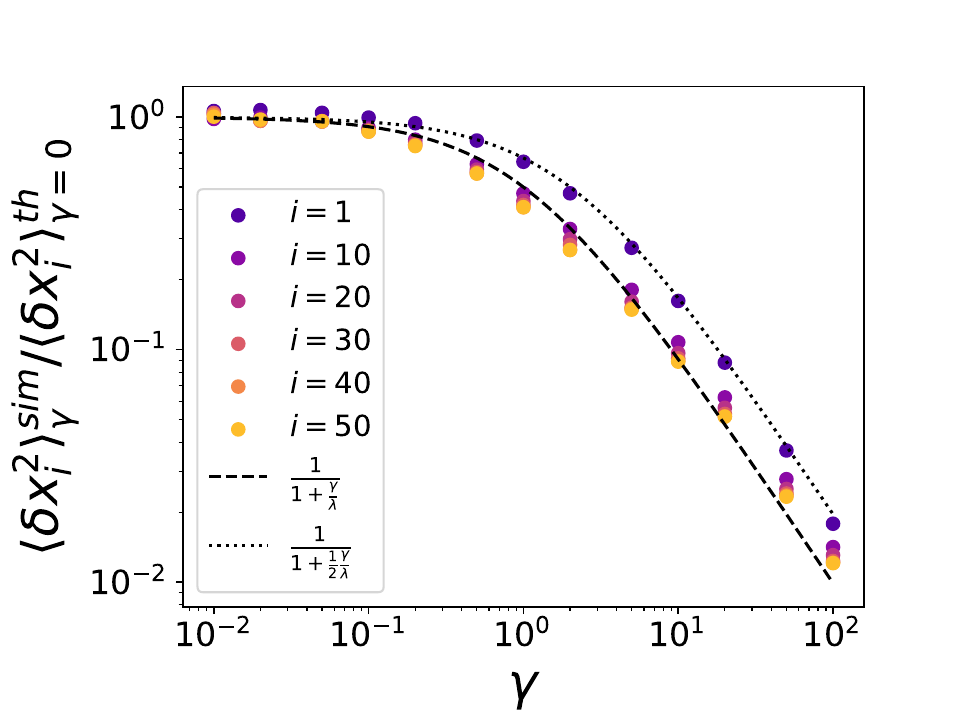}
    \includegraphics[width=0.325\linewidth,trim={0cm 0 1cm 1cm},clip]{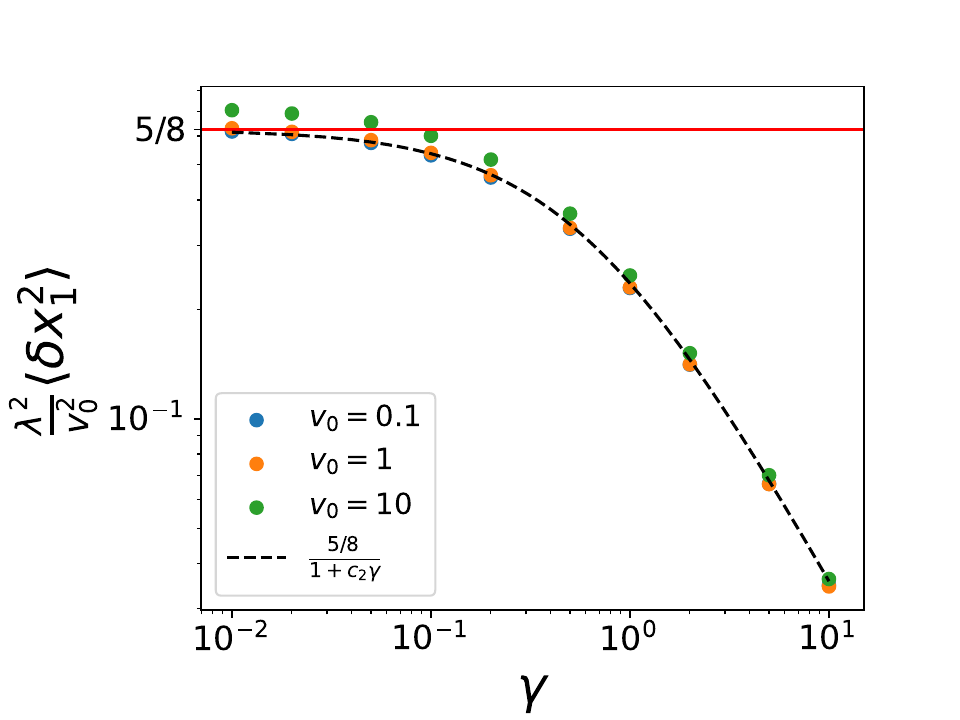}
    \includegraphics[width=0.325\linewidth,trim={0cm 0 1cm 1cm},clip]{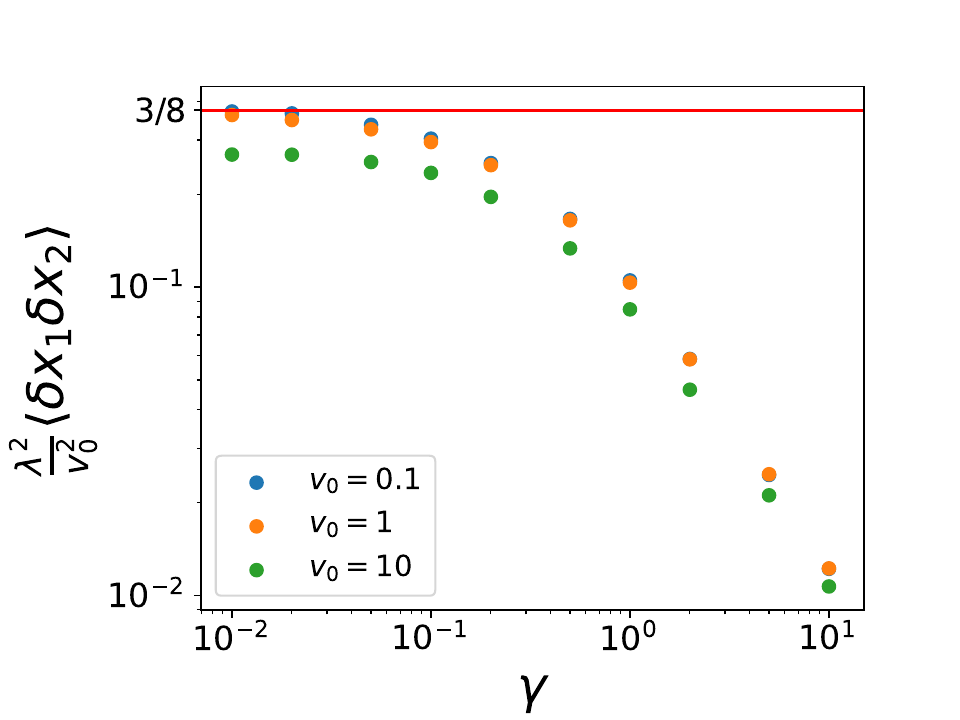}
    \caption{Test of the range of validity of the weak noise and small $\gamma$ regimes at finite $N$. 
    {\bf Left:} Ratio between the variance obtained from numerical simulations of Eq. \eqref{model0} at $T=0$
    with $N=100$ particles and the theoretical prediction \eqref{covHessian} at $\gamma=0$, as a function of $\gamma$, for different particle indices. We used $v_0=0.1$, $g=1$ and $\lambda=1$. For every particle the variance is a decreasing function of $\gamma$, and exhibits a plateau
    for $\gamma \leq 1$. More precisely, for the edge particle ($i=1$) it is very well fitted by a $(1+c\gamma)^{-1}$ decay with $c=\frac{1}{2}$, while for bulk particles a larger value of $c$ (of the order of $1$) gives a reasonable agreement. For a single particle, this functional form $(1+c\gamma)^{-1}$ with $c=2$ is exact for all $\gamma$ \cite{DKM19}. {\bf Center:} Variance of the position of particle $1$ for $N=2$, as a function of $\gamma$, for different values of $v_0$ (and $g$ and $\lambda=1$). The horizontal red solid line shows the prediction for $\gamma \to 0$ given in \eqref{cov_N2}. For small $v_0$ the data is fitted very well by the function $\frac{5/8}{1+c_2\gamma}$ with $c_2\simeq 1.648...$. {\bf Right:} Same plot for the covariance of particle $1$ and $2$.}
    \label{gammaEffect}
\end{figure}

First, we need to go back to the two assumptions that we made, namely $\delta x_i - \delta x_{i+1} \ll \sqrt{g/\lambda}/N$ and $\gamma \to 0$. The domain of validity of the first assumption was discussed in Section \ref{ADBM_main} and it is confirmed by numerical simulations, see Fig. \ref{gammaEffect}.
The second approximation seems quite restrictive. However, numerical simulations show that the covariance is a monotonically decreasing function of $\gamma$, so that \eqref{covHessian} gives an upper-bound on the covariance for generic $\gamma$ (see Fig.~\ref{gammaEffect} where a fit for the dependence in $\gamma$ is
also proposed). Although we were not able to prove this analytically, one can get a good intuition of why this is true by looking at the $N=1$ case \cite{DKM19}. Indeed, in this case the distribution of positions has a finite support $[x_-,x_+]$, where $x_\pm = \pm v_0/\lambda$ is the fixed point corresponding to $\sigma=\pm 1$ respectively. When $\gamma \ll \lambda$, the density is concentrated at the edges $x_\pm$, but as $\gamma$ increases it becomes more and more concentrated around $0$. For arbitrary $N$, this effect can be understood by the fact that, when $\gamma$ becomes large, particles will not have time to reach fixed points which are far away from their equilibrium position before $\vec{\sigma}$ changes again, and therefore they will become more and more localized. In the limit $\gamma \to +\infty$, the $\vec{\sigma}$ term in \eqref{small_dx_dynamics} averages to zero and $\delta \vec{x}$ simply relaxes to $0$.

{Finally, note that \eqref{covHessian} has a fairly general form and could also apply to other systems of interacting run-and-tumble particles in the weak noise, long persistence time limit. The only restrictive requirement is that the particles should remain close to their equilibrium position when $v_0$ is small. Keeping this in mind, this result should remain valid in some regime when considering other types of confining potential and other forms of diverging repulsive interactions (e.g. interaction forces of the form $\sgn(x_i-x_j) |x_i-x_j|^{-\alpha}$ with $\alpha>0$ as in the Riesz gas).}

\subsection{Calogero-Moser model}

We now turn to the CM model, for which a relation very similar to \eqref{covHessian} was proved in \cite{Agarwal2019}. For this model the potential reads
\begin{equation}
V^{CM}(\vec{X}) = \frac{\lambda}{2} \sum_i X_i^2 + \frac{4\tilde g^2}{N^2}\sum_{i<j} \frac{1}{(X_i-X_j)^2}\; .
\label{CalogeroPotential}
\end{equation}
Recalling that $X_{\rm eq,i}  = \frac{1}{\lambda^{1/4}}\sqrt{\frac{2 \tilde g}{ \, N}}\, y_i$, where the $y_i$ are the zeros of the Hermite polynomial $H_N(y)$, the Hessian of this potential is
\begin{eqnarray}
\frac{\partial^2 V^{CM}}{\partial X_i \partial X_j} (\vec{X}_{{\rm eq}}) &=& \delta_{ij}\left(\lambda+\frac{24 \tilde g^2}{N^2} \sum_{k\neq i} \frac{1}{(X_{{\rm eq},i}-X_{{\rm eq},k})^4}\right) - (1-\delta_{ij}) \frac{24 \tilde g^2}{N^2} \frac{1}{(X_{{\rm eq},i}-X_{{\rm eq},j})^4} \\
&=& \lambda\left[\delta_{ij}\left(1+ \sum_{k\neq i} \frac{6}{(y_i-y_k)^4}\right) - (1-\delta_{ij}) \frac{6}{(y_i-y_j)^4}\right] = \lambda (\mathcal{H}^2)_{ij}\; .
\label{calogeroHessian}
\end{eqnarray}
The last equality is non trivial and uses the fact that the $y_i$ are the zeroes of the Hermite polynomials.
It was proved in \cite{Agarwal2019}. 

As discussed in detail below in Section \ref{Sec:Hessian}, it turns out that the eigenvalues of ${\cal H}$ are simply the $N$ first strictly positive integers $k=1,2,\cdots, N$. We denote by $\psi_k$ the corresponding normalized eigenvectors, with components $(\psi_k)_i$ with $i=1,2,\cdots, N$ such that
\bea
\sum_{j=1}^N {\cal H}_{ij} (\psi_k)_j = k\, (\psi_k)_i \;.
\eea
For small displacements $\delta X_i = X_i -X_{{\rm eq},i}$ one can linearize the equations of motion (\ref{Calogero}) around the equilibrium positions, which gives
\be 
\frac{d}{dt} \delta X_i(t) = - \lambda \sum_{j=1}^N(\mathcal{H}^{2})_{ij} \, \delta X_j + \sqrt{\frac{2 T}{N}} \xi_i(t) \;.
\ee 
Taking the Fourier transform with respect to time one thus obtains by inversion in the frequency domain
\be 
\delta \hat X_i(\omega) = \sqrt{\frac{2T}{N}} \sum_{j=1}^N  [i \omega \mathbb{1}_N + \lambda \mathcal{H}^{2}]^{-1}_{ij} \hat \xi_j(\omega) \;,
\ee 
where $\mathbb{1}_N$ is the $N \times N$ identity matrix, $\delta \hat X_i(\omega) = \int_{-\infty}^{\infty} e^{i \omega t} \delta \hat X_i(\omega)\,dt$ and $\hat \xi_j(\omega)$ is a Gaussian white noise with correlations $\langle \hat \xi_i(\omega) \hat \xi_j(\omega') \rangle= 2 \pi \delta_{ij}\,\delta(\omega+\omega')$. 
We thus obtain the correlations at equilibrium
\begin{eqnarray}
\langle \delta X_i(t) \delta X_j(t') \rangle &=& \frac{2 T}{N} \int \frac{d\omega}{2 \pi} e^{-i \omega (t-t')} [\omega^2 \mathbb{1}_N + (\lambda \mathcal{H}^2)^2]^{-1}_{ij} \nonumber \\
&=& \frac{2 T}{N} \sum_{k=1}^N (\psi_k)_i (\psi_k)_j \int \frac{d\omega}{2 \pi} \frac{e^{-i \omega (t-t')}}{\omega^2+(\lambda k^2)^2} \nonumber \\
&=& \frac{T}{\lambda N} \sum_{k=1}^N \frac{e^{-\lambda k^2 |t-t'|}}{k^2} (\psi_k)_i (\psi_k)_j \;. \label{correl_time}
\label{CM_Cov2}
\end{eqnarray}
In the following sections, we will discuss the equal time correlations (hence forgetting the time dependence $\delta X_i(t) \to \delta X_i$). We will come back to the time-dependent correlations in Sec.~\ref{sec:dyn}. In the case of equal time correlations, $t=t'$, \eqref{correl_time} simply becomes (see also \cite{Agarwal2019})
\begin{equation}
\langle \delta X_i \delta X_j \rangle = \frac{T}{\lambda N} (\mathcal{H}^{-2})_{ij} \;.
\label{CalogeroCov}
\end{equation}
Surprisingly, this is the same as \eqref{covHessian} for the active DBM with only a different prefactor. Therefore the fluctuations in the two models can be studied simultaneously by inverting the matrix $\mathcal{H}$. Note however that this result is more general in the case of the CM model, since here we only assumed that the parameter $T/\tilde g \sqrt{\lambda}$ is small. We will see that \eqref{CalogeroCov} actually gives quite accurate results up to relatively high temperatures.

\subsection{Inverting the Hessian: a general formula}\label{Sec:Hessian}

From now on we will present the derivation of the results for the active DBM. Everything that follows can be easily transposed to the overdamped CM model by replacing the prefactor $v_0^2/\lambda^2$ by $T/(\lambda N)$.

It turns out that the matrix $\mathcal{H}$ can be diagonalized exactly using Hermite polynomials (see \cite{eigenvectors, Agarwal2019}). Its eigenvalues are simply the integers from 1 to $N$, and the normalized eigenvector $\psi_k$ associated to the eigenvalue $k$ has components given by
\begin{equation}
    (\psi_k)_i = \frac{u_k(y_i)}{\sqrt{\sum_{j=1}^N u_k(y_j)^2}} \quad , \quad u_k(y) = \frac{H_N^{(k)}(y)}{H_N'(y)} 
    = 2^{k-1} \frac{(N-1)!}{(N-k)!} \frac{H_{N-k}(y)}{H_{N-1}(y)} \;.
    \label{Hermite_eigenvectors}
\end{equation}
The proof is recalled in Appendix \ref{ProofEigenvectors}. Using the eigenvector decomposition
\begin{equation}
    (\mathcal{H}^{-2})_{ij} = \sum_{k=1}^N \frac{(\psi_k)_i (\psi_k)_j}{k^2} \;,
    \label{H_inversion}
\end{equation}
we obtain, in the case of the active DBM 
\begin{equation}
\langle \delta x_i \delta x_j \rangle = \frac{v_0^2}{\lambda^2} \sum_{k=1}^N \frac{1}{k^2} \frac{u_k(y_i)u_k(y_j)}{\sum_{l=1}^N u_k(y_l)^2} \quad {\rm with} \quad u_k(y)=\frac{H_N^{(k)}(y)}{H_N'(y)} \;.
\label{cov_dx}
\end{equation}
In particular the variance of the displacement for a single particle is given by
\begin{equation}
\langle \delta x_i^2 \rangle = \frac{v_0^2}{\lambda^2} \sum_{k=1}^N \frac{1}{k^2} \frac{u_k(y_i)^2}{\sum_{l=1}^N u_k(y_l)^2} \; .
\label{var_dx}
\end{equation}
Note that, since $H_N(-x)=(-1)^N H_N(x)$, we have the symmetry $\langle \delta x_i \delta x_j \rangle = \langle \delta x_{N-i+1} \delta x_{N-j+1} \rangle$. From \eqref{var_dx} we can already deduce the average of the variance over all particles for any $N$, as well as its large $N$ limit
\begin{equation}
\frac{1}{N} \sum_{i=1}^N \langle \delta x_i^2 \rangle = \frac{v_0^2}{\lambda^2 N} \sum_{k=1}^N \frac{1}{k^2} \simeq \frac{\pi^2}{6} \frac{v_0^2}{\lambda^2 N}\; .
\label{avg_var_dx}
\end{equation}
The goal of the next sections is to simplify the expressions (\ref{cov_dx})-(\ref{var_dx}) in different limits. A convenient starting point is the differential equation satisfied by the Hermite polynomials
\begin{equation}
    H_N''(x) = 2x H_N'(x) - 2N H_N(x)\; .
    \label{Hermite_equation}
\end{equation}
Differentiating $k$ times the above equation, evaluating it at $x=y_i$ and dividing both sides by $H_N'(y_i)$ we obtain the recurrence relation
\begin{equation}
    u_{k+2}(y_i) = 2y_i u_{k+1}(y_i) - 2(N-k) u_{k}(y_i) \;,
    \label{recursion_exact}
\end{equation}
with initial conditions $u_0(y_i)=0$ and $u_1(y_i)=1$, which allows to determine $u_k(y_i)$ (note that the recursions stops since 
$u_{N+1}(y_i)=0$). Although this recursion relation does not have a simple solution in general, we will see in the next section how an approximate version of this equation can be used to obtain an asymptotic expression for (\ref{cov_dx})-(\ref{var_dx}) in the large $N$ limit.

As a side remark, a different approach would consist in diagonalizing $\mathcal{H}$ approximately for large $N$, as done e.g. in \cite{bouchaud_book} (chap. 5.4) for the DBM, by assuming the density to be uniform in the bulk, using plane waves, and computing the inverse of the Hessian (very much as a calculation of displacements using phonons in a solid). This approach gives less accurate results than the one presented below, but it is somewhat simpler to implement and also more general, so we give the main ideas in Appendix \ref{Bouchaud_approach}.

\vspace*{0.5cm}
\noindent{\bf The special cases $N=1$ and $N=2$.}
Below we will mainly focus on large values of $N$. However, \eqref{cov_dx} is also valid at small $N$. In particular for $N=1$ it gives $\langle \delta x_1^2 \rangle = \frac{v_0^2}{\lambda^2}$. This is indeed the $\gamma \to 0$ limit of the result obtained in \cite{DKM19}, $\langle \delta x_1^2 \rangle = \frac{v_0^2}{\lambda^2}\frac{1}{1+2\frac{\gamma}{\lambda}}$. Note that the variance is a decreasing function of $\gamma$, as discussed previously. For $N=2$ one obtains
\begin{equation}
\langle \delta x_i^2 \rangle = \frac{5}{8}\frac{v_0^2}{\lambda^2} \quad (i=1,2) \quad , \quad
\langle \delta x_1 \delta x_2 \rangle = \frac{3}{8}\frac{v_0^2}{\lambda^2} \;.
\label{cov_N2}
\end{equation}
These results are compared to simulations in Fig.~\ref{gammaEffect}. Both the variance and covariance seem to converge to the predicted result as $\gamma \to 0$ for small values of $v_0$.

\section{Large $N$ limit in the bulk} \label{sec:bulk}

\begin{figure}
    \centering
    \includegraphics[width=0.325\linewidth,trim={0cm 0 1cm 1cm},clip]{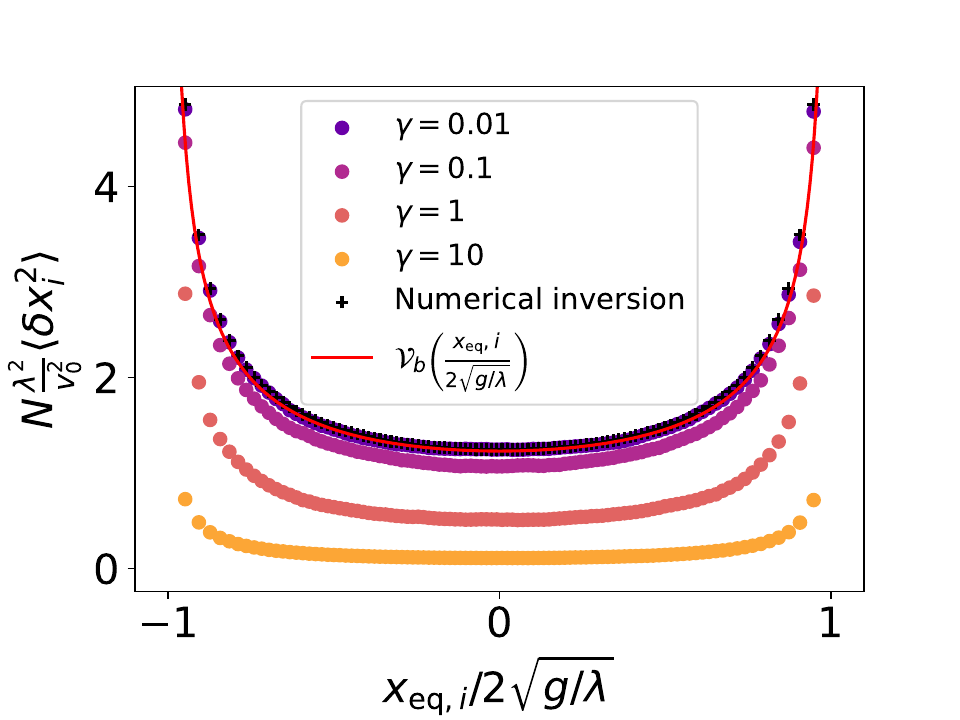}
    \includegraphics[width=0.325\linewidth,trim={0cm 0 1cm 1cm},clip]{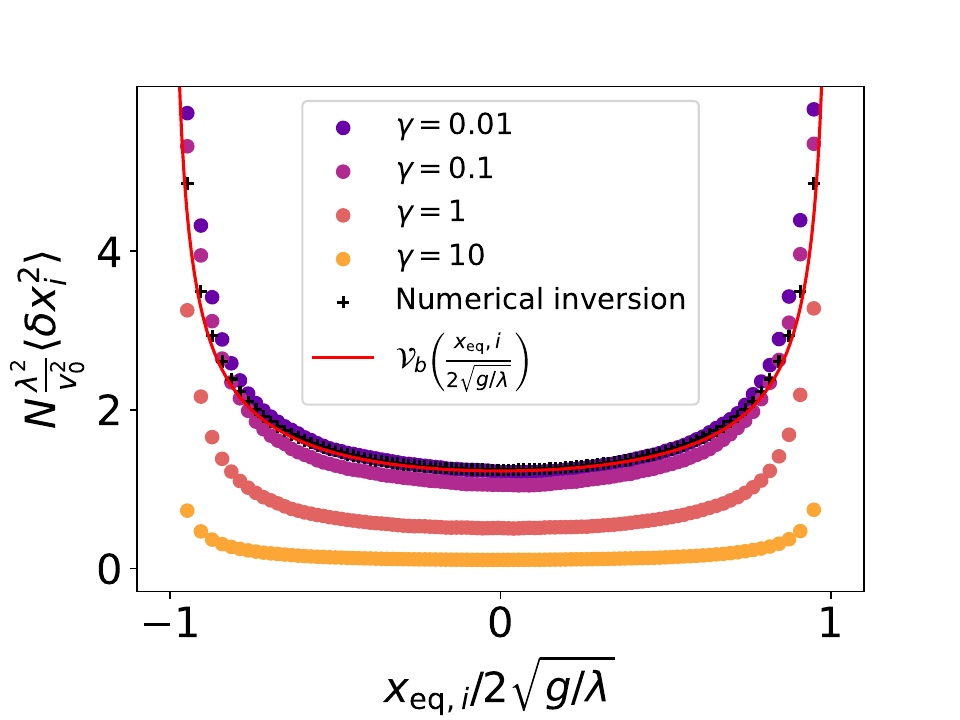}
    \includegraphics[width=0.325\linewidth,trim={0cm 0 1cm 1cm},clip]{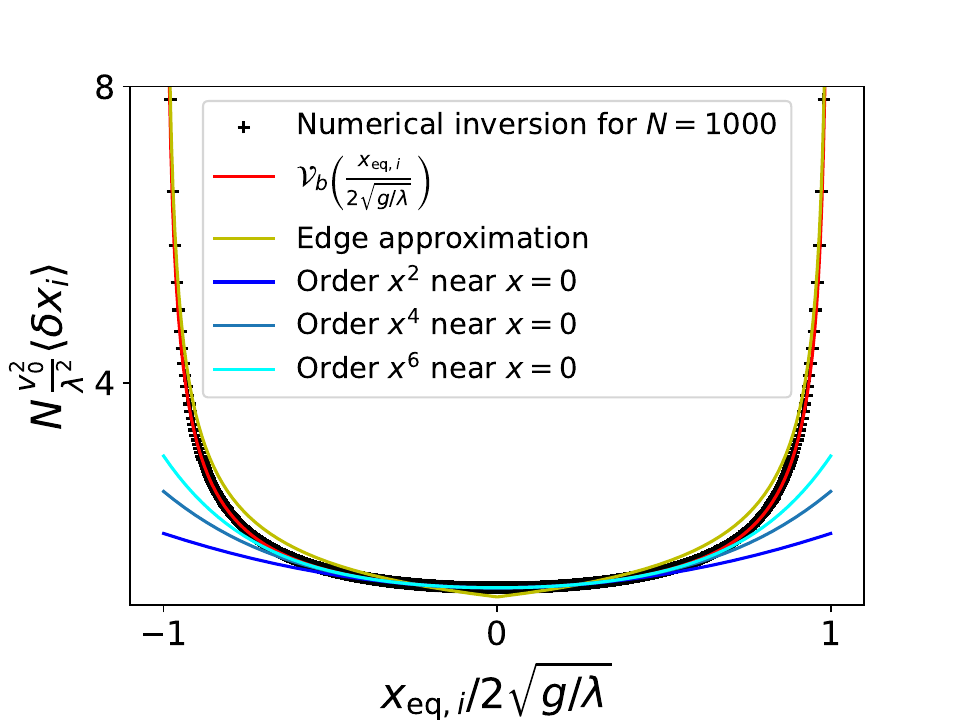}
    \caption{Results for the active DBM: comparison of the numerical simulations of Eq. \eqref{model0} at $T=0$, with our analytical predictions. 
    {\bf Left}: Rescaled variance of the particle position $x_i$ as a function of $x_i/2\sqrt{g}$ for $N=100$, $\lambda=1$, $g=1$, $v_0=0.1$ and different values of $\gamma$. The results of the simulations are compared with those (i) obtained by exact inversion of the Hessian matrix and (ii) with the asymptotic expression for large $N$ given in \eqref{var_largeN}. As expected the agreement is good for small values of $\gamma$. 
    {\bf Center}: Same plot for the large value $v_0=1$. The agreement decreases near the edges but remains good in the bulk. {\bf Right}: Expansion of $\mathcal{V}_b(x)$ around $x=0$ [see Eq. (\ref{v_smallx})] and around $x=1$ the edges [see Eq. (\ref{Vb_1})].}
    \label{variancefig}
\end{figure}

We first notice that the factor $u_k(y_i)u_k(y_j)/\sum_{l=1}^N u_k(y_l)^2$ in \eqref{cov_dx} is always smaller than 1 (this can be shown using $u_k(y_i)u_k(y_j)<\frac{1}{2}(u_k(y_i)^2+u_k(y_j)^2)$). Hence the sum over $k$ in \eqref{cov_dx} is convergent and bounded 
by $\sum_{k\geq 1} 1/k^2$. As a result we can obtain an asymptotic expression for $N \to +\infty$ by focusing on values of $k$ such that $k\ll N$. 
In fact, we will see that it is dominated by $k=O(1)$ in the bulk, and
$k=O(N^{1/3})$ at the edge. 

We will now simplify Eqs. (\ref{cov_dx}) and (\ref{var_dx}) in the limit of large $N$. 
There is a systematic method to study the recursion relation \eqref{recursion_exact} at large $N$, which is presented in Appendix \ref{app:expansion}. 
We give here only the leading order
which is sufficient for our present considerations. The results below can also be obtained using the Plancherel-Rotach formula
(see Appendix \ref{Plancherel}).

We start by rescaling \eqref{recursion_exact}, introducing $y_i=\sqrt{2N}r_i$ and $u_k(y_i)=(2N)^{\frac{k-1}{2}}v_k(r_i)$ to obtain
\begin{equation}
    v_{k+2}(r_i) = 2r_i v_{k+1}(r_i) - \left(1 - \frac{k}{N} \right) v_{k}(r_i) \; .
    \label{recursion_rescaled}
\end{equation}
Since we can focus on $k \ll N$,  we can simplify the recurrence relation \eqref{recursion_rescaled} and {get rid of the term proportional to $k/N$ in \eqref{recursion_rescaled}, leading to}
\begin{equation}
    v_{k+2}(r_i) = 2r_i v_{k+1}(r_i) - v_{k}(r_i) \; .
    \label{recursion_approx}
\end{equation}
It turns out that this scaling together with this approximation are adapted to the bulk, $i,j \sim N$, to which we now focus on (and we
recall that the relevant values of $k$ are of order unity). One can check that this approximation leads to a relative error of order $O(\frac{1}{N})$ in the bulk. We then recognize that the simplified equation \eqref{recursion_approx} is the recursion relation satisfied by the Chebyshev polynomials of the second kind $U_k(r_i)$ \cite{ChebyWiki}. Since $v_1(r_i)=1=U_0(r_i)$ and $v_2(r_i) = 2r_i = U_1(r_i)$, we obtain $v_k(r_i)=U_{k-1}(r_i)$ for all $k\geq 1$, i.e.
\begin{equation}
    u_k(y_i) \simeq (2N)^{\frac{k-1}{2}} U_{k-1} \left( \frac{y_i}{\sqrt{2N}} \right)  = (2N)^{\frac{k-1}{2}} \frac{\sin(k \arccos(\frac{y_i}{\sqrt{2N}}))}{\sqrt{1-\frac{y_i^2}{2N}}} \;,
    \label{uk_chebyshev}
\end{equation}
up to an error of order $O(\frac{1}{N})$ (for the second identity we have used the fact that $|y_i|<\sqrt{2N}$ for all $i$). 
When plugging this result into Eq. \eqref{cov_dx} we see that the factors $(2N)^{\frac{k-1}{2}}$ simplify between the numerator and 
the denominator. Additionally, for large $N$, the density of the roots of Hermite polynomials converges to the Wigner semi-circle density. Therefore, we can replace the sum over all Hermite roots in the denominator of Eq. \eqref{cov_dx} by an integral over the semi-circle density to obtain
\begin{equation}
    \sum_{l=1}^N U_{k-1} \left( \frac{y_l}{\sqrt{2N}} \right)^2 \simeq N \int_{-1}^1 dx \frac{2\sqrt{1-x^2}}{\pi} U_{k-1}(x)^2 = N \;,
    \label{normalization}
\end{equation}
(the error made by replacing the sum by an integral is again of order $O(\frac{1}{N})$). The last equality comes from the fact that the Chebyshev polynomials of the second kind are orthonormal with respect to the Wigner semicircle measure (but it can also be shown through an explicit computation using the second expression in \eqref{uk_chebyshev}).

This leads to the following expressions for the covariances, valid in the large $N$ limit away from the edges (extending the sum to $+\infty$ instead of $N$ again leads to an error of order $O(\frac{1}{N})$)
\begin{equation}
\langle \delta x_i \delta x_j \rangle \simeq \frac{v_0^2}{\lambda^2 N} \mathcal{C}_b\left( \frac{x_{{\rm eq},i}}{2\sqrt{g/\lambda}}, \frac{x_{{\rm eq},j}}{2\sqrt{g/\lambda}} \right) \ , \ \mathcal{C}_b(x,y) = \sum_{k=1}^\infty \frac{1}{k^2} U_{k-1}(x) U_{k-1}(y) = \sum_{k=1}^\infty \frac{1}{k^2} \frac{\sin(k \arccos x)}{\sqrt{1-x^2}} \frac{\sin(k \arccos y)}{\sqrt{1-y^2}} \;.
\label{cov_largeN}
\end{equation}
This can also be written (see Appendix \ref{app:alternative} for a derivation)
\begin{equation}
\mathcal{C}_b(x,y) = \frac{\pi \arccos(\max(x,y)) - \arccos (x) \arccos (y)}{2\sqrt{1-x^2} \sqrt{1-y^2}} \;.
\label{newCb}
\end{equation}
For $x\geq y$ this reads
\begin{equation}
\mathcal{C}_b(x,y) = \frac{\arccos(x)(\pi - \arccos (y))}{2\sqrt{1-x^2} \sqrt{1-y^2}} \;.
\label{newCb2}
\end{equation}
For the variance of the position for a single particle, one finds
\begin{equation}
\langle \delta x_i^2 \rangle \simeq \frac{v_0^2}{\lambda^2 N} \mathcal{V}_b\left(\frac{x_{{\rm eq},i}}{2\sqrt{g/\lambda}}\right) \ , \ \mathcal{V}_b(x) = \sum_{k=1}^\infty \frac{1}{k^2} U_{k-1}(x)^2 = \sum_{k=1}^\infty \frac{1}{k^2} \frac{\sin^2(k \arccos x)}{1-x^2} = \frac{\arccos (x) (\pi-\arccos (x))}{2(1-x^2)}  \;,
\label{var_largeN}
\end{equation}
the total relative error being of order $O(\frac{1}{N})$ in both cases. We recall that $x_{{\rm eq},i} = \sqrt{\frac{2g}{\lambda \, N}}\, y_i \in (-2\sqrt{g/\lambda}, 2\sqrt{g/\lambda})$ and therefore $\mathcal{C}_b(x,y)$ and $\mathcal{V}_b(x)$ are defined on $(-1,1)^2$ and $(-1,1)$ respectively. 
Near $x=0$, the function $\mathcal{V}_b(x)$ has a minimum and behaves as
\begin{equation} \label{v_smallx}
    \mathcal{V}_b(x) = \frac{\pi^2}{8} + \left( \frac{\pi^2}{8} -\frac{1}{2} \right) x^2 + O(x^4) \;.
\end{equation}
For particles which are close to the edge (i.e. $i \ll N$), $x$ is very close to 1 and therefore $\mathcal{C}_b(x,y)$ and $\mathcal{V}_b(x)$ diverge. For instance one finds
\begin{eqnarray} \label{Vb_1}
\mathcal{V}_b(1-\epsilon) = \frac{\pi}{2\sqrt{2\epsilon}} - \frac{1}{2} + \frac{7\pi}{24\sqrt{2}} \sqrt{\epsilon} - \frac{\epsilon}{3} + O(\epsilon^{3/2}) \;.
\end{eqnarray}
For $\mathcal{C}_b(x,y)$ with $x=1-\epsilon \geq y$, one finds
\begin{equation} \label{Cb_1bulk}
\mathcal{C}_b(1-\epsilon,y) = \frac{(\pi-\arccos y)}{2\sqrt{1-y^2}} + O(1) \;,
\end{equation}
which remains finite at the edge when considering only one of the particle near the edge. By contrast,
if one considers two particles near the edge, i.e. 
if in addition $y=1-\delta$, one finds the diverging expression
\begin{equation} \label{Cb_1}
\mathcal{C}_b(1-\epsilon,1-\delta) = \frac{\pi}{2\sqrt{2\max(\epsilon,\delta)}} + O(1) \;.
\end{equation}
We will come back to this limit in the next section where we study the edge region.

One can also write the expansion for $\mathcal{C}_b(x,0)$ for $x\ll 1$, i.e. for the covariance between a particle at $x=0$ and a particle close to $x=0$. In this case one gets
\be
\mathcal{C}_b(x,0) = \begin{cases} &\frac{\pi \arccos x}{4\sqrt{1-x^2}} \ , \quad \ x\geq 0 \\ 
&\\
&\frac{\pi (\pi - \arccos x)}{4\sqrt{1-x^2}} \, \ x\leq 0 \end{cases} \hspace{0.5cm} = \frac{\pi^2}{8} -\frac{\pi}{4} |x| + \frac{\pi^2}{16} x^2 - \frac{\pi}{6} |x|^3 + O(x^4) \;,
\ee
which explains the cusp observed in Figs. \ref{AKDcomparison3} and \ref{FigcovarianceADBM}. In general when $x$ and $y$ are small and of the same order, one has
\be
\mathcal{C}_b(x,y) = \frac{\pi^2}{8} -\frac{\pi}{4} |x-y| + \frac{\pi^2}{16} (x^2+y^2) - \frac{xy}{2} 
+ {\rm higher \ order} \;.
\ee

In the bulk these expressions are in very good agreement with the results obtained by diagonalizing the Hessian numerically, as well as with numerical simulations 
for the active DBM for small $\gamma$, even for large values of $v_0$, of order $O(1)$
(see Fig. \ref{variancefig} for the one-particle variance and the left panel of Fig. \ref{FigcovarianceADBM} for the covariance). As one gets closer to the edges, the large $N$ approximation above becomes less accurate. 

\begin{figure}
    \centering
    \includegraphics[width=0.325\linewidth,trim={0cm 0 1cm 1cm},clip]{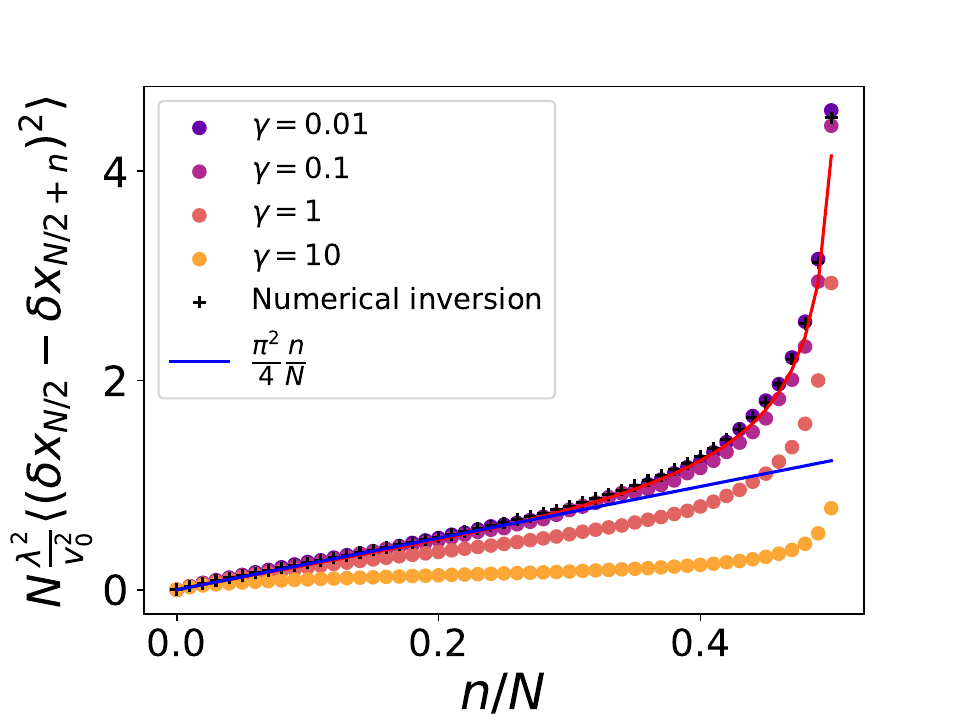}
    \includegraphics[width=0.325\linewidth,trim={0cm 0 1cm 1cm},clip]{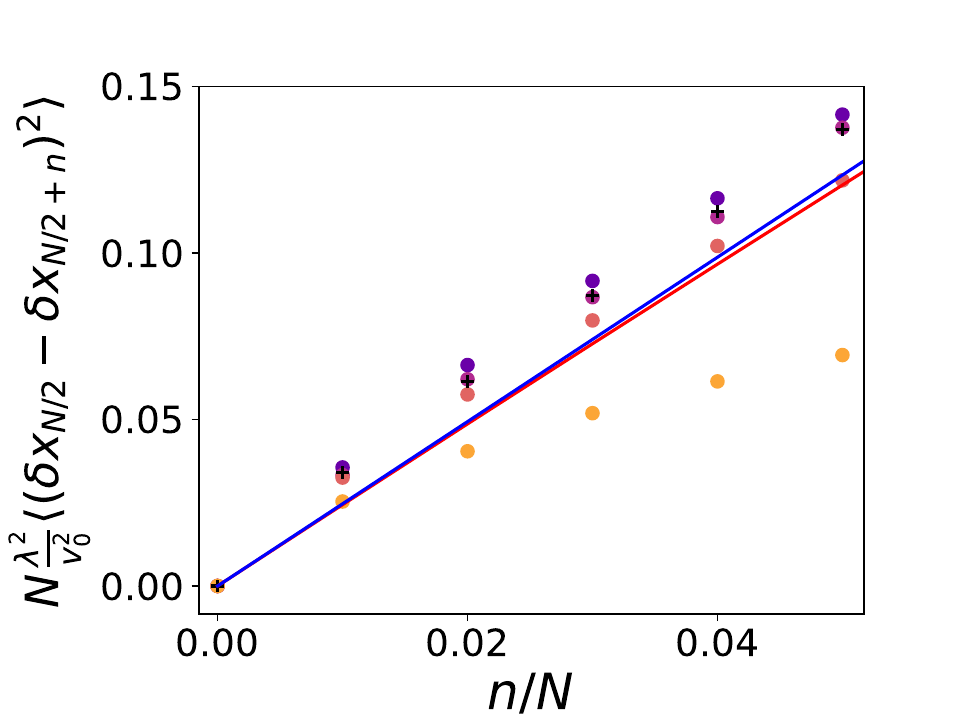}
    \includegraphics[width=0.325\linewidth,trim={0cm 0 1cm 1cm},clip]{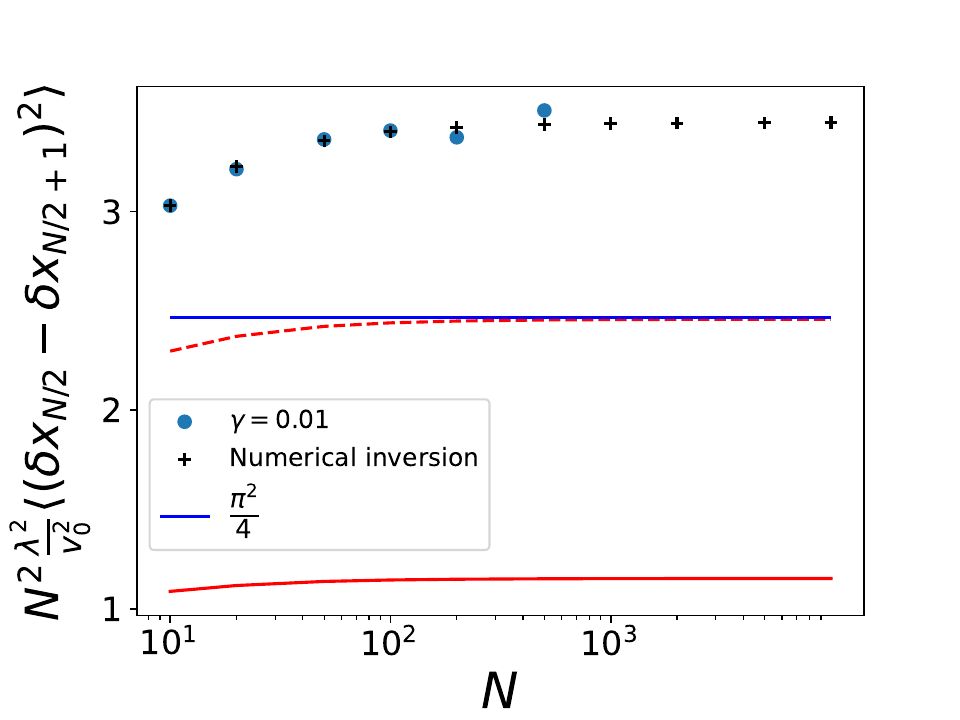}
    \caption{{\bf Left}: Variance of the distance between the central particle $i=N/2$ and the particle $i=N/2+n$ as a function of $n$ for $\lambda=1$, $g=1$, $v_0=1$ and $N=100$.
    The numerical simulations are performed for different values of $\gamma$. They are compared with (i) the numerical inversion of the Hessian (black crosses),
    (ii) the analytical prediction $\mathcal{D}_b\left(\frac{x_{{\rm eq},N/2}}{2\sqrt{g/\lambda}}, \frac{x_{{\rm eq},N/2+n}}{2\sqrt{g/\lambda}}\right)$ from \eqref{var_chn} at large $N$ (red line) and, (iii) the plane wave (linear)approximation 
    \eqref{gapvariance_largeN} (blue line). The behavior of the variance is linear in $n$ in the range $1\ll n \ll N$.
     {\bf Center:} zoom of the left figure on the small values of $n$. {\bf Right:} Variance of the central gap as a function of $N$ computed using the exact inverse of the Hessian (black crosses) versus using the large $N$ bulk approximation \eqref{var_chn}, $N \mathcal{D}_b\left(\frac{x_{{\rm eq},N/2}}{2\sqrt{g/\lambda}}, \frac{x_{{\rm eq},N/2+1}}{2\sqrt{g/\lambda}}\right)$, with the sum truncated at $N$ (full red line) or at $100 \; N$ (dashed red line). This approximation does not converge to the exact result as $N$ increases.}
    \label{variancefig2}
\end{figure}

Finally the expression of the covariance can be used to obtain the variance of the distance between two particles. Indeed
\begin{equation} \label{corr1} 
    \langle (\delta x_i - \delta x_{i+n})^2 \rangle = \langle \delta x_i^2 \rangle + \langle \delta x_{i+n}^2 \rangle - 2 \langle \delta x_i \delta x_{i+n} \rangle \;.
\end{equation}
For $1 \ll n \ll N$ it turns out that one can use the above results to estimate {the r.h.s of} \eqref{corr1}. 
This leads to 
\be
\langle (\delta x_i - \delta x_{i+n})^2 \rangle \simeq \frac{v_0^2}{\lambda^2 N} \mathcal{D}_b\left(\frac{x_{{\rm eq},i}}{2\sqrt{g/\lambda}}, \frac{x_{{\rm eq},i+n}}{2\sqrt{g/\lambda}}\right) \quad , \quad \mathcal{D}_b(x,y)= \mathcal{V}_b(x) + \mathcal{V}_b(y) - 2 \mathcal{C}_b(x,y)
\label{var_chn} \;.
\ee
This expression is valid for $n=\alpha N$ with $\alpha = O(1)$, i.e. on mesoscopic scales
in the bulk. 
%
%
%
In the limit where $\alpha \to 0$, i.e. when $x-y \ll 1$, one has
\be
\mathcal{D}_b(x,y) = \frac{\pi}{2} \frac{|x-y|}{(1-x^2)^{3/2}} + O\left( (x-y)^2 \right) \;.
\label{D_asympt}
\ee
Using the expression for the semi-circle density $\rho(x)= \sqrt{\frac{\lambda}{g}} \frac{\sqrt{1-(x/(2\sqrt{g/\lambda}))^2}}{\pi}$ together with the fact that $x_{{\rm eq},i}-x_{{\rm eq},i+n}\simeq n/(N\rho(x_{{\rm eq},i}))$, Eqs. \eqref{var_chn} and \eqref{D_asympt} lead to
\begin{equation}
    \langle (\delta x_i - \delta x_{i+n})^2 \rangle \simeq \frac{v_0^2}{4 \pi^2 g^2 \rho(x_{{\rm eq},i})^4} \frac{n}{N^2} \;.
    \label{gapvariance_largeN_general}
\end{equation}
As shown in Appendix \ref{Bouchaud_approach}, this expression coincides with the result obtained by a plane wave approximation given in Eq. \eqref{gapvariance_fourier}. 
%
In the case where $x$ and $y$ are close to $0$, one can easily get the next order of the expansion, namely 
\be
\mathcal{D}_b(x,y) = \frac{\pi}{2} |x-y| - \frac{1}{2}(x-y)^2 + {\rm higher \ order} \;,
\ee
which leads to
\begin{equation} \label{gapvariance_largeN}
    \langle (\delta x_i - \delta x_{i+n})^2 \rangle \simeq \frac{\pi^2}{4} \frac{v_0^2}{\lambda^2 N} \left( \frac{n}{N} - \frac{1}{2} \left( \frac{n}{N} \right)^2 \right) \quad , \quad i = N/2  \;.
\end{equation}


Let us recall that the above results are valid for $n=\alpha N$
with $\alpha \ll 1$. 
However it is important to note that it is {\it incorrect} for $n=1$, i.e. to compute the 
variance of the gaps, and 
more generally for $n=O(1)$. This is because we have studied the recursion \eqref{recursion_rescaled} only
to leading order at large $N$. This is sufficient to obtain the covariance at large $N$, but in the
calculation of \eqref{var_chn} for $n=O(1)$ there are cancellations of the leading order of each term and one needs a more precise estimate (which we have not obtained here). A more detailed analysis of the problems encountered to compute the gaps is given in Appendix \ref{app:gaps}. 
These considerations are in agreement with numerical simulations, see Fig.~\ref{variancefig2}.


In the range $1 \ll n \ll N$ however, the prediction \eqref{gapvariance_largeN} agrees very well with numerical simulations for small values of $\gamma$ (see the left panel of Fig. \ref{variancefig2}). The result in \eqref{gapvariance_largeN_general} gives us two important pieces of information. First we see that the variance of the distance between two particles increases linearly with the distance. This is to be compared with the standard DBM for which it increases logarithmically~\cite{bouchaud_book}. Second, while the variance in the position of a particle scales as $1/N$, the variance of the gap between two neighboring particles scales as $1/N^2$. This means that the particles fluctuate collectively.
\\

\noindent {\bf Remark on linear statistics in the stationary state of the ADBM}. Our method allows to obtain the leading order at large $N$ and for weak noise of the variance of the linear statistics. Defining, for any smooth function $f(x)$, the random variable
\be 
{\cal L}_N = \sum_{i=1}^N f(x_i) \;.
\ee 
To compute its variance one can expand around the equilibrium positions, which leads to
\be 
{\rm Var} {\cal L}_N = \sum_{i,j=1}^N f'(x_{{\rm eq}, i}) f'(x_{{\rm eq},j}) \langle \delta x_i \delta x_j \rangle 
\ee 
In the large $N$ limit, in the bulk and in the steady state, using Eq.~\eqref{cov_largeN} this leads to the leading order estimate
\bea   \label{var_linear}
{\rm Var} {\cal L}_N &\simeq& \frac{v_0^2 N}{\lambda^2} \frac{4}{\pi^2} \int_{-1}^1 dx \int_{-1}^1 dy \sqrt{1-x^2} \sqrt{1-y^2}  f'\left(2 \sqrt{\frac{g}{\lambda}} \, x\right) f'\left(2 \sqrt{\frac{g}{\lambda}} \, y\right) C_b(x,y) \\
&=& \frac{v_0^2 N}{\lambda^2} \frac{8}{\pi^2} \int_{-1}^1 dx \int_{-1}^x dy  f'\left(2 \sqrt{\frac{g}{\lambda}} \, x\right) f'\left(2 \sqrt{\frac{g}{\lambda}} \, y\right)
\arccos(x)(\pi - \arccos (y)) 
\eea  
Note that the first line in (\ref{var_linear}) can also be written using \eqref{cov_largeN} as
\bea \label{variance}
{\rm Var} {\cal L}_N \simeq \frac{v_0^2 N}{\lambda^2}  \sum_{k=1}^\infty \hat f_k^2 \quad , \quad k \hat f_k = \frac{2}{\pi}\int_{-1}^{1} dx \sqrt{1-x^2} f'\left(2 \sqrt{\frac{g}{\lambda}} \, x\right) U_{k-1}(x) \;.
\eea
Let us notice that since the $U_{k-1}(x)$ form an orthonormal basis w.r.t to the measure $\frac{2}{\pi} dx \sqrt{1-x^2}$ (for $x \in [-1,1]$)
one has the decomposition $f'(2 \sqrt{\frac{g}{\lambda}} \, x) = \sum_{k \geq 1} k \hat f_k U_{k-1}(x)$. Hence the coefficients $\hat f_k$ are the so-called Chebyshev-Fourier coefficients of $f(2 \sqrt{\frac{g}{\lambda}} \, x)$. These coefficients also appear in the linear statistics of the Gaussian $\beta$-ensembles in the bulk \cite{lambert}. In this case, the variance is proportional to $\sum_{k \geq 1} k f_k^2$. In fact, using the results for the correlations in the forthcoming sections, the same method allows to obtain the variance of the linear statistics in the case of the CM model and of the DBM at equilibrium.

\begin{figure}
    \centering
    \includegraphics[width=0.325\linewidth,trim={0cm 0 1cm 1cm},clip]{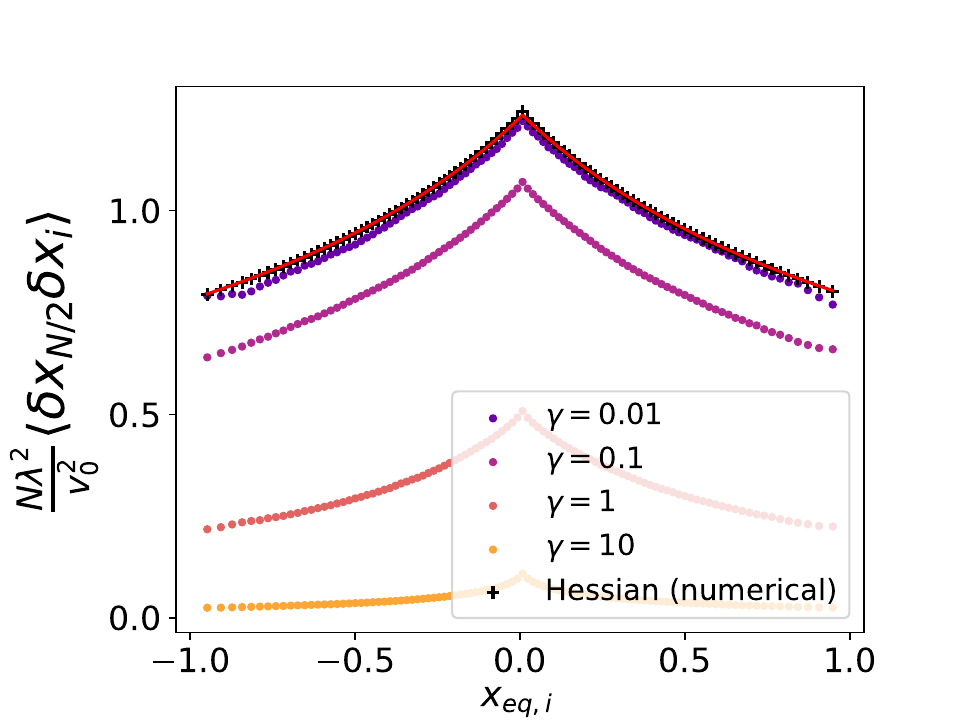}
    \includegraphics[width=0.325\linewidth,trim={0cm 0 1cm 1cm},clip]{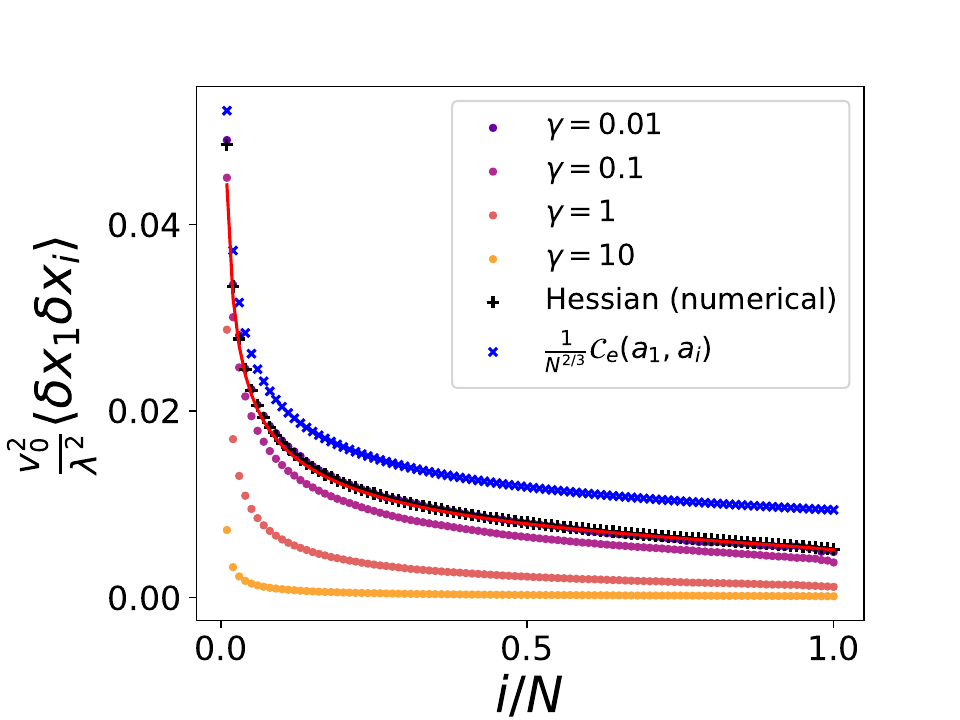}
    \caption{{\bf Left:} Correlations between the central particle and particle $i$ for $N=100$, $v_0=0.1$, $g=1$ and $\lambda=1$. The red line shows the prediction for $\mathcal{C}_b\left( \frac{x_{{\rm eq},N/2}}{2\sqrt{g/\lambda}}, \frac{x_{{\rm eq},i}}{2\sqrt{g/\lambda}} \right)$ from \eqref{cov_largeN}. {\bf Right:} Correlations between the edge particle and particle $i$ for the same parameters. The results from simulations and numerical inversion of the Hessian are compared both with the edge-edge expression \eqref{cov_edge_integral} and the edge-bulk expression \eqref{cov_edge_bulk}, $\frac{1}{N} \frac{\pi - \arccos \big( \frac{x_{{\rm eq},i}}{2\sqrt{g/\lambda}} \big)}{2\sqrt{1-\big( \frac{x_{{\rm eq},i}}{2\sqrt{g/\lambda}} \big)^2}}$ (red line).}
    \label{FigcovarianceADBM}
\end{figure}

\section{Large $N$ limit at the edge} \label{sec:edge}

In the previous section we focused on particles which are inside the bulk of the distribution. We now want to look at particles which are located close to the edges, i.e. with a label $i \ll N$. Obtaining the behavior of the variance and covariance near the edge is more difficult since $\mathcal{V}_b(x)$ and  $\mathcal{C}_b(x,y)$ diverge as $x \to \pm 1$ [see Eq. (\ref{Vb_1})]. This implies that the scaling $\delta x_i^2 \sim 1/N$ breaks down in this limit. A first way to see this is to start from the bulk result for the covariance and use the asymptotic expansion for the largest roots of the Hermite polynomials \cite{Hermite_asymptotics,DeiftAsymptotics1999}. The
equilibrium positions near the edge read
\begin{eqnarray}
x_{{\rm eq},i} = 2\sqrt{\frac{g}{\lambda}} \left( 1  + \frac{a_i}{2} N^{-2/3} + O(N^{-1}) \right) \;,
\label{hermite_roots_edge}
\end{eqnarray}
where $a_i$ is the $i^{th}$ zero of the Airy function, which for large $i$ is given by $a_i = - (\frac{3 \pi}{8} (4i-1))^{2/3} + O(i^{-4/3})$. For simplicity we focus on the case of the variance. Inserting this expansion in \eqref{cov_largeN} and using the asymptotic behavior 
in Eq. \eqref{Cb_1}, we obtain for $i\leq j$,
\begin{equation}
\langle \delta x_i \delta x_j \rangle \simeq \frac{\pi}{2\sqrt{-a_j}} \frac{v_0^2}{\lambda^2 N^{2/3}} \underset{j \gg 1}{\simeq} \left( \frac{\pi^2}{3(4j-1)} \right)^{1/3} \frac{v_0^2}{\lambda^2 N^{2/3}} \;.
\label{rightmost_cov}
\end{equation}
Therefore we find that the variance of the position of the rightmost particle scales as $N^{-2/3}$ for the active DBM, and as $N^{-5/3}$ for the Calogero-Moser model (versus $N^{-4/3}$ for the standard DBM). This is confirmed by numerical simulations for the active DBM for small values of $v_0$ and $\gamma$ (see Fig. \ref{variancefig2}). For the Calogero-Moser model this result once again agrees with \cite{Agarwal2019}.

The expression \eqref{rightmost_cov} works very well in the intermediate regime $1 \ll i \ll N$. However, for $i$ of order 1 it is possible to obtain a more precise formula. Let us go back to \eqref{Hermite_eigenvectors}-\eqref{cov_dx} and again use the $k \ll N$ approximation to compute the normalization factor, but this time keeping the Hermite polynomials at the numerator and using \eqref{normalization} to evaluate the denominator (which is identical to its value in the bulk)
\begin{equation}
\langle \delta x_i \delta x_j \rangle \simeq \frac{v_0^2}{\lambda^2 N} \sum_{k=1}^{\infty} \frac{1}{k^2} \frac{1}{(2N)^{k-1}} \left(2^{k-1} \frac{(N-1)!}{(N-k)!} \right)^2 \frac{H_{N-k}(y_i)}{H_{N-1}(y_i)} \frac{H_{N-k}(y_j)}{H_{N-1}(y_j)}\; .
\label{cov_edge_step1}
\end{equation}
We then use the asymptotic expression of Hermite polynomials near the edge \cite{ForresterHermiteEdge},
\begin{equation}
    e^{-x^2/2} H_n(x) = \pi^{1/4} 2^{n/2+1/4} \sqrt{n!} \ n^{-1/12} (\Ai(t) + O(n^{-2/3})) \label{Airy_approx}
\end{equation}
with $x=\sqrt{2n}(1+n^{-2/3}t/2)$. Denoting $t_{n,i}=2n^{2/3}(y_i/\sqrt{2n}-1)$ we get
\begin{equation}
    \frac{H_{N-k}(y_i)}{H_{N-1}(y_i)} \simeq 2^{(1-k)/2} \sqrt{\frac{(N-k)!}{(N-1)!}} \ \left(\frac{N-k}{N-1}\right)^{-1/12} \frac{\Ai(t_{N-k,i})}{\Ai(t_{N-1,i})}\; .
\end{equation}
Using Stirling's formula \eqref{cov_edge_step1} simplifies to (at first order in $1/N$)
\begin{equation}
\langle \delta x_i \delta x_j \rangle \simeq \frac{v_0^2}{\lambda^2 N} \sum_{k=1}^{\infty} \frac{1}{k^2} \frac{\Ai(t_{N-k,i})}{\Ai(t_{N-1,i})} \frac{\Ai(t_{N-k,j})}{\Ai(t_{N-1,j})}\; .
\label{cov_edge_step2}
\end{equation}
We can then apply \eqref{hermite_roots_edge} ($y_{i} \simeq \sqrt{2N} \left( 1  + \frac{1}{2} N^{-2/3} a_i \right)$) to obtain
\begin{equation}
    t_{N-k,i} \simeq a_i + kN^{-1/3}\; .
\end{equation}
Until now we only assumed $k=o(N)$. For $k=1$, we can Taylor expand the Airy function to first order and write
%
\begin{equation}
    \Ai(t_{N-1,i}) \simeq N^{-1/3} \Ai'(a_i)\; .
\end{equation}
This allows to estimate the denominator in Eq. (\ref{cov_edge_step2}). 
In the numerator we need to take into account the terms up to $k \sim N^{1/3}$ and we can approximate the sum
as a Riemann integral
\begin{eqnarray}
\langle \delta x_i \delta x_j \rangle &\simeq& \frac{v_0^2}{\lambda^2 N^{1/3}} \frac{1}{\Ai'(a_i) \Ai'(a_j)} \sum_{k=1}^{\infty} \frac{\Ai(a_i + kN^{-1/3})\Ai(a_j + kN^{-1/3})}{k^2} \\
&\simeq& \frac{v_0^2}{\lambda^2 N^{2/3}} \frac{1}{\Ai'(a_i) \Ai'(a_j)} \int_0^{+\infty} dx \ \frac{\Ai(a_i + x)\Ai(a_j + x)}{x^2}\;,
\label{cov_edge_integral}
\end{eqnarray}
where we recall that $a_i$ is the $i^{th}$ zero of the Airy function. We recover the $N^{-2/3}$ scaling discussed above (\ref{rightmost_cov}). Note that the leading correction term is of order $1/N$ since going from the integral to the sum introduces a relative error of order $N^{-1/3}$. Once again, this result directly applies to the Calogero-Moser model after replacing $v_0^2$ by $T/N$. This result is very similar to the expression proved for the DBM in \cite{GorinInfiniteBeta} where the factor $1/k^2$ is replaced by $1/k$. Note that here it is obtained via different methods. 
For the one-particle variance, equation \eqref{cov_edge_integral} reads
\begin{equation}
\langle \delta x_i^2 \rangle \simeq \frac{v_0^2}{\lambda^2 N^{2/3}} \frac{1}{\Ai'(a_i)^2} \int_0^{+\infty} dx \ \frac{\Ai(a_i + x)^2}{x^2}\; .
\label{var_edge_integral}
\end{equation}

\begin{figure}
    \centering
    \includegraphics[width=0.325\linewidth,trim={0cm 0 1cm 1cm},clip]{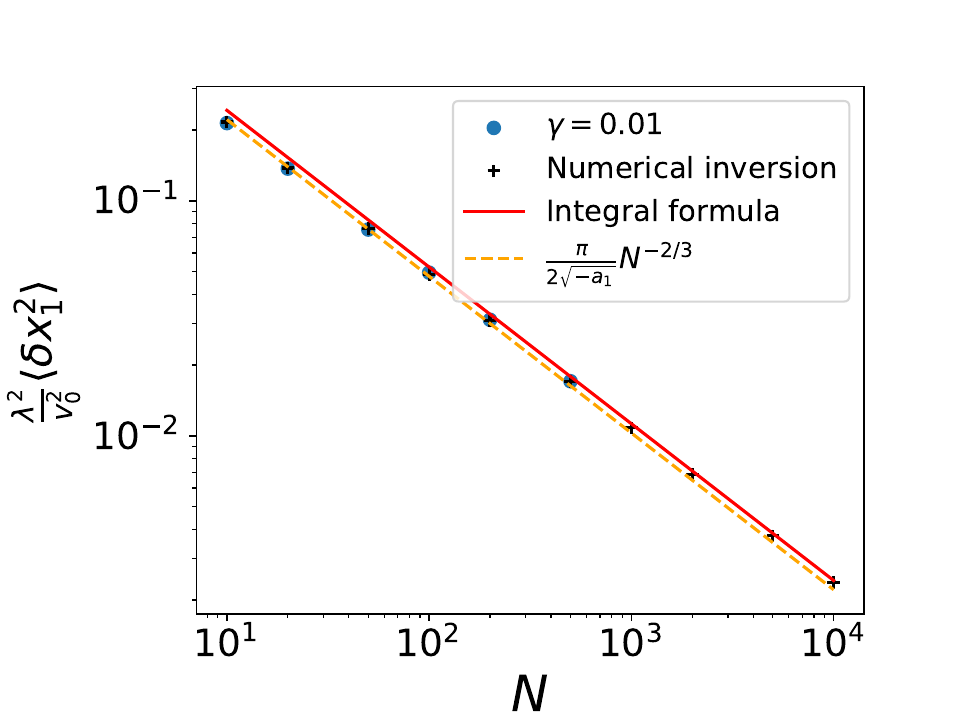}
    \includegraphics[width=0.325\linewidth,trim={0cm 0 1cm 1cm},clip]{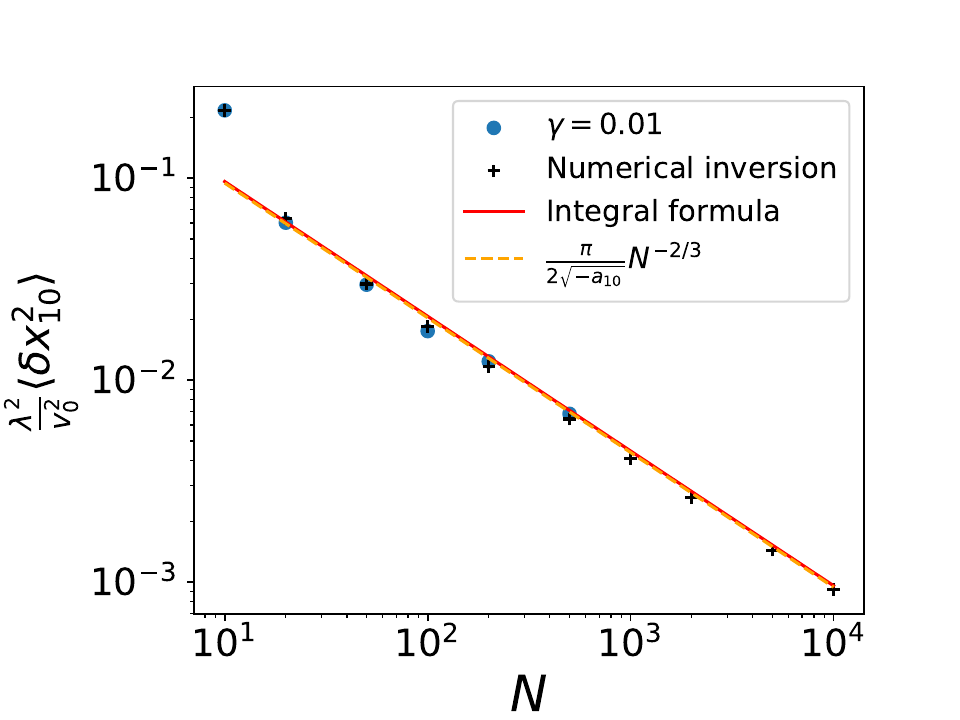}
    \includegraphics[width=0.325\linewidth,trim={0cm 0 1cm 1cm},clip]{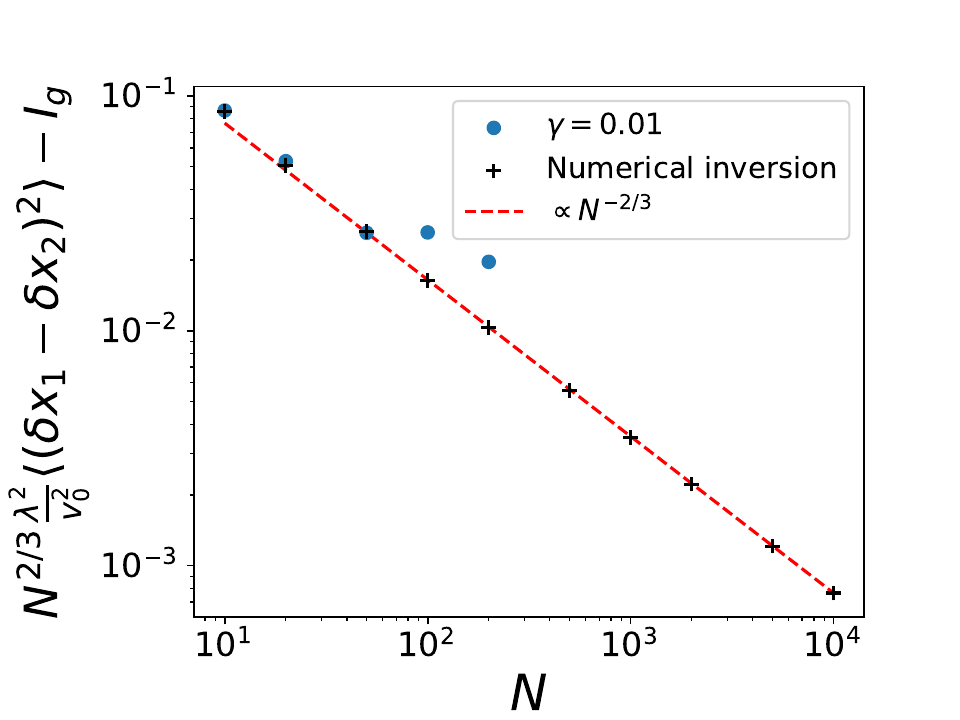}
    \caption{{\bf Left}: Variance of the position of the rightmost particle $x_1$ as a function of $N$, for $\lambda=1$, $g=1$, $v_0=0.1$ and $\gamma=0.01$. The simulation results are in good agreement with the approximate expression for large $N$ given in \eqref{var_edge_integral} which scales as $N^{-2/3}$. {\bf Center}: Same plot for the $10^{th}$ particle $x_{10}$. The result is in good agreement with both \eqref{var_edge_integral} and \eqref{rightmost_cov}. {\bf Right:} Difference between the scaled variance of the gap between particles 1 and 2, computed by exact numerical inversion of the Hessian matrix, and the predicted value $I_g = \int_0^{+\infty} \frac{dx}{x^2} \ \left[ \frac{\Ai(a_1 + x)}{\Ai'(a_1)} - \frac{\Ai(a_2 + x)}{\Ai'(a_2)} \right]^2 \simeq 0.345302...$ (see \eqref{gap_var_edge}), as a function of $N$. The approximation error seems to decrease as $N^{-2/3}$ at large $N$. Simulation results for $\gamma=0.01$ are also plotted for moderate values of $N$.}
    \label{variancefigedge}
\end{figure}

As in the bulk regime, one can directly deduce from \eqref{cov_edge_integral} an expression for the variance of inter-particle distances near the edge
\begin{equation}
\langle (\delta x_i - \delta x_{i+n})^2 \rangle = \langle \delta x_i^2 \rangle + \langle \delta x_{i+n}^2 \rangle - 2 \langle \delta x_i \delta x_{i+n} \rangle \simeq
\frac{v_0^2}{\lambda^2 N^{2/3}} \int_0^{+\infty} \frac{dx}{x^2} \ \left[ \frac{\Ai(a_i + x)}{\Ai'(a_i)} - \frac{\Ai(a_{i+n} + x)}{\Ai'(a_{i+n})} \right]^2 \;.
\label{gap_var_edge}
\end{equation}
This expression is valid for $i$ and $n$ of order 1. Contrary to the bulk, the variance of the gap between two particles is of the same order as the variance of $\delta x_i$. This shows that the correlations are much weaker at the edge than in the bulk, as one would expect. In this case the leading order terms do not cancel and this expression is valid even for $n=1$, as can be seen in Fig. \ref{variancefigedge} right panel. Interestingly, the leading relative error seems to be of order $N^{-2/3}$ (versus $N^{-1/3}$ when looking at $\langle \delta x_i^2 \rangle$, see Fig. \ref{AKDcomparison2}). 

\vspace*{0.5cm}
\noindent {\bf Matching from the edge to the bulk.} 
In the limit of large $i$, the expression \eqref{var_edge_integral} for the one-point variance should reduce to the result \eqref{rightmost_cov} obtained above in the bulk. In this limit, $-a_i$ is large and one can use the asymptotic expression of the Airy function to write
\begin{equation}
    \Ai'(a_i) \simeq -\frac{(-a_i)^{1/4}}{\sqrt{\pi}} \cos \left( \frac{2}{3}(-a_i)^{3/2} + \frac{\pi}{4} \right) \simeq (-1)^{i+1} \frac{(-a_i)^{1/4}}{\sqrt{\pi}} \;,
\end{equation}
where in the second step we used $a_i \simeq - (\frac{3 \pi}{8} (4i-1))^{2/3}$. To evaluate the integral we further approximate $x\ll -a_i$ (since the integral decays as $1/x^2$ for $x \ll -a_i$ and then exponentially for $x \gg -a_i$)
\begin{equation}
    \Ai(a_i+x) \simeq \frac{(-a_i-x)^{-1/4}}{\sqrt{\pi}} \sin \left( \frac{2}{3}(-a_i-x)^{3/2} + \frac{\pi}{4} \right) \simeq (-1)^{i} \frac{\sin(\sqrt{-a_i} \ x)}{\sqrt{\pi} \ (-a_i)^{1/4}}\; .
\end{equation}
Thus we obtain
\begin{equation}
    \int_0^{+\infty} dx \frac{\Ai(a_i + x)^2}{x^2} \simeq \frac{1}{\pi} \int_0^{+\infty} dx \frac{\sin^2(\sqrt{-a_i} \ x)}{\sqrt{-a_i} \ x^2} \simeq \frac{1}{\pi} \int_0^{+\infty} du \frac{\sin^2(u)}{u^2} = \frac{1}{2} \;,
\end{equation}
and we indeed recover \eqref{rightmost_cov}. This shows that \eqref{var_edge_integral} is a refinement of \eqref{rightmost_cov} valid for any $i\ll N$. It is more precise than the previous formula when $i$ is of order 1. In addition 
\begin{eqnarray}
    \frac{1}{\Ai'(a_i)\Ai'(a_j)} \int_0^{+\infty} dx \frac{\Ai(a_i + x)\Ai(a_j+x)}{x^2} &\simeq& \frac{1}{\sqrt{-a_i}\sqrt{-a_j}} \int_0^{+\infty} dx \frac{\sin(\sqrt{-a_i} \ x) \sin(\sqrt{-a_j} \ x)}{x^2} \\
    &\simeq& \frac{\pi}{4} \frac{\sqrt{-a_i}+\sqrt{-a_j} - |\sqrt{-a_i}-\sqrt{-a_j}|}{\sqrt{-a_i}\sqrt{-a_j}} \;,
\end{eqnarray}
and thus for $i\leq j$, \eqref{cov_edge_integral} becomes
\be
\langle \delta x_i \delta x_j \rangle \simeq \frac{\pi}{2\sqrt{-a_j}} \frac{v_0^2}{\lambda^2 N^{2/3}}
\ee
and we indeed recover \eqref{rightmost_cov}.
\\
%

{\bf Correlation between the edge and the bulk}. One may also want to consider the covariance between an edge particle and a bulk particle (or between two particles at opposite edges). For this we can go back to the bulk expression \eqref{cov_largeN}, in the limit \eqref{Cb_1bulk}. Although one particle is at the edge, using this expression does not lead to any divergence and we obtain accurate results. One finds
\be \label{cov_edge_bulk}
\langle \delta x_i \delta x_j \rangle \simeq \frac{v_0^2}{\lambda^2 N} \frac{\pi - \arccos \big( \frac{x_{{\rm eq},j}}{2\sqrt{g/\lambda}} \big)}{2\sqrt{1-\big( \frac{x_{{\rm eq},j}}{2\sqrt{g/\lambda}} \big)^2}} \;.
\ee
Two cases of particular interest are the covariance between an edge particle and a particle at the center ($x_{{\rm eq},j}=0$), for which $\langle \delta x_i \delta x_j \rangle \simeq \frac{v_0^2}{\lambda^2 N} \frac{\pi}{4}$, and the covariance between two particles at opposite edges, (obtained by taking the limit $\frac{x_{{\rm eq},j}}{2\sqrt{g/\lambda}} \to -1$), which gives, for $i$ and $j=O(1)$,
\be
\langle \delta x_i \delta x_{N+1-j} \rangle \simeq \frac{v_0^2}{2\lambda^2 N} \;.
\ee
This also provides a lower bound for the covariance between two arbitrary particles in the system.

We have tested the prediction \eqref{var_edge_integral} for the variance of edge particles with numerical simulations, see Fig. \ref{variancefigedge} left and middle panels. 
In addition we have also studied the correlations between the rightmost particle and the particle $x_i$ both at the edge
and in the bulk. This is shown in the right panel of Fig. \ref{FigcovarianceADBM}. The edge prediction \eqref{cov_edge_integral} is only valid when both particles are in the bulk, and for very large values of $N$ (due to the $N^{-1/3}$ error) which explains the discrepancy. The bulk prediction \eqref{cov_edge_bulk} on the other hand gives quite accurate results as soon as the particle $i$ is far enough from the edge.



\section{Finite time correlations for the Calogero-Moser model}\label{sec:dyn}

In this section, we extend the previous analysis to the study of time correlations in the CM model at low temperature, i.e., $T/(\tilde g \sqrt{\lambda}) \ll N$, see  
Fig. \ref{phase_diagram_CM}. The starting point of our analysis is the formula given in Eq. (\ref{correl_time}). Substituting the explicit expression of the eigenvectors $(\psi_k)_i$ given in (\ref{Hermite_eigenvectors}) into (\ref{correl_time}) one obtains
\begin{equation}
\langle \delta X_i(t) \delta X_j(t') \rangle = \frac{T}{\lambda N} \sum_{k=1}^N \frac{e^{-k^2 \lambda |t-t'|}}{k^2} \frac{u_k(y_i)u_k(y_j)}{\sum_{l=1}^N u_k(y_l)^2} \quad {\rm with} \quad u_k(y)=\frac{H_N^{(k)}(y)}{H_N'(y)} \;.
\label{cov_dx_time}
\end{equation}

We now analyse this formula in the large $N$ limit, both in the bulk and at the edge.

\vspace*{0.5cm}
\noindent{\bf Bulk.} In the bulk, the extension of the derivation for equal-time correlations presented in Sec.~\ref{sec:bulk} is rather straightforward. It yields (keeping the same notation for the scaling function $\mathcal{C}_b$ for simplicity)
\begin{equation}
\langle \delta X_i(t) \delta X_j(t') \rangle \simeq\frac{T}{\lambda N^2} \mathcal{C}_b\left( \frac{ \lambda^{1/4} X_{{\rm eq},i}}{2\sqrt{\tilde g}}, \frac{\lambda^{1/4} X_{{\rm eq},j}}{2\sqrt{\tilde g}}, \lambda |t-t'| \right) \quad {\rm with} \quad \mathcal{C}_b(x,y,\tilde t) = \sum_{k=1}^\infty \frac{e^{-k^2 \tilde t}}{k^2} U_{k-1}(x) U_{k-1}(y) \;.
\label{covCM_largeN_time}
\end{equation}
In particular, at large times, $|t-t'| \gg \lambda^{-1}$, one has (keeping only the term $k=1$ in Eq. (\ref{covCM_largeN_time})) 
\begin{equation} \label{exp_decay}
    \langle \delta X_i(t) \delta X_j(t') \rangle \simeq\frac{T}{\lambda N^2} e^{-\lambda |t-t'|} \;,
\end{equation}
where we have used $U_0(x)= 1$. Interestingly, this large time behavior of correlations is independent of the position, which indicates
that it arises from a collective mode.

One can study the motion of the $i$-th particle. Let us recall that $\lambda {\cal H}^2$
has $N$ eigenvalues $\lambda k^2$, $k=1, \dots N$. Hence there are $N$ relaxation times
which are the inverses $1/(\lambda k^2)$. One has the exact formula
\be 
\langle (\delta X_i(t)-\delta X_i(0))^2 \rangle = 2 \frac{T}{\lambda N^2}  
\sum_{k=1}^N \frac{(1- e^{-k^2 t})}{k^2} \frac{u_k(y_i)^2}{\sum_{l=1}^N u_k(y_l)^2} \;.
\ee 
For $t \ll 1/(\lambda N^2)$, i.e. for
extremely short time, one can expand the exponential and one obtains 
\be  \label{single_diff_bulk}
\langle (\delta X_i(t)-\delta X_i(0))^2 \rangle \simeq 2 \frac{T}{N} t \;,
\ee 
i.e., the single particle diffusion dominates. For larger times,
$t \gg 1/(\lambda N^2)$ one has, using \eqref{uk_chebyshev} 
\be \label{discretesum} 
\langle (\delta X_i(t)-\delta X_i(0))^2 \rangle \simeq  2 \frac{T}{\lambda N^2}  
\sum_{k=1}^{+\infty}  \frac{(1-e^{-k^2 \lambda t})}{k^2} \frac{\sin^2(k \theta_i) }{\sin^2 \theta_i} \;,
\ee 
where $\theta_i = {\rm arccos}  \frac{\lambda^{1/4} X_{{\rm eq},i}}{2\sqrt{\tilde g}}$. 
If $1/(\lambda N^2) \ll t \ll 1/\lambda$ one can approximate the sum by an integral.
Denoting $p=k \sqrt{\lambda t}$
one obtains
\be 
\langle (\delta X_i(t)-\delta X_i(0))^2 \rangle \simeq 2 \frac{T}{\lambda N^2}  \frac{\sqrt{\lambda t}}{\sin^2 \theta_i} 
\int_0^{+\infty} dp  \frac{1-e^{-p^2}}{p^2} \sin^2 \left(\frac{ p \, \theta_i}{\sqrt{\lambda t} } \right) \;.
\ee 
Since $\lambda t \ll 1$ one can replace $\sin^2 \left(\frac{ p \, \theta_i}{\sqrt{\lambda t} } \right) \to \frac{1}{2}$
and one obtains
\bea \label{anomalous} 
\langle (\delta X_i(t)-\delta X_i(0))^2 \rangle &\simeq&  \frac{T}{N^2\sin^2 \theta_i}  \sqrt{\frac{\pi t}{\lambda}} = \frac{T}{N^2 \left(1-(\frac{\lambda^{1/4} X_{{\rm eq},i}}{2\sqrt{\tilde g}})^2\right)}  \sqrt{\frac{\pi t}{\lambda}} \\
&=& \pi^{-3/2} \frac{T}{\tilde g \tilde \rho_i^2 } \sqrt{t} \;,
\eea
with $\tilde \rho_i =N \rho_{eq}(X_{{\rm eq},i})$ and where
$\rho_{eq}(X)$ 
is the equilibrium density at the position of particle $i$, normalized to $N$, given in the r.h.s. of Eq. \eqref{CM_Wigner}.
Thus in this time window the diffusion of the particle is anomalous, with $\delta X_i(t)-\delta X_i(0) \sim t^{1/4}$.
This behavior is similar to the anomalous diffusion of a tracer in single file diffusion problems, see
\cite{Krapivsky2015,MajumdarBarma} and references therein. The prefactor shows a $1/\rho^2$ behavior as a function of the density
which is in agreement with the prediction for non-crossing Brownian particles with sufficiently short range interactions
at equilibrium, see \cite{SpohnTracer}. This result is in agreement with the fact that 
the interaction force in the CM model is short range, i.e. it decays faster than $1/x^2$, 
(for longer range interactions
the exponent is modified, see \cite{Krapivsky2023} for the Riesz gas and \cite{SpohnTracer} for the DBM). 

The anomalous diffusion holds until time $t \sim 1/\lambda$, at which the displacement saturates at
its asymptotic value $\langle (\delta X_i(t)-\delta X_i(0))^2 \rangle \simeq 
2 \langle \delta X_i^2 \rangle \sim T/(\lambda N^2)  $ 
given in Eq. \eqref{cov_largeN_CM}. The crossover behavior between the $t^{1/4}$ diffusion
and the saturation is described by the discrete sum in Eq. \eqref{discretesum}. Note that the approximations
made in this section does not allow us to 
obtain the crossover between the diffusive and anomalous $t^{1/4}$ regime.

\vspace*{0.5cm}
\noindent{\bf Edge.} In the edge regime, using again the approximation \eqref{Airy_approx} in Eq. (\ref{cov_dx_time}) and performing the same computations as in Sec.~\ref{sec:edge} leads to
\begin{eqnarray} \label{cov_edge_integral_CM_time_sum}
\langle \delta X_i(t) \delta X_j(t') \rangle &\simeq& \frac{T}{\lambda N^{4/3}} \frac{1}{\Ai'(a_i) \Ai'(a_j)} \sum_{k=1}^{\infty} \frac{\Ai(a_i + kN^{-1/3})\Ai(a_j + kN^{-1/3})}{k^2} e^{-k^2 \lambda |t-t'|} \\
&\simeq& \frac{T}{\lambda N^{5/3}} \frac{1}{\Ai'(a_i) \Ai'(a_j)} \int_0^{+\infty} dx \ \frac{\Ai(a_i + x)\Ai(a_j + x)}{x^2} e^{-x^2 N^{2/3} \lambda |t-t'|} \;.
\label{cov_edge_integral_CM_time}
\end{eqnarray}
Hence we see that it is natural to rescale the times by a scale ${\cal O}(N^{-2/3})$, which leads to the scaling form
\begin{equation}
\langle \delta X_i(t) \delta X_j(t') \rangle \simeq \frac{T}{\lambda N^{5/3}} \mathcal{C}_e(a_i,a_j, N^{2/3} \lambda |t-t'|) \;\;, \;\; {\cal C}_e(a_i,a_j, \tau) =\frac{1}{\Ai'(a_i) \Ai'(a_j)} \int_0^{+\infty} dx \ \frac{\Ai(a_i + x)\Ai(a_j + x)}{x^2} e^{-x^2 \tau} \;.
\label{cov_edge_integral_CM_time_result}
\end{equation}

Our results allow for a general discussion of the relaxation mechanisms and regimes. 
From \eqref{cov_edge_integral_CM_time_result} we see that, at the edge, the typical time-scale of the correlations $t_e$ is of order $O(N^{-2/3})$, which is much smaller than its counterpart in the bulk, which is of order $O(1)$ -- see Eq. (\ref{covCM_largeN_time}). The relaxation time scale $t_e= O(N^{-2/3})$ can be understood by comparing the effect of free diffusion
with the equilibrium fluctuations, i.e., 
$D_N t_e \sim\langle \delta X_i^2 \rangle \sim N^{-5/3}$, where $D_N \sim T/N$ is the single particle diffusion coefficient (see Eq. (\ref{Calogero})). By contrast the relaxation in the bulk has a collective nature, resulting in longer relaxation time scale of $O(1)$. To test
this picture let us compute the large time asymptotics of the correlations at the edge. There are
actually two regimes. The first regime corresponds to $\tau=O(1)$ but large $\tau\gg 1$ (i.e. $|t-t'|\gg N^{-2/3}\lambda ^{-1}$).
Then one can perform the change of variable $u = x \sqrt{\tau}$ and expand for large $\tau$. This yields, to leading order,
\begin{equation} \label{sqrt_time_decay}
    {\cal C}_e(a_i,a_j, \tau) \underset{\tau\to\infty}{\simeq} \frac{1}{\sqrt{\tau}}\int_0^{+\infty} du \ e^{-u^2} = \frac{1}{2} \sqrt{\frac{\pi}{\tau}} \;,
\end{equation}
which is also independent of the positions, as in the bulk (\ref{cov_dx_time}). This algebraic decay of the scaling function at the edge is 
consistent with pure diffusion and is very different for the exponential tail found in the bulk (\ref{cov_dx_time}).
There is however a second large time regime for $\tau\gg N^{2/3}$ (i.e, $|t-t'|=O(1)$ in $N$ but $|t-t'| \gg \lambda^{-1}$).
In that regime the Riemann integral approximation in \eqref{cov_edge_integral_CM_time_result} breaks down 
(the integral becomes dominated by very small values of $x$) 
and
the sum in \eqref{cov_edge_integral_CM_time_sum} is then dominated by the $k=1$ term. This thus leads to the same large time exponential decay as in the bulk~\eqref{exp_decay}. 
Once again one sees that the large time correlations are uniform in space, with no difference between the edge and the bulk.

The variance of the displacement of a particle between time $0$ and $t$ can be computed in the same way. One has
with $t = \tau/(\lambda N^{2/3}) $, 
\be 
\langle (\delta X_i(t) - \delta X_i(0))^2 \rangle \simeq 
\frac{2T}{\lambda N^{5/3}} \frac{1}{\Ai'(a_i)^2} \int_0^{+\infty} dx \ \frac{\Ai(a_i + x)^2}{x^2} (1-e^{-x^2 \tau}) \;.
\ee 
For $\tau \ll 1$ one can expand the exponential and one finds simply
\be \label{single_diff_edge}
\langle (\delta X_i(t) - \delta X_i(0))^2 \rangle \simeq \frac{2 T}{N} t 
\ee 
where we used the identity $\int_0^{+\infty} dx \, \Ai(a_i + x)^2 = \Ai'(a_i)^2$. 
Hence one recovers the single particle diffusion at small time, as in the bulk [see Eq. (\ref{single_diff_bulk})]. 
For large $\tau$ the variance of the displacement saturates and converges to
its asymptotic value $\sim T/(\lambda N^{5/3})$ as $\langle (\delta X_i(t) - \delta X_i(0))^2 \rangle \simeq  2 
\langle \delta X_i^2 \rangle - \frac{T}{\lambda N^{5/3}} \sqrt{\frac{\pi}{\tau}}$, from \eqref{sqrt_time_decay}.
Thus the intermediate regime with anomalous diffusion obtained in the bulk does not exist at the edge.

\begin{figure}
    \centering
    \includegraphics[width=0.325\linewidth,trim={0cm 0 1cm 1cm},clip]{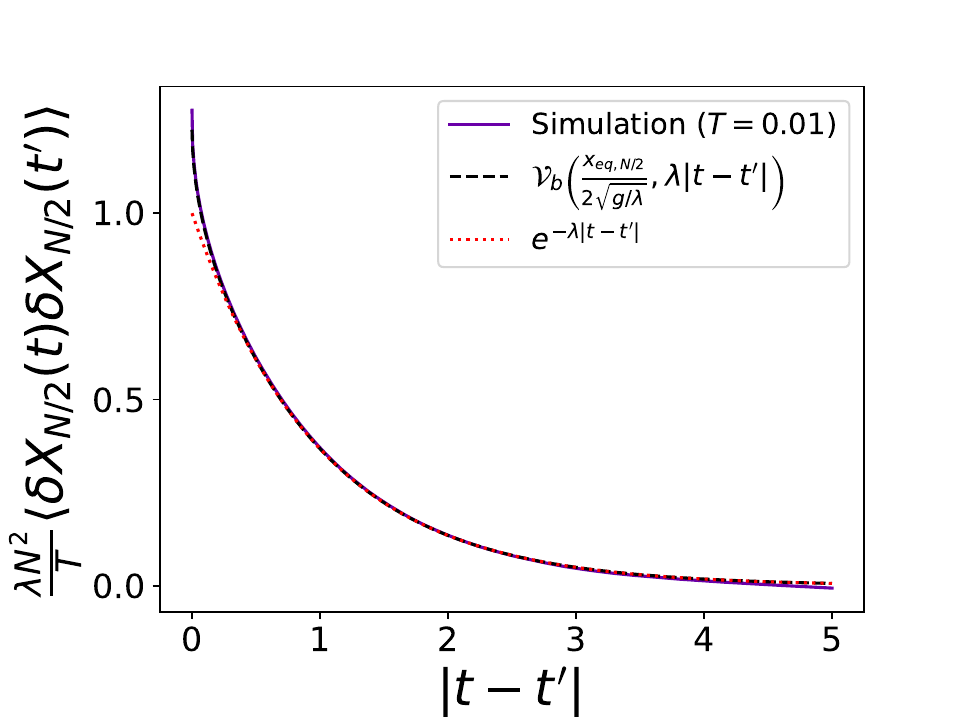}
    \includegraphics[width=0.325\linewidth,trim={0cm 0 1cm 1cm},clip]{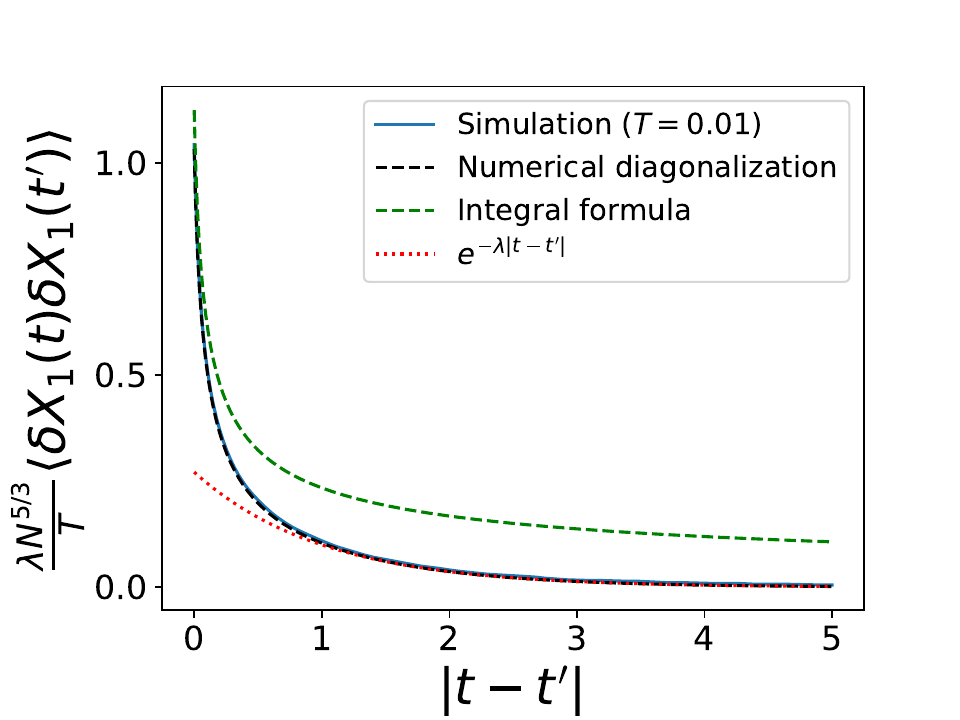}
    \includegraphics[width=0.325\linewidth,trim={0cm 0 1cm 1cm},clip]{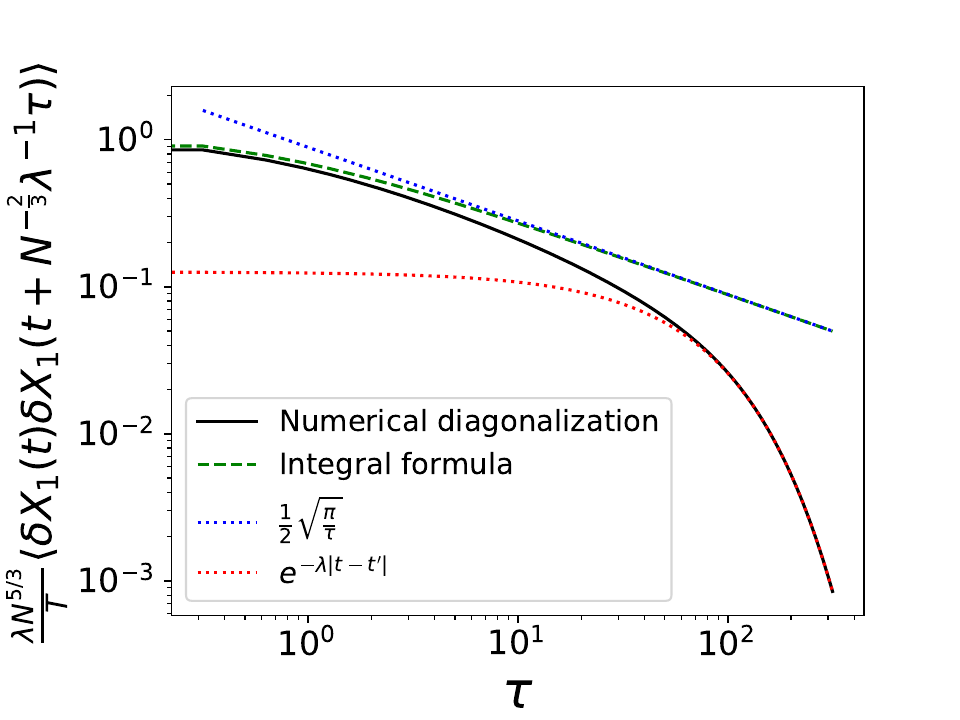}
    \caption{{\bf Left}: Time correlations of the position of the central particle in the CM model as a function of the time difference. The results obtained by simulating the Langevin dynamics for $N=50$ particles at $T=0.01$ ($\lambda=1$ and $g=1$) are in excellent agreement with the large $N$ expression for the bulk \eqref{covCM_largeN_time}. {\bf Center}: Same plot for the rightmost particle. The simulation result is in good agreement with the exact analytical formula \eqref{cov_dx_time} and the large time asymptotics \eqref{exp_decay}, but the comparison with the integral expression \eqref{cov_edge_integral_CM_time_result} would require larger values of $N$, and as discussed in the text is valid only for times of order $N^{-2/3}$. {\bf Right:} Same plot (rightmost particle) for $N=500$ in log-log scale. Here the exact analytical formula \eqref{cov_dx_time} is compared with the integral expression \eqref{cov_edge_integral_CM_time_result} (valid for $\tau \ll N^{2/3}$), the intermediate $\tau^{-1/2}$ regime \eqref{sqrt_time_decay} (valid for $1 \ll \tau \ll N^{2/3}$, but which is not very visible for this value of $N$), and the large time exponential decay \eqref{exp_decay} (valid for $\tau \gg N^{2/3}$). The two
    relaxation regimes at the edge discussed in the text are thus clearly visible.}
    \label{variancefigtime}
\end{figure}

We have checked numerically some of the above predictions, see Fig. \ref{variancefigtime}. 
The agreement is perfect in the bulk, see the left panel. At the edge, one clearly sees
the two regimes in time discussed above for $N=50$ and $N=500$, see the middle and right panel.

Note that, contrary to the above sections, the results of this section are only valid for the CM model. Indeed, since we have taken the limit $\gamma \to 0$ it is not obvious how to treat the exact dynamics for the active DBM to obtain a similar result.




\section{Comparison with the (passive) DBM} \label{sec:dbm}

Let us consider now the standard DBM, to which our method can also be applied. 
It is obtained by setting $v_0=0$ in \eqref{model0}, which corresponds to parameters
$\beta_{\rm DBM}=2 g/T$, and with a support at large $N$ with edges at $\pm 2 \sqrt{g/\lambda}$. 
The fluctuations at low temperature at the edge have been studied before in 
\cite{GorinInfiniteBeta}. There are also some results for $\beta_{\rm DBM}=1,2,4$, e.g. in Ref. \cite{ORourke2010}.
We will compare our results to these works. 

We start by noting that for the DBM 
the stationary correlations at low temperature $T$ read (by the same calculation as in \eqref{CM_Cov2})
{
\begin{equation}
\langle \delta x_i(t) \delta x_j(t') \rangle = \frac{T}{\lambda N} \sum_{k=1}^N \frac{e^{-k \lambda |t-t'|}}{k} \frac{u_k(y_i)u_k(y_j)}{\sum_{l=1}^N u_k(y_l)^2} \quad {\rm with} \quad u_k(y)=\frac{H_N^{(k)}(y)}{H_N'(y)} \;,
\label{DBM_Cov}
\end{equation}
and in particular for equal time correlations one has 
\begin{equation}
\langle \delta x_i \delta x_j \rangle = \frac{T}{\lambda N} (\mathcal{H}^{-1})_{ij} \quad , \quad (\mathcal{H}^{-1})_{ij} = \sum_{k=1}^N \frac{1}{k} \frac{u_k(y_i)u_k(y_j)}{\sum_{l=1}^N u_k(y_l)^2} \;.
\label{DBMcov_dx}
\end{equation}
}
Starting from this point we can perform the same large $N$ approximations as before. The difference is that now the series does not converge, and therefore we expect the approximation to be worse in this case. Indeed, we find that, inside the bulk, the variance and covariance can be given by equivalents of the above formulas, namely
\begin{equation}
\langle \delta x_i(t) \delta x_j(t') \rangle \simeq \frac{T}{\lambda N^2} \mathcal{\tilde C}_b\left( \frac{x_{{\rm eq},i}}{2\sqrt{g/\lambda}}, \frac{x_{{\rm eq},j}}{2\sqrt{g/\lambda}}, \lambda |t-t'| \right) \quad {\rm with} \quad \mathcal{\tilde C}_b(x,y,\tilde t) = \sum_{k=1}^{+\infty}  \frac{e^{-k\tilde t}}{k} U_{k-1}(x) U_{k-1}(y) \;.
\label{DBMcov_largeN_time}
\end{equation}
This formula is valid for $\tilde t=O(1)$ in $N$ and $\tilde t>0$. It is important to note that for equal time correlations, i.e.
$\tilde t=0$, the sum over $k$ has a logarithmic divergence and the formula breaks down. It is possible to obtain an approximate formula
for $\tilde t=0$ and large $N$ using the proper cutoff $k_{\max}=N$. This leads to (see calculation in the Appendix \ref{app:dbm})

\begin{equation}
\langle \delta x_i^2 \rangle \simeq \frac{T}{\lambda N^2} \frac{\ln N}{2(1-(\frac{x_{{\rm eq},i}}{2\sqrt{g/\lambda}})^2)} + O(N^{-2}) \;.
\label{DBMvar_largeN0}
\end{equation}
To obtain the next term $O(N^{-2})$ requires a priori to study the recursion \eqref{recursion_rescaled}
at large $N$ in more details and goes beyond the scope of this paper. 
Our prediction Eq. \eqref{DBMvar_largeN0} can be compared with the result in \cite{ORourke2010}. In that paper the stationary DBM for eigenvalues $\lambda_i$ is considered. 
It is proved in theorem 5 of \cite{ORourke2010} that for $\beta_{\rm DBM}=1,2,4$ the distribution of the centered variable $\lambda_i- y_i$ 
converges to a Gaussian distribution at large $N$ with
\be \label{ORourke} 
{\rm Var} \lambda_i   \simeq \frac{\log N}{2 \beta N (1 - \frac{y_i^2}{2 N}) } \;.
\ee 
If we take into account the connection with our notations, one has
\be \label{mappingDBM}
x_i = 2 \sqrt{\frac{g}{\lambda}} \frac{\lambda_i}{\sqrt{2 N}}  \;,
\ee 
and one can check that \eqref{ORourke} is identical to \eqref{DBMvar_largeN0}. 
This is remarkable since our result is a priori derived for $\beta_{\rm DBM} \gg 1$,
as can be seen from the simple estimate of the dimensionless ratio (for $n=O(1)$) 
given in \eqref{bouchaudDBM}. It is thus tempting to conjecture that in the
case of the DBM this formula (\ref{ORourke}) is valid for any $\beta$.

At the edge our method applied to the DBM gives the stationary two time correlations as
\begin{equation}
\langle \delta x_i(t) \delta x_j(t') \rangle \simeq \frac{T}{\lambda N^{4/3}} {\cal \tilde C}_e(a_i,a_j, N^{1/3}\lambda |t-t'|) \;\;, \;\; {\cal \tilde C}_e(a_i,a_j,\tau) =\frac{1}{\Ai'(a_i) \Ai'(a_j)} \int_0^{+\infty} dx \ \frac{\Ai(a_i + x)\Ai(a_j + x)}{x} e^{-x\tau}
\label{DBMcov_edge_integral_time}
\end{equation}
and in that case there is no problem to extend this formula to $|t-t'|=0$. The equal-time correlations
are thus given~by
\begin{equation}
\langle \delta x_i \delta x_j \rangle \simeq \frac{T}{\lambda N^{4/3}} {\cal \tilde C}_e(a_i,a_j) \;\;, \;\; {\cal \tilde C}_e(a_i,a_j) =\frac{1}{\Ai'(a_i) \Ai'(a_j)} \int_0^{+\infty} dx \ \frac{\Ai(a_i + x)\Ai(a_j + x)}{x} \;.
\label{DBMcov_edge_integral}
\end{equation}
Our predictions \eqref{DBMcov_edge_integral_time} and \eqref{DBMcov_edge_integral} have the same form as 
 the results obtained by different methods in Ref. \cite{GorinInfiniteBeta}, where the non-stationary version of the 
DBM was studied. 

It is interesting to note that the result \eqref{DBMcov_edge_integral} for the DBM can also be obtained 
by yet another completely different method, 
using the stochastic Airy operator and perturbation theory, in an alternative form \eqref{DBM_cov_alternative}. The derivation is detailed in Appendix \ref{app:airy_op}. 
 
{\bf Remark}. Note also that Theorem 6 in
\cite{ORourke2010} gives an estimate of the variance of the DBM for $\beta_{\rm DBM}=1,2,4$ at the edge, 
which behaves as
\be \label{DBM_edge_previous}
{\rm Var} \,\lambda_i \simeq \left(\frac{1}{12 \pi}\right)^{2/3} \frac{2 \log i}{\beta_{\rm DBM} i^{2/3} N^{1/3}} \;.
\ee 
This result can be recovered by taking the limit of large $i$ in \eqref{DBMcov_edge_integral} with $i=j$. The computation is the same as in Sec.~\ref{sec:edge}, except that one needs to add a cutoff at $x=-a_i$ corresponding to the exponential decay of the Airy function for positive values, since the integral diverges otherwise. This gives (for $1 \ll i \ll N$)
\begin{eqnarray}
\langle \delta x_i^2 \rangle &\simeq& \frac{T}{\lambda N^{4/3}} \frac{1}{\Ai'(a_i)^2} \int_0^{-a_i} dx \ \frac{\Ai(a_i + x)^2}{x}
\simeq \frac{T}{\lambda N^{4/3}} \frac{1}{-a_i} \int_0^{(-a_i)^{3/2}} dx \ \frac{\sin^2 x}{x} \\
&\simeq& \frac{3}{4} \frac{T}{\lambda N^{4/3}} \frac{\ln (-a_i)}{-a_i}
\simeq \left(\frac{1}{12\pi}\right)^{2/3} \frac{T}{\lambda N^{4/3}} \frac{\ln i}{i^{2/3}}
\end{eqnarray}
which after a rescaling using \eqref{mappingDBM} coincides with \eqref{DBM_edge_previous}.
Once again, it is tempting to conjecture the formula (\ref{DBM_edge_previous}) is valid for any $\beta_{\rm DBM}$.

\section{Conclusion}

In this paper we have studied in detail the active DBM model introduced by us in \cite{ADBM1}. We have focused on 
the regime of weak active noise and large persistence time. In that regime we have obtained the covariance of the particle positions in the
non-equilibrium stationary state to
lowest order in $v_0^2/(g \lambda)$ and for $\gamma \to 0^+$, for arbitrary $N$. This was achieved first by relating 
the small displacements (the active phonons) to the Hessian matrix, and in a second stage using the exact spectrum of
this matrix to obtain an exact expression for the covariance matrix of these displacements in terms of
Hermite polynomials. Using the large order asymptotics of these polynomials we were able to show that in the
large $N$ limit these formulae for the covariance take nontrivial scaling forms which we obtained explicitly. 
We found two distinct regimes. In the bulk of the ``active crystal'' we find that the covariance
scales as $1/N$ and the scaling function involves polylogarithm functions. In addition,
we obtained a formula for the variance of the relative displacements between two particles
of rank difference $n \sim N$ (i.e., separated by $n-1$ particles). It remains a challenge
however to extend this formula for $n=O(1)$, in particular to obtain the statistics of the
gaps. At the edge, the covariance of the positions is larger, i.e. it scales as $N^{-2/3}$ and,
the scaling function involves the Airy function and its zeroes. In this case, we obtain a formula
for the variance of the relative displacements which is valid for any $n = O(1)$. 

These predictions are consistent with the existence of three regimes as
a function of $v_0^2/(g \lambda)$, which are displayed in Fig. \ref{phase_diagram}, confirming the results in \cite{ADBM1}.
We have performed detailed numerical simulations to confirm our finite $N$
and large $N$ predictions. In addition these simulations have allowed us to ascertain the range
of parameters where the weak active noise, and large persistence time is numerically
accurate. In particular, although our results are valid only for very small tumbling rates, the covariance appears to be a monotonously decreasing function of $\gamma$. Hence the results of this paper provide upper bounds for the case of arbitrary tumbling rates. It would be interesting in the future to understand better the effect of this parameter.

Interestingly we have unveiled a connection between the weak noise regime of the active DBM
and the low temperature regime of a priori unrelated equilibrium problem (i.e with passive noise), namely the overdamped dynamics 
of the CM model. This connection allowed us to use our predictions for the active DBM
to obtain directly the covariance matrix for the displacements in the CM model both
at finite $N$ and at large $N$. In particular, the bulk and edge scaling functions 
are identical in both models. We have compared these analytical predictions with
the results, mostly numerical, of Ref.~\cite{Agarwal2019}, and found very good agreement
even up to values of the inverse temperature parameter $\beta = O(1)$. 

The above connection between the active DBM and the CM model in the weak noise regime
was possible because the covariance matrices of the displacements in both models is proportional to the same matrix ${\cal H}^{-2}$
where ${\cal H}$ is the Hessian of the DBM. Since the exact diagonalization of this matrix is possible,
the covariance could be obtained in closed form. As an immediate extension, we also obtained the
covariance for the passive DBM which is proportional to ${\cal H}^{-1}$. This allowed us to
recover by a different method the result of \cite{GorinInfiniteBeta} for the passive DBM. 
Along similar lines, an interesting extension of the present work would be to consider the active CM model, 
i.e. the CM model with Brownian noise replaced by run-and-tumble noise.
Indeed in that case the covariance matrix will be proportional to ${\cal H}^{-4}$ 
and very similar formulae could be derived for the covariances at low temperature.
Because of the resulting $1/k^4$ spectrum, we anticipate that this will
lead to interesting so-called giant fluctuations in the number of particles in 
an interval, see e.g.~\cite{DasGiant2012}.
Such fluctuations are commonly observed in active systems, see e.g.~\cite{DasGiant2012},
but are suppressed in the case of the active DBM ($1/k^2$ spectrum) 
because of the rigidity of the logarithmic interaction. 

\bigskip

{\bf Acknowledgments}. {We thank Manas Kulkarni for stimulating discussions.}

\setcounter{section}{0}
\renewcommand{\thesection}{\Alph{section}}





\newpage

\appendix

\section{Proof of (\ref{Hermite_eigenvectors})}
\label{ProofEigenvectors}

For the sake of completeness we briefly recall the proof of the expression for the eigenvectors of the Hessian matrix $\mathcal{H}$ given in \cite{eigenvectors}, based on the residue theorem. The statement is that the matrix
\begin{equation}
A_{ij} = \mathcal{H}_{ij} - \delta_{ij} = \delta_{ij}\left(\sum_{k\neq i} \frac{1}{(y_i-y_k)^2}\right) - (1-\delta_{ij}) \frac{1}{(y_i-y_j)^2}
\end{equation}
where the $y_i$'s are the $N$ roots of the Hermite polynomial $H_N(x)$ verifies
\begin{equation}
    A\psi_k = (k-1)\psi_k \quad {\rm with} \quad (\psi_k)_i = \frac{H_N^{(k)}(y_i)}{H_N'(y_i)} \quad (i=1,\ldots,N)
    \label{eigvec_toprove}
\end{equation}
for $k=1,...,N$. For this we introduce the function
\begin{equation}
    V_i^{(k)}(z) = (z-y_i)^{-2}\frac{H_N^{(k)}(z)}{H_N(z)}, \quad i=1,\ldots,N \ {\rm and} \ k=1,\ldots,N.
\end{equation}
This is a meromorphic function of $z$ (with $N-1$ simple poles at $z=y_j$, $j=1,...,i-1,i+1,...,N$, and a triple pole at $z=y_i$) that vanishes at least as $|z|^{-3}$ when $|z| \to +\infty$, therefore the sum of its residues vanishes. Computing the residues yields:
\begin{eqnarray}
{\rm Res}_{y_j} V_i^{(k)} &=& (y_i-y_j)^{-2}\frac{H_N^{(k)}(y_j)}{H_N'(y_j)} \quad {\rm for} \ j \neq i \\
{\rm Res}_{y_i} V_i^{(k)} &=& \left[ -\sum_{j\neq i} (y_i-y_j)^{-2} + k-1 \right] \frac{H_N^{(k)}(y_i)}{H_N'(y_i)}
\end{eqnarray}
where we have used the differential equation satisfied by the Hermite polynomials \eqref{Hermite_equation} as well as the identity $\sum_{j\neq i} (y_i-y_j)^{-2} = \frac{2}{3} (N-1) - \frac{y_i^2}{3}$ (see \cite{eigenvectors}). Summing the residues and equating the sum to zero then yields the desired result \eqref{eigvec_toprove}. Finally, normalizing the eigenvectors and using the fact that $\mathcal{H}$ is symmetric, hence its eigenvectors are orthogonal, we obtain the formula for $\mathcal{H}^{-2}$ given in \eqref{H_inversion}.

\section{Explicit formula for the correlations in the bulk}
\label{app:alternative} 

In this Appendix, we derive the following formula, for $u$ and $v$ in $(0,\pi)$:
\begin{equation} \label{formulaLeo} 
\sum_{k=1}^\infty \frac{\sin(k u) \sin(k v)}{k^2} 
= \frac{1}{2} \left( \pi \min(u,v) - uv \right) \;,
\end{equation}
from which one obtains the formula (\ref{newCbIntro}) given in the text. Before providing the derivation of this relation, let us note that it is related to the correlator of the standard Brownian bridge $B(t)$, with ${B(1)}=B(0)=0$,
which reads 
\be 
\overline{ B(t) B(s) } = \min(s,t) - s t \;.
\ee 
Here one has $u=\pi t$, $v= \pi s$ and $\sum_{k=1}^\infty \frac{\sin(k u) \sin(k v)}{k^2} = \frac{\pi^2}{2} \overline{ B(t) B(s) }$.
To understand the connection, let us recall that $B(t)$ admits the
Fourier decomposition $B(t)= \sum_{k \geq 1} b_k \sin(\pi k t)$, where the $b_k$ are independent centered
Gaussian random variables of variance ${\rm Var} (b_k)= 2/(\pi^2 k^2)$ (which follows from the Brownian measure $\sim \exp(- 1/2 \int_0^1 dx (dB(x)/dx)^2 )$ \cite{BrownianBridgeWiki}.

Note that one can write, with $\theta_x=\arccos x$
\be
 \mathcal{C}_b(x,y) = \frac{d\theta_x}{dx} \frac{d\theta_y}{dy} 
 \frac{1}{2} \left( \pi \min(\theta_x , \theta_y) - \theta_x \theta_y \right) 
\ee
This formula is reminiscent of the ones obtained in \cite{Smithcountingstat} 
for the correlations of fermions in a quadratic well. 
Apart from the two density factors, it is the analog of the formula (16) in \cite{Smithcountingstat}, for the correlations of the Gaussian free field on a circle (with vanishing conditions at the boundary). Here this logarithmically correlated field
is replaced by a Brownian bridge, as discussed above (since $1/k$ there is replaced by $1/k^2$ here). Note also
the resemblance with the formula in Eq. (15) in \cite{Smithcountingstat} which likewise contains two density factors.

Let us first show for $\theta \in (0,2\pi)$
\begin{equation} \label{sums}
\sum_{k=1}^{\infty} \frac{\sin(k\theta)}{k} = {\rm Im} \left( \sum_{k=1}^{\infty} \frac{e^{ik\theta}}{k} \right) = {\rm Im}(-\ln (1-e^{i\theta})) = - {\rm Im} \left( \frac{i\theta}{2} + \ln (-2i\sin(\frac{\theta}{2})) \right) = -\frac{1}{2}(\theta-\pi) \;.
\end{equation}
There is a discontinuity at $\theta=2n\pi$ for any $n\in\mathbb{Z}$ (including $\theta=0$). In particular for $\theta \in (-2\pi,0)$ one should replace $\theta \to \theta + 2\pi$, leading to
\begin{equation}
\sum_{k=1}^{\infty} \frac{\sin(k\theta)}{k} = -\frac{1}{2}(\theta+\pi) \;.
\end{equation}

Then, for $u,v \in([0,\pi)$, one has
\begin{equation}
\sum_{k=1}^\infty \frac{\sin(k u) \sin(k v)}{k^2} = \frac{1}{2} \sum_{k=1}^\infty \frac{\cos(k(u-v))}{k^2} - \frac{1}{2} \sum_{k=1}^\infty \frac{\cos(k(u+v))}{k^2} \;.
\end{equation}
We thus need to compute $\sum_{k=1}^\infty \frac{\cos(k\theta)}{k^2}$ for $\theta\in(-2\pi,2\pi)$. But we know that
\begin{equation}
\partial_\theta \sum_{k=1}^\infty \frac{\cos(k\theta)}{k^2} = - \sum_{k=1}^\infty \frac{\sin(k\theta)}{k} = \begin{cases} \frac{1}{2} (\theta -\pi) \quad {\rm for } \ \theta \in(0,2\pi) \\
\frac{1}{2} (\theta +\pi) \quad {\rm for } \theta \ \in(-2\pi,0) \end{cases}
\end{equation}
Integrating over $\theta$ (the discontinuity at $\theta=0$ becomes a cusp) and using that the sum should be $\frac{\pi^2}{6}$ for $\theta=0$ to fix the integration constant, we obtain
\begin{equation}
\sum_{k=1}^\infty \frac{\cos(k\theta)}{k^2} = \frac{|\theta|}{2}\left(\frac{|\theta|}{2}-\pi\right) + \frac{\pi^2}{6} \;.
\end{equation}
Replacing $\theta$ by $u\pm v$, one finds for $u,v \in (0,\pi)$
\begin{equation}
\sum_{k=1}^\infty \frac{\sin(k u) \sin(k v)}{k^2} = \frac{1}{2} \left( \frac{(u-v)^2}{4} - \frac{\pi}{2}|u-v| - \frac{(u+v)^2}{4} - \frac{\pi}{2}(u+v) \right) = \frac{1}{2} \left( \pi \min(u,v) - uv \right) \;.
\end{equation}
which shows \eqref{sums}.

\section{Absence of particle crossings in the Calogero-Moser model}
\label{NoCrossingCalogero}
Consider the CM model for $N=2$. The difference between the positions of the two particles $y=x_2-x_1$ follows the equation
\begin{equation}
 \dot y(t) = - \lambda y(t) +  \frac{4\tilde g^2}{y^3} + \sqrt{2T} \xi(t)
\end{equation}
where $\sqrt{2} \xi(t) = \xi_1(t)+\xi_2(t)$, hence $\xi(t)$ is Gaussian white noise with zero mean and variance 1. We set $\lambda=1$ and $\tilde g=1$. Discretizing time with a time-step $\Delta t$ we obtain:
\begin{equation}
 y_{t+\Delta t} = (1-\Delta t)y_t + \frac{4\Delta t}{y_t^3} + \sqrt{2T \Delta t} \ \eta_t
\end{equation}
where $\eta_t$ is a Gaussian random variable with zero mean and unit variance. Let us assume that at time $t$ $y>0$. We want to know the probability that at time $t+\Delta t$, $y<0$, i.e. the two particles have crossed. This is given by:
\begin{eqnarray}
    \mathbb{P}(y_{t+\Delta t}<0|y_t>0) &=& \mathbb{P}\left(\eta_t < -\frac{(1-\Delta t)y_t}{\sqrt{2T\Delta t}} - \frac{4\Delta t}{\sqrt{2T\Delta t} \ y_t^3} \right) = \mathbb{P}\left(\eta_t > f_{\Delta t}(y_t) \right) = \frac{1}{2}{\rm erfc}\left(\frac{f_{\Delta t}(y_t)}{\sqrt{2}}\right) \\
    f_{\Delta t}(y_t) &=& \frac{(1-\Delta t)y_t}{\sqrt{2T\Delta t}} + \frac{4\sqrt{\Delta t}}{\sqrt{2T} \, y_t^3}
\end{eqnarray}
where we have used the parity of the distribution of $\eta_t$, and ${\rm erfc}(x)=\frac{2}{\sqrt{\pi}}\int_x^{+\infty}dt e^{-t^2}$. Since $f_{\Delta t}(y)$ has a minimum on $(0,+\infty)$, we can get an upper bound on the probability by simply minimizing $f_{\Delta t}(y_t)$ over $y_t$. This gives
\begin{eqnarray}
    y^* = \left(\frac{12\Delta t}{1-\Delta t}\right)^{1/4} \quad \Rightarrow \quad \min_{y>0} f_{\Delta t}(y) = f_{\Delta t}(y^*) = \frac{4}{3^{3/4}}\frac{(1-\Delta t)^{3/4}}{\sqrt{T} \, \Delta t^{1/4}}
\end{eqnarray}
Thus one has (reintroducing $\tilde g$ and $\lambda$)
\begin{equation}
    \mathbb{P}(y_{t+\Delta t}<0|y_t>0) \leq \frac{1}{2}{\rm erfc}\left(\frac{2\sqrt{2}}{3^{3/4}} \sqrt{\frac{\tilde g}{T}} \frac{(1-\lambda\Delta t)^{3/4}}{\ \Delta t^{1/4}}\right) \underset{\Delta t \to 0}{\sim} C_1 \sqrt{\Delta t} \ e^{-\frac{C_2}{\sqrt{\Delta t}}}
\end{equation}
with $C_1=\frac{3^{3/4}}{4\sqrt{2\pi}}\sqrt{\frac{T}{\tilde g}}$ and $C_2=\frac{8}{3^{3/2}}\frac{\tilde g}{T}$. Since the number of time steps during a fixed time interval only increases as $1/\Delta t$, this proves that in the continuous time limit $\Delta t \to 0$ the particles cannot cross. For larger values of $N$, when two particles become very close to each other the effect of the other particles can be neglected, and so the reasoning above still holds. Finally note that, performing the same computation for the Dyson Brownian motion, one finds that the minimum of $f_{\Delta t}(y)$ is independent of $\Delta t$, and therefore one cannot conclude in this case. More generally, if one takes an interaction force of the form $\sgn(x_i-x_j) |x_i-x_j|^{-\alpha}$ and an arbitrary external potential, one gets for small enough $\Delta t$
\begin{equation}
    \mathbb{P}(y_{t+\Delta t}<0|y_t>0) \leq \frac{1}{2}{\rm erfc}\left(K \Delta t^{\frac{1}{\alpha+1}-\frac{1}{2}}\right)
\end{equation}
with $K>0$, and therefore the probability of crossing goes to zero in the continuous time limit for any $\alpha>1$.

\section{Approximate plane wave diagonalization approach}
\label{Bouchaud_approach}

Another approach is to diagonalize $\mathcal{H}$ approximately for large $N$, as done e.g. in \cite{bouchaud_book} (chap. 5.4) for the DBM, by assuming the density to be uniform in the bulk, using plane waves, and computing the inverse of the Hessian (very much as a calculation of displacements using phonons in a solid). This yields with $\rho$ the mean density (normalized to unity), i.e. $\rho = 1/(N a)$ where $a$ is typical interparticle distance
\begin{equation}
    (\mathcal{H}^{-2})_{ij} \simeq \frac{1}{N} \left(1 + 2 \sum_{k=1}^{N/2-1} \frac{\cos \left( \frac{2\pi k |i-j|}{N} \right)}{(1+4\pi^2 g \rho^2 k/\lambda)^2} \right) \;.
\end{equation}
Using this result as well as \eqref{corr1} and \eqref{covHessian_ADBM}, and replacing the sum by an integral, one can first compute the
variance of the distance between two particles in the bulk 
\begin{equation}
    \langle (\delta x_i - \delta x_{i+n})^2 \rangle  
    = \frac{4 v_0^2}{\lambda^2 N} \sum_{k=1}^{N/2-1} \frac{1-\cos \left( \frac{2\pi k |i-j|}{N} \right)}{(1+4\pi^2 g \rho^2 k/\lambda)^2}
    \simeq \frac{v_0^2}{2\pi^2 g^2 \rho^4 N^2} \int_0^\pi \frac{dq}{\pi} \frac{1-\cos(nq)}{q^2} \simeq \frac{v_0^2 n}{4\pi^2 g^2\rho^4 N^2}
\label{gapvariance_fourier}
\end{equation}
where $\rho$ is the density both at $x_i$ and $x_{i+n}$, which are assumed to be close ($n\ll N$). In the last step we have assumed $n\gg 1$. Comparing
with the numerics one finds that it is indeed a good approximation at short distances, where it coincides with the result of the main text \eqref{gapvariance_largeN_general}. Note that the momentum integral in 
\eqref{gapvariance_fourier} behaves as $1/q^2$ at small $q$, at variance with the DBM where it is $1/q$. 
Note that for this quantity the momentum integral is convergent since the factor $(1-\cos q n)$ regularises it at small $q$. 
This results in deformations of the equilibrium crystal growing as $\sqrt{n}$ with the distance $n$, 
instead of logarithmically for the DBM. In both cases this can be related to the fluctuations
of the number of particles in an interval.

Another interesting quantity which usually measures the degree of translational order in a solid is the variance
\be
\langle \delta x_i^2 \rangle = \frac{v_0^2}{\lambda^2} (\mathcal{H}^{-2})_{ii} \simeq \frac{v_0^2}{N} \left(\lambda^{-2} + 2 \sum_{k=1}^{N/2-1} \frac{1}{(\lambda+4\pi^2 g \rho^2 k)^2} \right) \simeq  \frac{v_0^2}{N^2} \int_0^\pi \frac{dq}{\pi} \frac{1}{(\lambda N^{-1}+2\pi g\rho^2 q)^2}
    \simeq \frac{v_0^2}{2\pi^2 g \lambda \rho^2 N} 
\label{variance_fourier} 
\ee
where as a first step we have approximated the sum by an integral, which would be valid for $g \rho^2 \ll \lambda$ (the more realistic case 
$g \rho^2/\lambda = O(1)$ is treated below).
Here by contrast, for infinite $N$, the momentum integral is divergent at small $q$. This reflects the translational invariance 
of the system. This invariance however is broken by the quadratic well which leads to an additional term $\lambda/N$ in the denominator.
That term regularizes the integral at small $q$, leading to the above result. 
Although the scaling is correct, the amplitude turns out to be inaccurate when compared with numerical simulations. 
The reason for this is that this observable is dominated by large scales and is very sensitive to the details e.g. of the variations of the density 
(and of the confining potential) at large scale. 
As shown in the rest of the paper it is possible to make a more accurate calculation which includes these effects.

However, in the bulk $g \rho^2/\lambda = O(1)$ one needs to keep the sum in \eqref{variance_fourier}.
Extending this sum to $N=+\infty$ we obtain
\be
\langle \delta x_i^2 \rangle \simeq  \frac{v_0^2}{N} \left(\frac{1}{\lambda^2} + \frac{1}{8\pi^4 g^2 \rho^4} \psi'(1+\frac{\lambda}{4\pi^2 g \rho^2}) \right)
\ee
where $\psi(z)$ is the digamma function. Note that indeed, for $\lambda \gg g \rho^2$ this formula
matches the estimate \eqref{variance_fourier}. However this estimation, which is more accurate 
in the bulk, does not significantly improve the agreement with numerical simulations.

Let us recall that for the DBM this computation gives \cite{bouchaud_book} for $n = O(1) \gg 1$
\be \label{bouchaudDBM} 
\frac{ \langle (\delta x_i - \delta x_{i+n})^2 \rangle }{ \langle x_i - x_{i+n} \rangle^2 } \simeq \frac{2}{\beta_{\rm DBM} \pi^2 n^2} \log n 
\ee 
where in our notations $\beta_{\rm DBM}=2 g/T$. 



\section{Connection with the Plancherel-Rotach formula} \label{Plancherel}
An alternative way to obtain an approximate expression for $u_k(y)$ at large $N$, for $k \ll N$, is to use the Plancherel-Rotach formula (see e.g. \cite{ForresterHermiteEdge})
\begin{eqnarray}
    &&H_{N+m}(\sqrt{2N}X) = \left(\frac{2}{\pi}\right)^{1/4} \frac{2^{m/2+N/2}}{(1-X^2)^{1/4}} N^{m/2-1/4} (N!)^{1/2} e^{NX^2} \cos(\phi_N(X) - m \arccos(x)) \left(1+O(N^{-1})\right) \\
    &&\phi_N(X) = NX\sqrt{1-X^2} + (N+\frac{1}{2})\arcsin(X) - N\frac{\pi}{2}
\end{eqnarray}
Taking $m=-k$ (or $m=-1$) and $\sqrt{2N}X=y_i$, this leads to (using Stirling's formula to write $\frac{(N-1)!}{(N-k)!} \simeq N^{k-1}$)
\begin{equation}
u_k(y_i) = 2^{k-1} \frac{(N-1)!}{(N-k)!} \frac{H_{N-k}(y_i)}{H_{N-1}(y_i)} \simeq (2N)^{\frac{k-1}{2}} \frac{\cos(\phi_N(\frac{y_i}{\sqrt{2N}}) + k \arccos(\frac{y_i}{\sqrt{2N}}))}{\cos(\phi_N(\frac{y_i}{\sqrt{2N}}) + \arccos(\frac{y_i}{\sqrt{2N}}))}
\end{equation}
In addition, since $H_N(y_i)=0$ we must have $\cos(\phi_N(\frac{y_i}{\sqrt{2N}}))=0$, i.e. $\phi_N(\frac{y_i}{\sqrt{2N}})=(2p+1)\frac{\pi}{2}$ for some integer $p$. This implies
\begin{equation}
u_k(y_i) \simeq (2N)^{\frac{k-1}{2}} \frac{\sin(k \arccos(\frac{y_i}{\sqrt{2N}}))}{\sin(\arccos(\frac{y_i}{\sqrt{2N}}))} = (2N)^{\frac{k-1}{2}} \frac{\sin(k \arccos(\frac{y_i}{\sqrt{2N}}))}{\sqrt{1-\frac{y_i^2}{2N}}}
\end{equation}
which is the same as \eqref{uk_chebyshev}.

\section{Problems in the evaluation of the variance of the gaps}
\label{app:gaps} 

In this appendix we give an alternative derivation for the leading order of \eqref{gapvariance_largeN}, i.e. the variance of interparticle distance for two particles near $x=0$. This gives a better understanding of why our approximation fails when $n=O(1)$. We start again from \eqref{var_chn}, but we do not extend the sum to $+\infty$ for reasons that will become clear later. Since the denominator only gives subleading corrections in $\epsilon$ we have, writing $\psi = \theta - \epsilon$ (with $\epsilon=O(1/N)$)
\be
\mathcal{D}_{b,N}(x,y) = \sum_{k=1}^N \frac{1}{k^2} \left[\frac{\sin(k\theta)}{\sin\theta} - \frac{\sin(k\psi)}{\sin\psi} \right]^2 \simeq \frac{1}{\sin^2 \theta} \sum_{k=1}^N \frac{1}{k^2} \left[ \sin(k\theta) - \sin(k\theta) \cos(k\epsilon) + \cos(k\theta) \sin(k\epsilon) \right]^2
\ee
Specialising to $\theta=\frac{\pi}{2}$ (i.e. $x=0$) this simplifies to
\be
\mathcal{D}_{b,N}(x,y) \simeq \sum_{k=0}^{\lfloor N/2\rfloor} \frac{(1-\cos((2k+1)\epsilon))^2}{(2k+1)^2} + \sum_{k=1}^{\lfloor N/2\rfloor} \frac{\sin^2(2k\epsilon)}{4k^2}
\ee
We then use the fact that $\epsilon \ll 1$ to approximate the sums by Riemann integrals. This gives
\be
\mathcal{D}_{b,N}(x,y) \simeq \frac{|\epsilon|}{2} \left( \int_0^{N\epsilon} dx \ \frac{(1-\cos x)^2}{x^2} + \int_0^{N\epsilon} dx \ \frac{\sin^2 x}{x^2} \right) = |\epsilon| \int_0^{N\epsilon} dx \ \frac{1-\cos x}{x^2}
\ee
Since we look at particles around $x=0$ we have $\epsilon = \frac{\pi}{2} \frac{n}{N}$, which finally gives
\be
\mathcal{D}_{b,N}(x,y) \simeq \frac{\pi}{2} \frac{n}{N} \int_0^{\frac{n \pi}{2}} dx \ \frac{1-\cos x}{x^2} \xrightarrow[n\to+\infty]{} \frac{\pi^2}{4} \frac{n}{N}
\ee
And we get back \eqref{gapvariance_largeN} for $n \gg 1$. For $n=O(1)$ this result is not better than the one above, but it shows that, for the computation of the gap variance for $n=O(1)$, the sum is dominated by the terms with $k$ of order $N$, and this is why neglecting the $k/N$ term gives very bad results in this case. Qualitatively, this is due to the fact that $u_k(y)$ is a polynomial of order $k-1$, and so its oscillations become bigger and bigger as $k$ increases, so that the difference $u_k(y_i)-u_k(y_{i+1})$ becomes larger and larger. It is not even sure that going to higher order in $1/N$ will give significantly better results, in particular for $n=1$. One probably needs to find a different method or to solve the recursion exactly.

\section{Equal time correlations for the DBM}\label{app:dbm}

\begin{figure}
    \centering
    \includegraphics [width=0.325\linewidth,trim={0cm 0 1cm 1cm},clip]{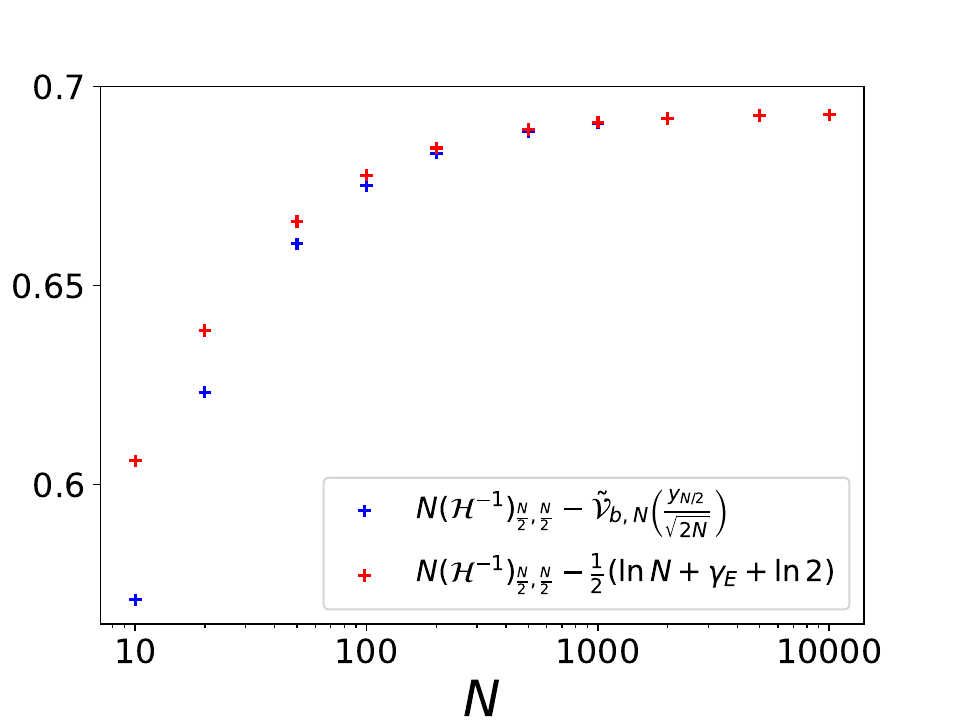}
    \caption{Difference between the scaled variance of the central particle in the DBM computed from the Hessian and the function prediction $\mathcal{\tilde V}_{b,N}\left(\frac{y_{N/2}}{\sqrt{2N}}\right)$ (blue) as well as with its approximation to order $o(1)$ (red). The difference converges to a constant, while the leading behaviour in each of the compared terms behaves as $\ln N$.}
    \label{DBM_var_fig}
\end{figure}

We give here some details on the equal time correlations for the position of bulk particles in the case of the DBM. As mentioned in the text, taking $t=t'$ in \eqref{DBMcov_largeN_time} leads to a logarithmically diverging sum. Thus in this case we do not extend the sum to infinity and we write instead

\begin{equation}
\langle \delta x_i \delta x_j \rangle \simeq \frac{T}{\lambda N^2} \mathcal{\tilde C}_{b,N}\left( \frac{x_{{\rm eq},i}}{2\sqrt{g/\lambda}}, \frac{x_{{\rm eq},j}}{2\sqrt{g/\lambda}} \right) \quad {\rm with} \quad \mathcal{\tilde C}_{b,N}(x,y) = \sum_{k=1}^N \frac{1}{k} U_{k-1}(x) U_{k-1}(y) \;,
\label{DBMcov_largeN}
\end{equation}
and for the variance
\begin{equation}
\langle \delta x_i^2 \rangle \simeq \frac{T}{\lambda N^2} \mathcal{\tilde V}_{b,N}\left(\frac{x_{{\rm eq},i}}{2\sqrt{g/\lambda}}\right) \quad {\rm with} \quad \mathcal{\tilde V}_{b,N}(x) = \sum_{k=1}^N \frac{1}{k} U_{k-1}(x)^2 = \frac{1}{1-x^2} \sum_{k=1}^N \frac{\sin^2(k \arccos x)}{k} \;.
\label{DBMvar_largeN}
\end{equation}
Note that in this case the functions $\mathcal{\tilde C}_{b,N}(x,y)$ and $\mathcal{\tilde V}_{b,N}(x)$ depend on $N$ (they are of order $\ln N$ at large $N$). As can be seen in Fig.~\ref{DBM_var_fig}, these results are still correct to leading order, but the leading relative error is now of order $1/N^2$ (i.e. logarithmic w.r.t. the leading order).

In the case of the variance, equation \eqref{DBMvar_largeN} can be easily approximated to leading order, yielding the result \eqref{DBMvar_largeN0} mentioned in the text. Indeed, one has
\begin{equation}
\sum_{k=1}^\infty \frac{(\sin^2(k \theta)-1/2)}{k} = \frac{1}{4} \ln(4(\sin^2(\theta))
\end{equation}
and thus
\begin{equation}
\sum_{k=1}^N \frac{\sin^2(k \arccos x)}{k} = \frac{1}{2} \sum_{k=1}^N \frac{1}{k} + O(1) = \frac{1}{2} \ln N + O(1) 
\end{equation}
which leads to
\begin{equation}
\langle \delta x_i^2 \rangle \simeq \frac{T}{\lambda N^2} \frac{\ln N}{2(1-(\frac{x_{{\rm eq},i}}{2\sqrt{g/\lambda}})^2)} + O(N^{-2}) \;.
\label{DBMvar_largeN_leading}
\end{equation} 

{Note that the presence of the $O(N^{-2})$ correction can be understood, e.g. by assuming that neglecting the $k/N$ term in the recursion is a good approximation for $k<\alpha_c N$ ($0<\alpha_c<1$) which leads to such a correction.}
\\




{\bf Remark:} One can show that for any $\beta$ and any $N$ (see e.g. 6.18 in \cite{bouchaud_book}, see \cite{VW22} for generalizations) the following expectation
value 
\be 
\langle \prod_i (z - x_i) \rangle = \sqrt{\frac{2 g}{\lambda N}} H_N \left(z \sqrt{\frac{\lambda N}{2 g}}\right)
\ee 
is independent of $\beta$. This implies that the expectation values of all the elementary symmetric polynomials,
$\langle e_p(x_1,\dots,x_N) \rangle$,
are independent of $\beta$.
This implies a series of identities for the moments of $\delta x_i = x_i - x_i^{\rm eq}$ such as
\bea   
&& \sum_i \langle \delta x_i \rangle = 0 \\
&& 2 \sum_{i \neq j} x_i^{\rm eq} \langle \delta x_j \rangle + \sum_{i \neq j} \langle \delta x_i \delta x_j \rangle = 0
\eea

\section{Generating function for the recursion relation \eqref{recursion_exact}} \label{app:expansion}

Let us start from the recurrence relation \eqref{recursion_rescaled} for the rescaled series $v_k(r_i)$ and write down the differential equation satisfied by its generating function (in this section we drop the index $i$ in $r_i$ and the $r_i$ dependence in $v(r_i)$)
\begin{equation}
    G_N(z) = \sum_{k=1}^N v_k z^k \;.
\end{equation}
Multiplying \eqref{recursion_rescaled} by $x^{k+2}$, summing over $k$ for $k=1$ to $N$ and using $v_1=1$, $v_2=2r$ and $v_{N+1}=v_{N+2}=0$, we obtain
\begin{equation} \label{GN_equation}
    \frac{z^3}{N} \partial_z G_N(z) + (2rz - z^2 -1) G_N(z) + z = 0.
\end{equation}
This differential equation can be solved exactly, yielding
\begin{equation}
G_N(z) = N z^N e^{N\left(-\frac{1}{2z^2}+\frac{2r}{z}\right)} \int_a^{1/z} dt \ t^N e^{N\left(\frac{t^2}{2}-2rt\right)}
\end{equation}
with $a$ some constant. Alternatively, we can compute an expansion of the solution in $1/N$, $v_k = v_k^0+\frac{1}{N} v_k^1+...$ Omitting the $1/N$ term leads to
\begin{equation} \label{GN_order0}
    G_N(z) = \frac{z}{1+z^2-2rz} + O \left( \frac{1}{N} \right)
\end{equation}




from which we can get back the solution $v_k^0(r) = U_{k-1}(r)$, starting from
\begin{eqnarray}
    G_N^0(z) &=& z \sum_{k=0}^{\infty} z^k (2r-z)^k = z \sum_{k=0}^{\infty} z^k \sum_{i=0}^k \binom{k}{i} (-z)^i (2r)^{k-i} = z \sum_{k=0}^{\infty} (2rz)^k \sum_{i=0}^k \binom{k}{i} \left( -\frac{z}{2r} \right)^i \\
    &=& \sum_{k=0}^{\infty} z^{k+1} \sum_{i=0}^{\lfloor n/2 \rfloor} (-1)^i \binom{k-i}{i} (2r)^{k-2i} = \sum_{k=0}^{\infty} z^{k+1} U_k(r)
    \end{eqnarray}
(see \cite{ChebyWiki} for the last equality). Computing the derivative of \eqref{GN_order0} and feeding back into \eqref{GN_equation}, we find
\begin{equation} \label{GN_order1}
    G_N(z) = \frac{z}{1+z^2-2rz} \left( 1 + \frac{z^2}{N} \frac{(1-z)(1+2r-z)}{(1+z^2-2rz)^2} \right) + O \left( \frac{1}{N^2} \right)
\end{equation}
from which we get, using Cauchy products
\begin{eqnarray}
    G_N^1(z) &=& z^3 (1-z) (1+2r-z) \left( \sum_{k=0}^{\infty} z^k U_{k}(r) \right)^3 \\
&=& z^3 (1+2r - 2(1+r)z + z^2) \sum_{k=0}^{\infty} z^k \sum_{0 \leq i \leq j \leq k} U_i(r) U_{j-i}(r) U_{k-j}(r) \\
    &=& \sum_{k=3}^{\infty} z^k [ (1+2r) S_{k-3}(r) - 2(1+r) S_{k-4}(r) + S_{k-5}(r) ]
\end{eqnarray}
and thus
\begin{equation}
v_k^1(r) = (1+2r) S_{k-3}(r) - 2(1+r) S_{k-4}(r) + S_{k-5}(r) \quad , \quad S_k(r) = \sum_{0 \leq i \leq j \leq k} U_i(r) U_{j-i}(r) U_{k-j}(r)
\end{equation}
with $S_k(r)=0$ if $k<0$. This result could in theory be used to compute $1/N$ corrections to \eqref{cov_largeN} and \eqref{var_largeN}.

\section{Covariance at the edge for the (passive) Dyson Brownian motion from the stochastic Airy operator: large $\beta$ expansion} \label{app:airy_op}

It is useful to reconsider the stationary measure of the DBM and study the fluctuations of the particle positions
at the edge of the gas, by the method of the stochastic Airy operator (SAO), see
\cite{EdelmanSutton,RamirezRiderVirag,Virag}. In a second stage one considers the large $\beta$ limit,
which, in that framework, amounts to perform standard perturbation theory in quantum mechanics.

At the edge, the position of particle $i$, for $i \geq 1$, can be rescaled as
\be 
x_i = \frac{x_e}{2} (2 + {\sf a}_i N^{-2/3}) 
\ee 
{ where $x_e$ is the position of the edge, here $x_e=2 \sqrt{g/\lambda}$ 
and in the limit of large $N$ the ${\sf a}_i$ form the Airy process with Dyson parameter $\beta=2 g/T$. We now 
use the fact that the Airy process ${\sf a}_i$ has the same statistics as ${\sf a}_i = - \epsilon_i$ 
where the $\epsilon_i$ are the eigenvalues of the SAO defined as 
\be 
{\cal H}_{\rm SAO} = - \partial_y^2 + y + \frac{2}{\sqrt{\beta}} w(y)
\ee 
for $y>0$ and Dirichlet boundary conditions at $y=0$. Here $w(y)$ is a standard Gaussian white noise in $y$,
i.e. with correlations $\langle w(y) w(y') \rangle = \delta(y-y')$. 
For $\beta=+\infty$ the normalized eigenfunctions
of ${\cal H}_{\rm SAO}$ are given by
\be 
\psi_i(x) = \frac{{\rm Ai}(x+a_i)}{{\rm Ai}'(a_i)} 
\ee 
where the $a_i$ is the $i$-th zero of the Airy function. One can thus perform standard perturbation
expansion to obtain, to second order 
\bea \label{perturbation} 
- {\sf a}_i = \epsilon_i = - a_i + \frac{2}{\sqrt{\beta}} \int_0^{+\infty} dy \left(\frac{{\rm Ai}(y+a_i)}{{\rm Ai}'(a_i)}\right)^2  w(y) 
+ { \frac{4}{\beta} \sum_{k \neq i} \frac{ (\int_0^{+\infty} dy {\rm Ai}(y+a_i) {\rm Ai}(y+a_k) w(y))^2}{{\rm Ai}'(a_i)^2 {\rm Ai}'(a_k)^2 (a_k-a_i)}
+ O(w^3) }
\eea 
Let us first obtain the covariances to lowest order in $1/\beta$. For this one does not need the terms $O(w^2)$. One obtains
\be 
\langle x_i x_j \rangle_c = \frac{x_e^2}{4} N^{-4/3} \langle {\sf a}_i {\sf a}_j \rangle_c  
\ee 
and one has
\be 
\langle {\sf a}_i {\sf a}_j \rangle_c = \frac{4}{\beta} \int_0^{+\infty} dy \left(\frac{{\rm Ai}(y+a_i)}{{\rm Ai}'(a_i)}\right)^2 
\left(\frac{{\rm Ai}(y+a_j)}{{\rm Ai}'(a_j)}\right)^2 
\ee 
Hence we find, using $x_e=2 \sqrt{g/\lambda}$ and $\beta = 2 g/T$
\be \label{DBM_cov_alternative}
\langle x_i x_j \rangle_c = \frac{2 T}{\lambda  N^{4/3}} \int_0^{+\infty} dy \left(\frac{{\rm Ai}(y+a_i)}{{\rm Ai}'(a_i)}\right)^2 
\left(\frac{{\rm Ai}(y+a_j)}{{\rm Ai}'(a_j)}\right)^2 
\ee 
If we compare with formula \eqref{DBMcov_edge_integral}, i.e. 
$\langle x_i x_j \rangle_c = \frac{T}{\lambda  N^{4/3}}  {\cal \tilde C}_e(a_i,a_j)$
for the two formula to agree we need the following identity, which we have checked numerically on some values of $i$ and $j$ using Mathematica
\begin{equation}
    2\int_0^{+\infty} dy \left(\frac{{\rm Ai}(a_i+y)}{{\rm Ai}'(a_i)}\right)^2 \left(\frac{{\rm Ai}(a_j+y)}{{\rm Ai}'(a_j)}\right)^2 = \frac{1}{\Ai'(a_i) \Ai'(a_j)} \int_0^{+\infty} dy \ \frac{\Ai(a_i + y)\Ai(a_j + y)}{y}
\end{equation}
It would be interesting to prove this identity. 
}
The formula \eqref{perturbation} allows to compute easily 
the first correction to the mean. One finds
\bea \label{mean_perturbatio} 
\langle - {\sf a}_i \rangle =  - a_i 
+ \frac{4}{\beta} \sum_{k \neq i} \frac{\int_0^{+\infty} dy {\rm Ai}(y+a_i)^2 {\rm Ai}(y+a_k)^2}{{\rm Ai}'(a_i)^2 {\rm Ai}'(a_k)^2 (a_k-a_i)}
\eea 
Leading to
\be 
\langle \delta x_i \rangle = \frac{2 x_e}{ \beta N^{2/3}} \sum_{k \neq i} 
\frac{\int_0^{+\infty} dy {\rm Ai}(y+a_i)^2 {\rm Ai}(y+a_k)^2}{{\rm Ai}'(a_i)^2 {\rm Ai}'(a_k)^2 (a_i-a_k)}
\ee 
For $i=1$ one finds $\langle \delta x_1 \rangle>0$. Hence the size of the gas increases as $T$ increases (eventually
at high $T \sim N$ the support of the density extends to infinity \cite{BouchaudGuionnet}). Since
${\sf a}_1$ is distributed according to the $\beta$-Tracy Widom distribution
\cite{EdelmanSutton,RamirezRiderVirag,Virag}, the quantities
$\langle \delta x_1 \rangle$ and $\langle \delta x_1^2 \rangle^c$ can be related to the 
mean and variance of that distribution. 

Finally, to obtain the covariance to the next order $O(T^2)$ one would need to push the
perturbation theory to order $O(w^3)$.

\section{High temperature limit of the Calogero-Moser model}
\label{app_CM_highT}

In this appendix we study, in a more general setting, the high temperature limit of the CM model (denoting here $x_i$ and $z_i$ the variables $\tilde X_i$ and $\zeta_i$ used in the text). Consider the joint PDF 
\be 
P(x_1,\dots,x_N) = N! \, p(x_1) \dots p(x_N) \theta(x_1>x_2>...>x_N) 
\ee 
where $\int_{-\infty}^{+\infty} dx \, p(x)=1$. For $p(x)=\frac{e^{-\frac{x^2}{2}}}{\sqrt{2\pi}}$ this corresponds to the joint distribution of the rescaled positions of particles in the CM model in the large temperature limit. This PDF can be seen as drawing $N$ i.i.d random variables from the distribution $p(x)$ and ordering them such that $x_1>...>x_N$. Then the marginal distribution of $x_i$ (for $i=1,\dots,N$) is given by
(see e.g. \cite{sm14})
\be 
q_i(x) = \frac{N!}{(i-1)! (N-i)!} p(x) \left( \int_{x}^{+\infty} dy \,p(y) \right)^{i-1} 
\left( \int_{-\infty}^x dy \,p(y) \right)^{N-i} \;.
\ee 
Similarly, for $1 \leq i<j \leq N$ the two-point marginal is
\bea  
&& q_{ij}(x,z) =  \frac{N!}{(i-1)! (j-i-1)! (N-j)!}  p(x) p(z) \theta(z<x) 
\left( \int_{x}^{+\infty} dy \,p(y) \right)^{i-1}
\left( \int_{z}^{x} dy \,p(y) \right)^{j-i-1}
\left( \int_{-\infty}^z dy \,p(y) \right)^{N-j}
\eea 
The cumulants of $q_i(x)$ can be obtained in the large $N$ limit, for $i=O(N)$, by computing the large deviation
generating function
\bea
&&\frac{1}{N}\log \langle e^{N \lambda x_i} \rangle \simeq \max_x( G_u(x) + \lambda x) - ((1-u) \log(1-u) + u \log u) \\
&&G_u(x) = u \log Q(x) + (1-u) \log(1-Q(x)) \quad , \quad
Q(x)= \int_x^{+\infty} dy \ p(y)  \quad , \quad u=i/N \;.
\eea
In the Gaussian case one has 
\be 
Q(x)= \frac{1}{2} {\rm erfc}\left(\frac{x}{\sqrt{2}}\right)
\ee
Let us define 
\be 
\phi_u(\lambda)= \max_x( G_u(x) + \lambda x) = G_u(x_\lambda) + \lambda x_\lambda \quad , \quad G_u'(x_\lambda) = - \lambda \;.
\ee 
One has
\be 
\phi_u'(\lambda) = x_\lambda \quad , \quad \phi_u''(\lambda) = - \frac{1}{G_u''(x_\lambda)} \;.
\ee
Let us denote $x^*=x_0$, which is given by the equation $G'(x^*)=0$. We obtain
\be \label{avgHighT_bulk}
\langle x_i \rangle = \phi_u'(0) = x^* \quad , \quad x^* = Q^{-1}(u) \quad , \quad u = i/N
\ee
and for the second cumulant 
\be  \label{varHighT_bulk}
N \langle x_i^2 \rangle_c = \phi_u''(0) = \frac{u(1-u)}{[Q'(Q^{-1}(u))]^2} \quad , \quad u = i/N \;.
\ee
These results are in agreement with \cite{TheseBertrand} (up to a factor $2\pi$ which seems to be missing there).
The higher cumulants are given by $N^{p-1} \langle x_i^k \rangle_c=\phi_u^{(p)}(0)$, hence
the marginal distribution of the rescaled position $\tilde x_i = N^{1/2} (x_i - \langle x_i \rangle)$
of a bulk particle is Gaussian at large $N$.

Similarly one can obtain the two-point covariance for two particles in the bulk separated by a distance of order $N$ by computing, for $1 \leq i<j \leq N$, with $i$ and $j=O(N)$ and $j-i=O(N)$,
\bea
&&\frac{1}{N} \ln e^{N(\lambda_i x_i + \lambda_j x_j)} = \max_{x,z|x>z} [H_{u,v}(x,z) + \lambda_i x + \lambda_j z] - c_{u,v} \\
&& H_{u,v}(x,z) = u \ln Q(x) +(v-u) \ln (Q(z)-Q(x))+(1-v)\ln(1-Q(z))
\label{highT_twopoints1}
\eea
where $u=i/N < v=j/N$, and $c_{u,v}=u \log u + (v-u) \log(v-u) + (1-v) \log(1-v)$. We introduce
\bea
&&\psi_{u,v}(\lambda_i,\lambda_j) = \max_{x,z|x>z} [H_{u,v}(x,z) + \lambda_i x + \lambda_j z] = H_{u,v}(x_{\lambda_i,\lambda_j},z_{\lambda_i,\lambda_j}) + \lambda_i x_{\lambda_i,\lambda_j} + \lambda_j z_{\lambda_i,\lambda_j} \\
&&\partial_x H_{u,v}(x_{\lambda_i, \lambda_j}, z_{\lambda_i, \lambda_j}) = - \lambda_i \quad , \quad \partial_z H_{u,v}(x_{\lambda_i, \lambda_j}, z_{\lambda_i, \lambda_j}) = - \lambda_j \label{22} 
\eea
 We find in particular
\beq
Q(x_{0,0})=u \quad , \quad Q(z_{0,0})=v
\eeq
We will denote $x^*=x_{0,0}$ and $z^*=z_{0,0}$. One has
\bea
&&\partial_{\lambda_i} \psi_{u,v}(\lambda_i, \lambda_j) = x_{\lambda_i, \lambda_j} \quad , \quad \partial_{\lambda_j} \psi_{u,v}(\lambda_i, \lambda_j) = z_{\lambda_i, \lambda_j}
\eea
from which one can show that
\be  
\partial_{\lambda_i} \partial_{\lambda_j} \psi_{u,v}(0,0) = 
\partial_{\lambda_j} x_{\lambda_i, \lambda_j} \Bigr|_{\substack{\lambda_i=0,\lambda_j=0}} =
\frac{\partial_x \partial_z H_{u,v}}{\partial^2_x H_{u,v} \partial^2_z H_{u,v} - (\partial_x \partial_z H_{u,v})^2} \Bigr|_{\substack{x^*,z^*}} = \frac{u (1-v)}{Q'(x^*)Q'(z^*)}
\ee 
where to obtain the second equality one takes a derivative w.r.t. $\lambda_j$ of both equations in \eqref{22} 
and eliminate $\partial_{\lambda_j} z_{\lambda_i, \lambda_j}$. Inserting the explicit form of 
$H_{u,v}(x,z)$ given in \eqref{highT_twopoints1} we finally obtain
\be 
N \langle x_i x_j \rangle_c = \frac{u(1-v)}{Q'(x^*)Q'(z^*)} \quad , \quad Q(x^*) = u = \frac{i}{N} \quad , \quad Q(z^*) = v = \frac{j}{N}
\label{covHighT_bulk}
\ee
which yields back \eqref{varHighT_bulk} when $j-i \ll N$.

We now focus on the right edge regime, i.e. $i=O(1)$. In this regime one can apply standard results from extreme value theory 
\cite{galambos,nagaraja}.
We restrict ourselves to the Gaussian case $p(x)=e^{-x^2/2}/\sqrt{2 \pi}$, which falls into the Gumbel universality class. One has the standard result at large $N$
\be 
x_j = \sqrt{2 \log N} (1 + \frac{z_j +c_N}{2 \log N} + \dots ) \quad , \quad c_N = - \log( \sqrt{4 \pi \log N} ) \;.
\ee 
The general case for the Gumbel class is obtained simply from the change of variable $N Q(x) \simeq e^{-z}$. The JPDF of the $k$
largest $z_j$'s, denoted here $w_k$, is then universal and given by \cite{galambos,nagaraja,sm14}
\be 
w_k(z_1,\dots,z_k) = \theta(z_k<\dots<z_1) \, e^{- \sum_{j=1}^k z_j } e^{- e^{-z_k}} 
\ee 
which is normalized to unity. The marginal of $z_k$ is then
\be 
q_k(z) = \frac{1}{(k-1)!} e^{- k z - e^{- z}} \;.
\ee 
Note that this JPDF can be rewritten as
\be \label{independent} 
w_k(z_1,\dots,z_k) = \theta(z_k<\dots<z_1) \, \prod_{\ell=1}^{k-1} \ell e^{- \ell (z_{\ell}-z_{\ell+1})}
\times \frac{1}{(k-1)!} e^{- k z_k - e^{-z_k}} \;.
\ee 
Hence to generate the $k$ largest points, one first chooses $z_k$ and then the successive 
gaps as independent exponentially distributed variables, with distinct parameters.
It is then easy to compute the generating function
\be 
\langle e^{ \lambda_1 z_1 + \dots + \lambda_k z_k} \rangle_k = \frac{\Gamma(k - \lambda_1 - \dots - \lambda_k) }{(1-\lambda_1)(2-\lambda_1-\lambda_2) \dots (k-1-\lambda_1-\dots - \lambda_{k-1}) } 
\ee 
where $\langle \dots \rangle_k$ denotes an average w.r.t. $w_k$ in (\ref{independent}). From that formula
all joint moments and cumulants can be obtained. For instance one obtains that for $j\leq k$ 
\be 
\langle e^{\lambda z_j} \rangle_k = \frac{\Gamma(j  - \lambda) }{\Gamma(j)}  
\ee 
independently of $k$ as required. 
The two-point generating function with $j<k$ is then
\be 
\langle e^{ \lambda_j z_j + \lambda_k z_k} \rangle_k = \frac{\Gamma(j- \lambda_j)}{\Gamma(j)} 
\frac{\Gamma(k - \lambda_j  - \lambda_k) }{\Gamma(k- \lambda_j) } 
\ee 
from which the joint two point cumulants are obtained. For $\lambda_j=-\lambda_k=\lambda$ one obtains
\be \label{avgHighT_edge}
\langle z_k \rangle = - \psi_0(k) \quad , \quad \langle z_k^2 \rangle_c = \psi_1(k) \;,
\ee 
where $\psi_0(x) = \Gamma'(x)/\Gamma(x)$ and $\psi_1(x) = \psi_0'(x)$ are the digamma and trigamma functions respectively. The second cumulants at distinct points are given, for $j < k$, by (see also \cite{SM_unpub})
\be \label{varHighT_edge}
\langle z_j z_k  \rangle_c = \psi_1(k) = \langle z_k^2 \rangle_c
\ee 
which is compatible with $z_k$ and $z_{j}-z_k$ being independent variables (see the remark above). 

One can check that these results correctly match the results for the bulk at the boundary between the two regimes. Indeed for $k \ll N$ the first cumulant in the bulk \eqref{avgHighT_bulk} becomes
\beq
\langle x_k \rangle = x^*_k = Q^{-1} \left( \frac{k}{N} \right) \simeq \sqrt{2 \ln N} \left( 1 - \frac{\ln \sqrt{4 \pi \ln N} + \ln k}{2 \ln N} \right) \;,
\eeq
which matches the edge result \eqref{avgHighT_edge} for $k \gg 1$ (using that at large $k$, $\psi_0(k) = \ln(k) + O(1/k)$). Using the same asymptotic expression for $x^*_k$, we get for the covariance in the bulk \eqref{covHighT_bulk} with $k \ll N$ and $j \ll N$ with $j<k$
\beq
\langle x_j x_k \rangle_c \simeq \frac{j/N}{ Q'(x_k^*) Q'(x_j^*)}
\simeq \frac{1}{2k\ln N} \label{ee} 
\eeq
where we have used $Q'(x_k^*) \simeq \frac{k}{N} \sqrt{2 \log N}$. This expression (\ref{ee}) 
is the same as the one obtained from the edge expression \eqref{varHighT_edge} in the limit $k \gg 1$ (using the fact that at large $k$ $\psi_1(k)\simeq 1/k$).

Finally the distribution $\rho_k(d)$ of the gap between two successive particles $d=d_k=x_k-x_{k+1}$ in the large $N$ limit can be obtained (see e.g. \cite{TheseBertrand}). In the bulk region one has
\beq \label{gapb}
\rho_k(d) \simeq N Q'(x^*_k) \ e^{-N Q'(x^*_k) d}
\eeq
which is a universal result valid for any $p(x)$, while near the edge one has for the Gaussian case ({using the expression of $Q'(x^*_k)$ given below \eqref{ee}}, see also \cite{sm14}) 
\beq \label{gape}
\rho_k(d) \simeq k \sqrt{2 \ln N} \,  e^{-k \sqrt{2 \ln N} d} \;.
\eeq
One can again check that the two expressions (\ref{gapb}) and (\ref{gape}) match for $1 \ll k \ll N$.

\begin{figure}
    \centering
    \includegraphics[width=0.325\linewidth,trim={0cm 0 1cm 1cm},clip]{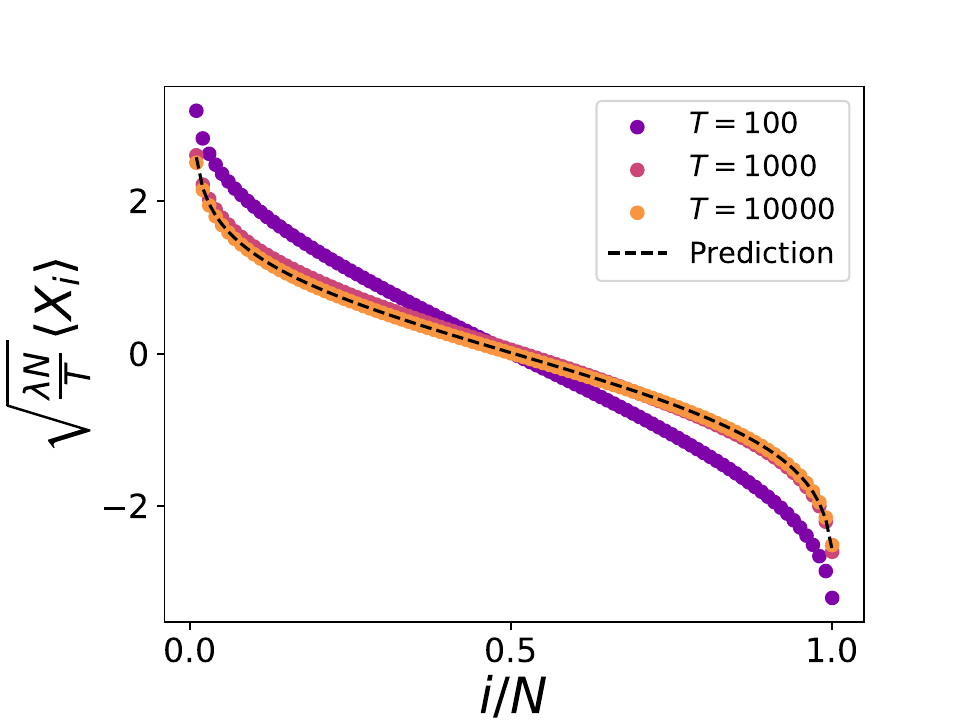}
    \includegraphics[width=0.325\linewidth,trim={0cm 0 1cm 1cm},clip]{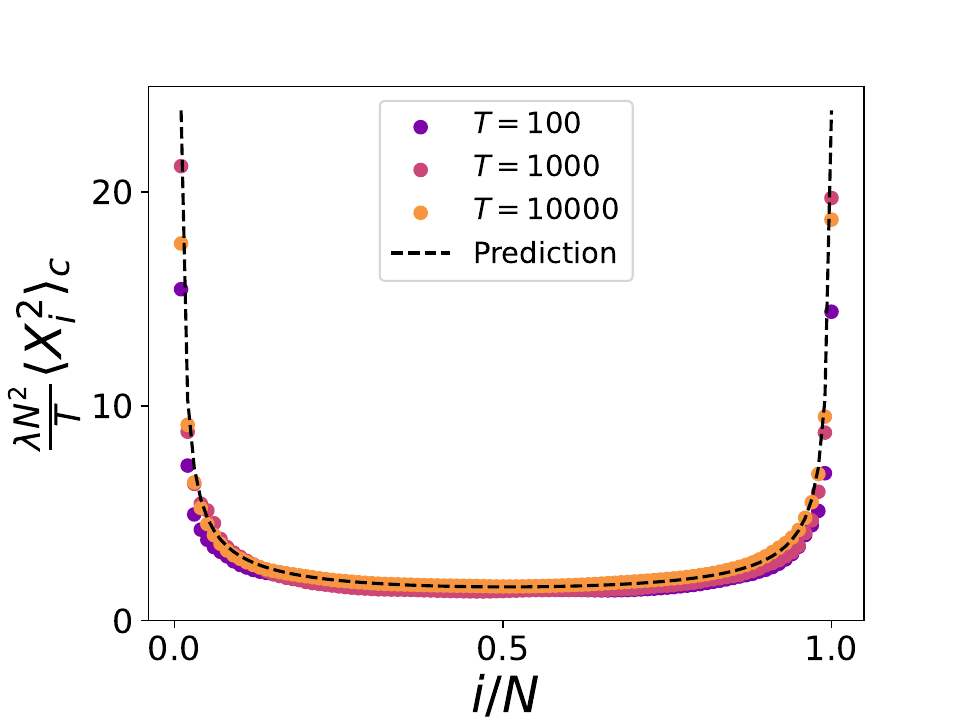}
    \caption{Rescaled average (left) and variance (right) of the position of particle $i$ as a function of $i/N$ at high temperatures. The dashed black lines show the infinite temperature predictions \eqref{CM_highT}. 
    }
    \label{CM_largeT}
\end{figure}

The results are compatible with simulations (see Fig. \ref{CM_largeT}, where we used $u=\frac{i-\frac{1}{2}}{N}$ for symmetry since $i=1,...,N$) and with an existing result for the variance of the central particle (see \cite{BarkaiSilbey2009}).


\end{document}